\documentclass[11pt,a4paper]{article}
\usepackage{jheppub}

\usepackage{tikz}
\def\checkmark{\tikz\draw[scale=0.4,fill=black](0,.35) -- (.25,0) -- (1,.7) -- (.25,.15) -- cycle;}
\def\halfcheckmark{\tikz\draw[scale=0.4,fill=black](0,.35) -- (.25,0) -- (1,.7) -- (.25,.15) -- cycle (0.75,0.2) -- (0.77,0.2) -- (0.6,0.7) -- cycle;}

\usepackage{array}
\usepackage{multirow}
\usepackage{subcaption}
\usepackage{bbm}
\usepackage{placeins}
\usepackage{bm}
\usepackage{pifont}
\usepackage{setspace}
\newcommand{\pl}{\;\;\,}
\DeclareMathOperator\sign{sign}
\DeclareMathOperator\arctanh{arctanh}
\DeclareMathOperator\BR{BR}

\begin{document}

\title{Prospects for measuring quark polarization and\\ spin correlations in $b\bar b$ and $c\bar c$ samples at the LHC}

\author{Yevgeny Kats}
\author{and David Uzan}
\affiliation{Department of Physics, Ben-Gurion University, Beer-Sheva 8410501, Israel}
\emailAdd{katsye@bgu.ac.il}
\emailAdd{daviduz@post.bgu.ac.il}

\abstract{
Polarization and spin correlations have been studied in detail for top quarks at the LHC, but have been explored very little for the other flavors of quarks. In this paper we consider the processes $pp\to q\bar{q}$ with $q = b$, $c$ or $s$. Utilizing the partial preservation of the quark's spin information in baryons in the jet produced by the quark, we examine possible analysis strategies for ATLAS and CMS to measure the quark polarization and spin correlations. We find polarization measurements for the $b$ and $c$ quarks to be feasible, even with the currently available datasets. Spin correlation measurements for $b\bar b$ are possible using the CMS Run~2 parked data, while such measurements for $c\bar c$ will become possible with higher integrated luminosity. For the $s$ quark, we find the measurements to be challenging with the standard triggers. We also provide leading-order QCD predictions for the polarization and spin correlations expected in the $b\bar b$ and $c\bar c$ samples with the cuts envisioned for the above analyses. Apart from establishing experimentally the existence of spin correlations in $b\bar b$ and $c\bar c$ systems produced in $pp$ collisions, the proposed measurements can provide new information on the polarization transfer from quarks to baryons and might even be sensitive to physics beyond the Standard Model.}
		
\maketitle
	
\section{Introduction}
\label{sec:intro}

Quark polarization and spin correlations are properties that have been extensively researched for top quarks at the LHC, both theoretically (e.g., refs.~\cite{Bernreuther:2001rq,Bernreuther:2004jv,Mahlon:2010gw,Baumgart:2011wk,Baumgart:2012ay,Baumgart:2013yra,Bernreuther:2013aga,Bernreuther:2015yna,Behring:2019iiv,Afik:2020onf,Severi:2022qjy}) and experimentally (e.g., refs.~\cite{ATLAS:2014abv,ATLAS:2016bac,CMS:2016piu,CMS:2019nrx,ATLAS:2019zrq,ATLAS:ttbar-entanglement}), but have not yet been explored much for the $b$, $c$, or $s$ quarks. Measurements of these quantities can provide interesting information on both Standard Model (SM) and Beyond the Standard Model (BSM) interactions. There exist proposals for methods to measure quark polarizations in samples of $pp\to t\bar{t}$, in which $b$ quarks are available from the $t\to W^+b$ decay, and $c$ and $s$ quarks from the subsequent decay $W^+\to \bar{s}c$~\cite{Galanti:2015pqa,Kats:2015cna,Kats:2023gul}, and using samples of $pp\to W^-c$ for $c$ quarks~\cite{Kats:2015zth}. The $b$, $c$, and $s$ quarks in these processes are expected to be highly polarized due to the electroweak interaction. On the other hand, in the current paper we want to examine quark-antiquark pair production, $pp\to q\bar{q}$, where $q$ can be either $b$, $c$ or $s$. These processes are dominated by QCD interactions, which produce the quarks unpolarized at the leading order. However, sizable spin correlations are expected, as we will quantify, similar to the top-quark case.

Unlike the top quark, whose spin information can be obtained from the angular distributions of its decay products, the $b$, $c$ and $s$ quarks are only observed as jets of hadrons, making it more challenging to obtain the spin information of the original quarks. It can nevertheless be done by measuring the polarization and spin correlations of baryons produced from these quarks. This approach was originally proposed for measuring the longitudinal polarization of the heavy quarks ($b$ and $c$) produced in $Z$ decays at LEP~\cite{Mannel:1991bs,Ball:1992fw,Falk:1993rf}. Such measurements were subsequently performed for the $b$ quark using $\Lambda_b$ baryons~\cite{ALEPH:1995aqx,OPAL:1998wmk,DELPHI:1999hkl}, confirming the expectation of a sizable polarization transfer from the $b$ quark to the $\Lambda_b$. Analogous measurements have shown that the $s$-quark longitudinal polarization is preserved in $\Lambda$ baryons that carry a significant fraction of the jet momentum~\cite{ALEPH:1996oew,ALEPH:1997an,OPAL:1997oem}. This method to access quark spin information has been analyzed in the context of the LHC in refs.~\cite{Galanti:2015pqa,Kats:2015cna,Kats:2023gul,Kats:2015zth}, where it was shown that a number of interesting measurements of longitudinal polarization of quarks produced in top-quark decays are possible even with the statistics of Run~2.\footnote{Run~2 is the period of LHC activity between 2015--2018, during which both ATLAS and CMS recorded about $140$~fb$^{-1}$ of $pp$ collisions at $\sqrt s = 13$~TeV. Since 2022, Run~3 is underway, with about $70$~fb$^{-1}$ of $pp$ collisions at $\sqrt s = 13.6$~TeV recorded to date.} In addition, attempts to measure the transverse polarization of $\Lambda_b$ in inclusive QCD samples, which is expected to be small, have been reported by LHCb~\cite{LHCb:2013hzx,LHCb:2020iux} and CMS~\cite{CMS:2018wjk}. When moving to spin correlations, the cost in statistics increases significantly since the prices for the fragmentation to baryons, the branching ratios of the useful decays, and the reconstruction efficiency, are squared. It is therefore an example of analyses that will benefit from the increase in statistics offered by the high-luminosity phase of the LHC.
	
The spin correlation measurements will allow quantifying the effect of the polarization transfer from quarks to baryons for longitudinal polarization (cross-checking the information that would presumably be obtained even earlier in the analyses proposed in refs.~\cite{Galanti:2015pqa,Kats:2015cna}) as well as transverse polarization. This will provide certain coarse-grained information about the spin-dependent fragmentation functions~\cite{Chen:1994ar,Adamov:2000is} of the $b$ and $c$ quarks hadronizing to the $\Lambda_b$ and $\Lambda_c$ baryons, respectively. Polarization and spin correlation measurements can also be sensitive to BSM contributions to $b\bar b$ or $c\bar c$ production. Similar ideas apply to $s\bar s$, but we will find the corresponding measurements to be challenging.

We will consider both the Run~2 and the High Luminosity LHC (HL-LHC) datasets of ATLAS and CMS, including the CMS parked $b$-hadron dataset~\cite{CMS-DP-2019-043,Bainbridge:2020pgi,CMS:2024syx}. The goal of the current paper is to do a broad survey of the possible analyses, considering a variety of baryon decay modes and selection schemes. We will therefore restrict ourselves to rough estimates of the expected sensitivity in each case, leaving more detailed simulations of individual analyses and the consideration of systematic uncertainties to future work. We also leave to future work the exploration of similar opportunities in LHCb. While limited in the integrated luminosity and acceptance, the LHCb detectors offer superior tracking and particle identification capabilities. It is therefore plausible that a complementary set of analyses, for low-$p_T$ quarks, will be possible in LHCb. It should be noted, however, that the assumption of the factorization between the quark production and its hadronization can break down at low $p_T$, making the result interpretation difficult.

The rest of the paper is organized as follows. Section~\ref{sec: polarization retention in baryons} reviews the essentials of quark polarization retention in baryons. Section~\ref{sec:decay chains} provides details on baryon production and discusses the baryon decay modes that will be relevant to us. Section~\ref{sec:spin correlations formalism} reviews the formalism for the description of polarization and spin correlations and presents the angular distributions through which these quantities can be measured. In section~\ref{sub:bb cc spin corr} we simulate the polarization and spin correlations for $b\bar b$ and $c\bar c$ expected in QCD after validating our simulation on $t\bar t$. In section~\ref{Sec: bcs}, we describe the various possible analysis channels in detail, discuss the most important backgrounds in each case, and estimate the expected sample purity and measurement precision. We summarize the conclusions in section~\ref{sec:summ and concl}. Appendix~\ref{app:uncertainty} presents the derivation of formulas for the statistical uncertainties of the polarization and spin correlation measurements.
		
\section{Polarization Retention in Baryons}
\label{sec: polarization retention in baryons}
	
For the heavy quarks, namely $b$ and (to some extent) $c$, the polarization is expected to be preserved through the hadronization (on timescales of order $1/\Lambda_{\rm QCD}$)~\cite{Mannel:1991bs,Ball:1992fw,Falk:1993rf}. This happens since $m_q \gg \Lambda_{\rm QCD}$ implies that the effect on the heavy quark spin via the chromomagnetic dipole moment, which scales as $\mu_q \propto 1/m_q$, is negligible.

If the heavy quark $q$ ($= b$ or $c$) ends up in a $\Lambda_q$ baryon, the baryon polarization is approximately equal to the quark polarization. It is so because the $\Lambda_q$ structure in the framework of the quark model is $qud$ with the $u$ and $d$ forming a spin singlet, so all the spin is on the $q$. If, on the other hand, the $q$ ends up in a $\Sigma_q$ or $\Sigma_q^\ast$ baryon, which are analogous to the $\Lambda_q$ but with the light quarks forming a spin and isospin triplet, the $\Lambda_q$ baryons produced in $\Sigma_q^{(\ast)}\to \Lambda_q \pi$ decays will not carry the same polarization as the original quark~\cite{Falk:1993rf}. The $\Lambda_q$ from these decay channels are hard to distinguish from the direct $\Lambda_q$ production and thus they lower the polarization retention. Due to similar reasons, it is essentially impossible to extract polarization information from meson decays~\cite{Falk:1993rf,Alonso:2021boj}. 

The polarization loss effect due to the contamination of the $\Lambda_q$ sample with $\Sigma_q^{(\ast)}\to \Lambda_q \pi$ decays has been analyzed in refs.~\cite{Falk:1993rf,Galanti:2015pqa} and it was found that the inclusive $\Lambda_q$ samples end up carrying between roughly $50\%$ and $80\%$ of the original quark polarization, and this number may differ between the cases of longitudinal and transverse polarization (with respect to the fragmentation axis). In the current paper we will not repeat the discussions on the various approaches to estimating these effects but only parameterize them in terms of the longitudinal and transverse \emph{polarization retention factors}, $r_L$ and $r_T$, defined as
\begin{align}
    r_{\hat{\mathcal{P}}} = \frac{\mathcal{P}(\Lambda_q)}{\mathcal{P}(q)} \;,
\label{eq:spin retention}
\end{align}
where $\mathcal{P}$ denotes polarization and $\hat{\mathcal{P}} = L$ or $T$ denotes whether its direction is longitudinal or transverse with respect to the fragmentation axis. We also note that $r_L$ for bottom and charm quarks can be measured by ATLAS and CMS in their Run~2 $t\bar t$ samples~\cite{Galanti:2015pqa}, and for charm quarks possibly also in $W$+$c$ samples~\cite{Kats:2015zth}. Measurements of both $r_L$ and $r_T$ for the different quark flavors using spin correlations would be one of the goals of the analyses proposed in the current paper.

The above heavy-quark argument does not apply to the $s$ quark, but LEP experiments have shown that $\Lambda$ baryons from $s$ quarks still preserve a large fraction of the polarization~\cite{ALEPH:1996oew,ALEPH:1997an,OPAL:1997oem}.

Describing the polarization transfer from a quark to a baryon in terms of two numbers, $r_L$ and $r_T$, is an approximation. More generally, the polarization transfer will depend on the fraction of the jet momentum carried by the baryon,\footnote{More precisely, $z$ is the component of the baryon momentum along the original quark direction of motion, divided by the quark momentum. See ref.~\cite{Metz:2016swz} for additional details.}
\begin{equation}
    z\simeq\frac{p_T^{\Lambda_q}}{p_T^{\rm jet}} \;,
\label{eq:z definition}
\end{equation}
and is described by the so-called \emph{spin-dependent} (or \emph{polarized}) \emph{fragmentation functions}~\cite{Metz:2016swz,Chen:1994ar,Stratmann:1996hn,deFlorian:1997zj,Adamov:2000is}. These functions vary slowly as a function of the energy scale of the process due to the renormalization group evolution~\cite{Stratmann:1996hn,deFlorian:1997zj}. Characterizing these effects and taking them into account will require high-statistics measurements that could follow up the measurements motivated here and in refs.~\cite{Galanti:2015pqa,Kats:2015cna,Kats:2015zth}. The only case where it is absolutely essential to take the $z$ dependence into account is that of the strange quark. Due to their low mass, soft strange quarks are copiously produced in parton showering. To reduce these contributions, it is essential to focus on $\Lambda$ baryons with high values of $z$. The dependence of the $\Lambda$ polarization on $z$ has been measured in $Z$ decays at LEP~\cite{ALEPH:1996oew,ALEPH:1997an,OPAL:1997oem} confirming the expectation~\cite{Gustafson:1992iq} that the polarization of the initial strange quark is preserved primarily in high-$z$ $\Lambda$ baryons. For example, as has been estimated in ref.~\cite{Kats:2015cna} based on the LEP measurements, roughly $60\%$ of the strange-quark polarization is preserved in $\Lambda$ baryons with $z > 0.3$. Additional information can be obtained using $t\bar t$ samples at the LHC~\cite{Kats:2015cna}. 
	
\section{Baryon Production and Relevant Decay Channels}
\label{sec:decay chains}

\begin{figure}
\centering
\includegraphics[width=0.6\linewidth]{"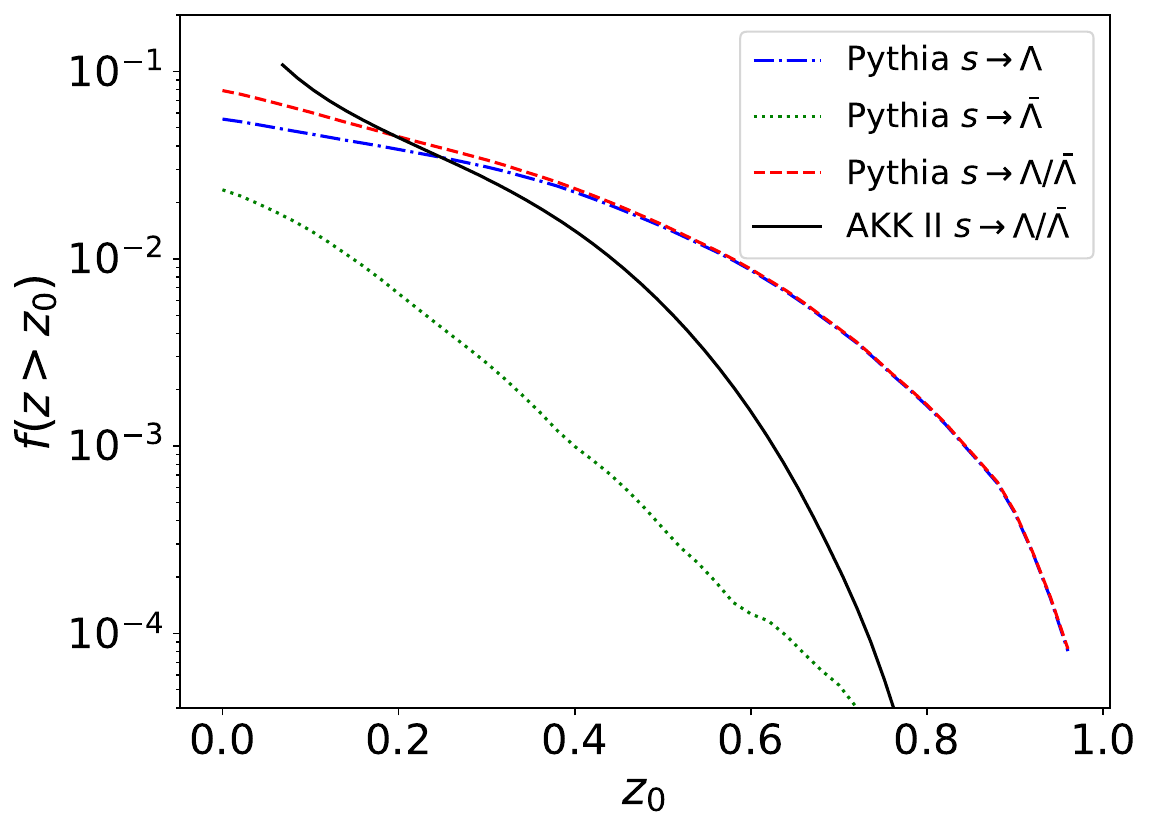"}
\caption{The fragmentation fraction for $z>z_0$, for $s\to\Lambda$, $s\to\bar{\Lambda}$, and $s\to\Lambda/\bar{\Lambda}$, where $\Lambda/\bar{\Lambda}$ denotes that either a $\Lambda$ or $\bar{\Lambda}$ is produced in the $s$-quark hadronization. The Pythia simulation was run at $\sqrt{s}=14$~TeV with $p_T>400$~GeV and $|\eta|<2.5$ cuts on the jets, which were clustered with the anti-$k_t$ jet algorithm with radius $R=0.4$~\cite{Cacciari:2008gp,Cacciari} and matched to the parton-level $s$ quarks. We also show the data-based AKKII results~\cite{Albino}.}
\label{fig:frag_ratio_v2}
\end{figure}

The fragmentation fractions for producing the baryons of interest at high energies are shown in table~\ref{tab:decay scheme}. While the numbers for $b\to\Lambda_b$ and $c\to\Lambda_c$ are available in the literature, for the $\Lambda$ we obtained the number from a Pythia~\cite{Sjostrand:2014zea} simulation. The simulation results are shown in figure~\ref{fig:frag_ratio_v2}, which presents the integrated fragmentation functions
\begin{equation}
    f(s\to\Lambda,\,z>z_0) = \int_{z_0}^1 f_{s\to\Lambda}(z)\,dz
\end{equation}
and similarly for $s\to\bar\Lambda$. This gives the following fragmentation fractions for $z_0 = 0.3$ (a cut motivated at the end of section~\ref{sec: polarization retention in baryons}):
\begin{align}
    f(s\to\Lambda,~z>0.3)=2.8\%\,,\qquad
    f(s\to\bar{\Lambda},~z>0.3)=0.3\%\,.
\end{align}
We also compute the sum of the $\Lambda$ and $\bar\Lambda$ numbers for a comparison with the AKKII data-based results~\cite{Albino}. A reasonable rough agreement is seen, except at the high-$z_0$ tail where neither of the methods can be trusted due to the scarcity of data at high $z$, and in our case also the limited statistics of the simulation. An accurate modeling of the small contributions at the tail is not essential for our purposes since we will not be using $z_0$ values higher than $0.3$.
	
\begin{table}
\centering
\begin{tabular}{c|c|c|c|c}\hline\hline
\multicolumn{2}{c|}{Fragmentation Fraction} & Decay Scheme & BR~\cite{Workman:2022ynf} & Spin analyzing power\\ \hline
    \multirow{3}{*}{$b\to\Lambda_b$} & \multirow{3}{*}{$7.0\%$~\cite{Galanti:2015pqa}} & $\Lambda_b \to X_c\mu^-\bar{\nu}_\mu$ &  $11\%$ & $\alpha_{\mu^-}\approx -0.26,\,\alpha_{\bar\nu_\mu}\approx 1$~\cite{Manohar:1993qn,Galanti:2015pqa} \\
     & & with $\Lambda \to p\pi^-$  & 2.7\% & \\
     & & with $\Lambda_c^+$ reco.\, & 2.0\% & \\\hline
 
    \multirow{3}{*}{$c\to\Lambda_c$} & \multirow{3}{*}{$6.4\%$~\cite{Lisovyi:2015uqa}} & $\Lambda_c^+ \to pK^-\pi^+$ & $6.3\%$ & $\alpha_{\rm eff}\approx 0.662$~\cite{LHCb:2022sck} \\\cline{3-5}
     & & $\Lambda_c^+ \to \Lambda\mu^+\nu_\mu$ & $3.5\%$ & $\alpha_{\mu^+}\approx 1$~\cite{Czarnecki:1994pu} \\
     & & with $\Lambda \to p\pi^-$ & $2.2\%$ & \\\hline
 
    $s\to\Lambda$ & \multirow{2}{*}{$2.8\%$~\cite{Sjostrand:2014zea,Albino}} & \multirow{2}{*}{$\Lambda \to p\pi^-$} & \multirow{2}{*}{$64\%$} & \multirow{2}{*}{$\alpha_p \approx 0.75$~\cite{Workman:2022ynf}} \\
    $(z>0.3)$ & & & & \\\hline\hline
\end{tabular}
\caption{Baryon fragmentation fractions, relevant decay schemes, branching ratios (BR) and spin analyzing powers (asymmetry parameters) of the decays.}
\label{tab:decay scheme}
\end{table}

Table~\ref{tab:decay scheme} also shows the decay channels we consider, selected based on their branching ratio (BR), spin analyzing power, and feasibility of identification and reconstruction. For the semileptonic $\Lambda_b$ decays, we consider three types of selections (similar to ref.~\cite{Galanti:2015pqa}):
\begin{itemize}
\item \emph{Inclusive Selection}, where $X_c$ represents any collection of particles containing a charmed hadron, which is usually a $\Lambda_c^+$.\footnote{The branching ratios for $\Lambda_b$ decays to charmed mesons are very small, e.g.\ $\BR(\Lambda_b \to D^0 p \pi^-) \approx 6 \times 10^{-4}$, $\BR(\Lambda_b \to D^{(\ast)+}p\pi^-\pi^-) \approx 8 \times 10^{-4}$~\cite{Workman:2022ynf}. We will neglect them in the following.} This selection, which will not apply any conditions to the particles produced along with the muon, will have high signal efficiency, but also unsuppressed backgrounds from semileptonic decays of $b$ mesons.
\item \emph{Semi-inclusive Selection}, where the $X_c$ is required to contain a $\Lambda \to p\pi^-$ decay, to reduce backgrounds from the semileptonic $b$-meson decays.
\item \emph{Exclusive Selection}, where the $X_c$ is required to contain a fully reconstructible $\Lambda_c^+$ decay (i.e., with charged products only), to reduce backgrounds from the semileptonic meson decays and facilitate the reconstruction of the $\Lambda_b$ decay kinematics for the polarization and spin correlation measurements. The full list of the $\Lambda_c^+$ decays we include here is provided in a later section, in table~\ref{fig:decay_scheme}.
\end{itemize}
In the semileptonic $\Lambda_c^+$ decay case, only the selection with the $\Lambda \to p\pi^-$ decay will be considered. Decays with electrons instead of muons, for both the $\Lambda_b$ and $\Lambda_c^+$, can be used as well. Their branching ratios and spin analyzing powers are approximately the same as for the decays with muons, and the trigger thresholds for electrons and muons are comparable. However, we will conservatively not take decays with electrons into account since reconstruction of electrons inside jets usually has low efficiency or high background~\cite{ATLAS:2016lxn,CMS:2021scf,CMS:2023aim}.

The last column in table~\ref{tab:decay scheme} indicates the decay products whose angular distributions are intended to be used for the polarization and spin correlation measurements. The spin analyzing power (or the asymmetry parameter) $\alpha$ is defined by writing the angular distribution of the decay as
\begin{equation}
    \frac{1}{\Gamma}\frac{d\Gamma}{d\cos\theta} = \frac12\left(1 + \alpha\mathcal{P}\cos\theta\right) \,,
\label{eq:polarization dist}
\end{equation}
where $\theta$ is the angle between the momentum of the decay product and the direction of the baryon polarization $\mathcal{\vec P}$, in the baryon rest frame.\footnote{To define $\alpha$ for the corresponding antibaryon decay, we will follow the common convention (as in refs.~\cite{Bernreuther:2001rq,Bernreuther:2004jv,Bernreuther:2015yna}, for example, but unlike in ref.~\cite{CMS:2019nrx}) of adding a minus sign in front of the $\alpha$ in eq.~\eqref{eq:polarization dist} for the antibaryon, such that for $CP$ conserving decays $\alpha$ has the same value for the baryon and antibaryon decay.\label{footnote:alpha-antibaryon}} This provides a handle for measuring the baryon and antibaryon polarizations and spin correlations.

The $\Lambda_c^+ \to pK^-\pi^+$ decay can proceed via various intermediate resonances, with different angular distributions in each case, and we quote the effective value of the spin analyzing power, $\alpha_{\rm eff}$, that corresponds to the sensitivity that can be obtained with a full amplitude analysis~\cite{LHCb:2022sck}.

Reconstructing the kinematics of the semileptonic decays, which is needed for the polarization and spin correlation measurements, is not straightforward because neutral particles cannot be assigned to a vertex and neutrinos are not observed at all. However, it can be done with certain approximations using the method described in detail in sections~4.2.2 and 4.2.3 of ref.~\cite{Galanti:2015pqa}, for example. It uses the fact that the energy fraction carried by a heavy-flavored hadron (relative to the original quark) has a relatively peaked distribution, with an average value around $70\%$ for the $b$ quark~\cite{ALEPH:2001pfo,DELPHI:2011aa,OPAL:2002plk,SLD:2002poq,ATLAS:2022miz,ATLAS:2021agf,CMS-PAS-TOP-18-012} and $50\%$ for the $c$ quark~\cite{OPAL:1997edj,ALEPH:1999syy}.\footnote{These numbers are appropriate for quark $p_T$ values of tens of GeV. They decrease slowly as a function of the energy scale due to renormalization group evolution~\cite{Cacciari:2005uk}.} To reconstruct the neutrino momentum, the vector pointing from the primary vertex to the baryon decay vertex is taken as the baryon flight direction. See also refs.~\cite{Dambach:2006ha,LHCb:2015eia,LHCb:2020ist,Ciezarek:2016lqu}. Of particular note is ref.~\cite{Dambach:2009wda}, which shows that reconstruction of a semileptonic $B$ hadron decay is feasible in CMS. Unfolding will be required to account for the approximations made.  

\section{Polarization and Spin Correlations}
\label{sec:spin correlations formalism}
	
We will now review the mathematical description of the polarization and spin correlations of $q\bar q$ pairs and describe how these properties are expected to manifest themselves in angular distributions of the baryon decays.
	
\subsection{Quark-Antiquark Pair Spin State Description}

The spin state of a quark and antiquark is described by a density matrix of the form~\cite{Fano:1983zz}
\begin{equation}
    \rho = \frac14\left(\mathbbm{1}\otimes\mathbbm{1}
    + \tilde{B}^+_i\,\sigma^i\otimes\mathbbm{1}
    + \tilde{B}^-_i\,\mathbbm{1}\otimes\sigma^i
    + \tilde{C}_{ij}\,\sigma^i\otimes\sigma^j
    \right) .
\label{eq:density mat}
\end{equation}
Here $\mathbbm{1}$ is a $2\times2$ unit matrix, $\sigma^i$ are the Pauli matrices, the indices (summation over which is implied) represent the coordinate axes, $\mathbf{\tilde B}^\pm$ are three-dimensional vectors characterizing the polarization of the quark and antiquark,\footnote{The quark polarization is $\mathbf{\tilde B}^+$, and the antiquark polarization is $-\mathbf{\tilde B}^-$.} and $\mathbf{\tilde C}$ is a $3\times3$ matrix that characterizes the spin correlations between them. The tilde symbol is used here to distinguish between these properties of the quark and antiquark and the measurable coefficients of the related angular distributions that will be defined in the next subsection.

As common in the $t\bar t$ literature~\cite{Bernreuther:2015yna,ATLAS:2016bac,CMS:2019nrx}, we will use the orthogonal set of axes $\{\hat{k},\hat{n},\hat{r}\}$. The axis $\hat{k}$ is defined as the direction of the outgoing quark's momentum in the center-of-mass (CM) frame of the produced $q\bar{q}$ pair. (In practice, the quark and antiquark momenta will be inferred from the momenta of the corresponding jets.) To define the other axes, we denote by $\hat{p}$ the positive direction of the beam axis in the laboratory frame,\footnote{For $2\to2$ parton reactions, it is also the momentum direction of one of the incoming partons in the $q\bar{q}$ pair CM frame (in the approximation that the parton momentum is collinear with the beam axis).} and by $\Theta$ the scattering angle of the outgoing quark defined by $\cos\Theta = \hat{k}\cdot\hat{p}$. Then the axes $\hat{n}$ and $\hat{r}$ are defined as
\begin{equation}
    \hat{n} = \sign(\cos\Theta)\,\frac{\hat{p}\times\hat{k}}{\sin\Theta} \;,\qquad
    \hat{r} = \sign(\cos\Theta)\,\frac{\hat{p}-\hat{k}\cos\Theta}{\sin\Theta} \;.
\label{eq:n-hat,r-hat}
\end{equation}

The $\sign(\cos\Theta)$ factor is included in eq.~\eqref{eq:n-hat,r-hat} to account for the Bose symmetry of the $gg$ initial state, meaning that without this sign factor the $gg$-initiated contributions to the polarizations and spin correlations of the sample as a whole will cancel between events with $\cos\Theta<0$ and $\cos\Theta>0$, as can also be seen explicitly in refs.~\cite{Dharmaratna:1996xd,Bernreuther:1995cx}. It is of note that the inclusion of the $\sign(\cos\Theta)$ factor leads to partial cancellation for events originating from $q\bar{q}$~\cite{Dharmaratna:1996xd,Bernreuther:1995cx}. It can be useful to also do a measurement without this factor to be more sensitive to $q\bar{q}$-initiated contributions. 

CMS measured the $\Lambda_b$ polarization along the $\hat{n}$ axis without the sign factor, using an amplitude analysis of the decay\footnote{The advantage of this decay chain is that every product is charged and therefore has a track, so one can reconstruct it exactly, leading to low backgrounds and precise kinematics. The downside of this channel is its tiny branching ratio ($\sim 3\times10^{-5}$).} $\Lambda_b\to J/\psi(\to\mu^+\mu^-)\Lambda(\to p\pi^-)$ at CM energies $\sqrt s = 7$ and $8$~TeV, finding $\mathcal{P} = 0.00 \pm 0.06\,(\text{stat.}) \pm 0.06\,(\text{syst.})$~\cite{CMS:2018wjk}. LHCb conducted $\Lambda_b$ polarization measurements using the same decay channel and found the polarization along $\hat n$ (without the sign factor) to be within $68\%$ credibility level intervals of $[-0.06,0.05]$, $[-0.04,0.05]$ and $[-0.01,0.07]$ at $\sqrt s = 7$, $8$ and $13$~TeV, respectively~\cite{LHCb:2020iux}. The absence of the sign factor in the LHCb measurement does not lead to a complete cancellation of the $gg$-initiated contributions since the LHCb detectors have coverage only in the forward direction.

Given the orthonormal basis $\{\hat{k},\hat{n},\hat{r}\}$, we can write the polarization vectors and spin correlation matrix as
\begin{equation}
    \mathbf{\tilde B}^\pm = b^\pm_k\hat{k} + b^\pm_n\hat{n} + b^\pm_r\hat{r}
\label{eq:B-tilde}
\end{equation}
and
\begin{align}
    \mathbf{\tilde C} =\; & c_{kk}\hat{k}\hat{k}+c_{nn}\hat{n}\hat{n}+c_{rr}\hat{r}\hat{r}\nonumber \\ &+c_{rk}(\hat{r}\hat{k}+\hat{k}\hat{r})+c_{nr}(\hat{n}\hat{r}+\hat{r}\hat{n})+c_{kn}(\hat{k}\hat{n}+\hat{n}\hat{k}) \label{eq:C-tilde} \\
    &+c_{n}(\hat{r}\hat{k}-\hat{k}\hat{r})+c_{k}(\hat{n}\hat{r}-\hat{r}\hat{n})+c_{r}(\hat{k}\hat{n}-\hat{n}\hat{k}) \nonumber
\end{align}
or equivalently
\begin{equation}
    \mathbf{\tilde C} =
    \begin{pmatrix}
    c_{kk}     &~~& c_{kn}+c_r &~~& c_{rk}-c_n\\
    c_{kn}-c_r &~~& c_{nn}     &~~& c_{nr}+c_k\\
    c_{rk}+c_n &~~& c_{nr}-c_k &~~& c_{rr}
    \end{pmatrix} .
\label{eq:C matrix}
\end{equation}
In this way of writing, the symmetric part of the matrix $\tilde{\textbf{C}}$ is described by the components
\begin{equation}
    c_{ij} = \frac12\left(\tilde C_{ij} + \tilde C_{ji}\right)
\label{c_ij}
\end{equation}
and the antisymmetric part by
\begin{equation}
    c_\ell = \frac12\epsilon_{ij\ell} \tilde C_{ij} \,.
\label{c_l}
\end{equation}
The range of possible values for $b_i^\pm$, $c_{ij}$, and $c_\ell$ is $[-1,1]$. These quantities in general depend on the production process, the partonic CM energy squared $\hat{s}$, and the quark's scattering angle $\Theta$ in relation to the proton going in the positive direction of the $\hat{z}$ axis.

Table~\ref{tab:P and CP of spin} classifies the polarization and spin correlation components according to their $P$ and $CP$ properties. The $P$ and $CP$ invariance of QCD (neglecting $\theta_{\rm QCD}$) allows spin correlations only in $c_{kk}$, $c_{rr}$, $c_{nn}$, and $c_{rk}$. For the polarizations, nonzero $b_n^+ = b_n^-$ are allowed, but expected to be small as these polarizations only appear at NLO QCD and are proportional to the quark mass~\cite{Dharmaratna:1996xd,Bernreuther:1995cx}. Small contributions involving the electroweak interactions are expected in many of the components (see, e.g., ref.~\cite{Bernreuther:2015yna}, where they were computed for the top quark), but we will not consider them in this paper. 
	
\begin{table}
\centering
\begin{tabular}{c|cc} \hline\hline
    Component & P & CP\\\hline
    $b_k^+ + b_k^-$ &Odd&Even\\
    $b_k^+ - b_k^-$ &Odd&Odd\\
    $b_r^+ + b_r^-$ &Odd&Even\\
    $b_r^+ + b_r^-$ &Odd&Odd\\
    $b_n^+ + b_n^-$ &Even&Even\\
    $b_n^+ - b_n^-$ &Even&Odd\\
    $c_{kk}$ &Even&Even\\
    $c_{rr}$ &Even&Even\\
    $c_{nn}$ &Even&Even\\
    $c_{rk}$&Even&Even\\
    $c_{n}$&Even&Odd\\
    $c_{nr}$&Odd&Even\\
    $c_{k}$&Odd&Odd\\
    $c_{kn}$&Odd&Even\\
    $c_{r}$&Odd&Odd\\
    \hline\hline
\end{tabular}
\caption{$P$ and $CP$ properties of the polarization and spin correlation components.}
\label{tab:P and CP of spin}
\end{table}

\subsection{Decay Angular Distributions}

Similar to the $t\bar t$ case~\cite{Bernreuther:2015yna}, if we denote the momentum vector of one of the baryon decay products in the baryon rest frame by $\mathbf{p}_1$, and the momentum vector of one of the antibaryon decay products in the antibaryon rest frame by $\mathbf{p}_2$, their angular distributions are given by
\begin{equation}
    \frac{1}{\sigma}\frac{d\sigma}{d\Omega_1 d\Omega_2} =
    \frac{1}{(4\pi)^2}\left(1 + \mathbf{B}^+\cdot\mathbf{\hat{p}}_1 + \mathbf{B}^-\cdot\mathbf{\hat{p}}_2 - \mathbf{\hat{p}}_1\cdot\mathbf{C}\cdot \mathbf{\hat{p}}_2\right) ,
\label{eq:total dist pre simple}
\end{equation}
where
\begin{equation}
    B^\pm_i=\alpha_\pm\,r_i\,f\,\tilde B^\pm_i\,,\qquad
    C_{ij}=\alpha_+\alpha_-\,r_i\,r_j\,f\,\tilde{C}_{ij}\,.
\label{eq:B,C}
\end{equation}
Here $\mathbf{\tilde B}^\pm$ and $\mathbf{\tilde{C}}$ are the quark and antiquark polarization vectors and their spin correlation matrix from eqs.~\eqref{eq:B-tilde}--\eqref{eq:C-tilde}. The factors $\alpha_+$ and $\alpha_-$, referring to the baryon and antibaryon, respectively, are the spin analyzing powers of their decays, as defined in eq.~\eqref{eq:polarization dist}. The factors $r_i$ are the polarization retention factors from eq.~\eqref{eq:spin retention}: $r_L$ for the $\hat k$ axis and $r_T$ for the $\hat n$ and $\hat r$ axes.\footnote{We work here in the approximation that the baryon momentum is parallel to the quark momentum and the only effect is the scaling of the polarization by the corresponding factor.}
The factor $f$ is the sample purity, namely the fraction of signal events out of the total number of selected events:
\begin{equation}
    f = \frac{N_{\rm sig}}{N_{\rm bg}+N_{\rm sig}} \;.
\end{equation}
The multiplication by $f$ in eq.~\eqref{eq:B,C} is only correct if we assume that the effect of the background is only to dilute the $B^\pm$ and $C$ coefficients. This would be the case, for example, for a background consisting of unpolarized and uncorrelated baryon-antibaryon pairs. If the background has a more general angular dependence, it will add a bias that will need to be subtracted. In some cases it will be possible to measure the bias using sidebands; in other cases it can be estimated through simulation.

It will also be useful for us to define
\begin{equation}
    C_{ij}^\pm = C_{ij} \pm C_{ji}
\end{equation}
and rewrite eq.~\eqref{eq:B,C} as
\begin{equation}
    B^\pm_i = \alpha_\pm r_i f b^\pm_i\,,\quad
    C_{ii} = \alpha_+\alpha_- r_i^2 f c_{ii}\,,\quad
    C_{ij}^+ = 2\alpha_+\alpha_- r_i r_j f c_{ij}\,,\quad
    C_{ij}^- = 2\alpha_+\alpha_- r_i r_j f c_\ell\,,
\label{eq:B,Cii,Cij+-}
\end{equation}
where we used eqs.~\eqref{c_ij}--\eqref{c_l}. In the expression for $C_{ij}^-$, the axes corresponding to the indices $i$, $j$ and $\ell$ are related via $\hat i \times \hat j = \hat\ell$.

Defining sets of spherical coordinates around the axes of interest for the baryon and antibaryon and integrating eq.~\eqref{eq:total dist pre simple} over the azimuthal angles gives
\begin{equation}
    \frac{1}{\sigma}\frac{d\sigma}{d\cos\theta^+_i d\cos\theta^-_j} =
    \frac{1}{4} \left(1 + B^+_i\cos\theta^+_i + B^-_j\cos\theta^-_j
                        - C_{ij}\cos\theta^+_i\cos\theta^-_j\right) ,
\label{eq:total dist}
\end{equation}
where $\theta^+_i$ ($\theta^-_j$) is the angle between the direction of the decay product and the chosen axis, in the baryon (antibaryon) rest frame.\footnote{For the antibaryon, we take the reference axes to be opposite to the $\{\hat k, \hat n, \hat r\}$ axes defined for the baryon, following the convention of ref.~\cite{Bernreuther:2015yna}. In combination with the convention described in footnote~\ref{footnote:alpha-antibaryon}, this makes our signs for $B^\pm_i$ and $C_{ij}$ agree with both ref.~\cite{Bernreuther:2015yna} and ref.~\cite{CMS:2019nrx}.} Integrating over one of the $\theta$s in eq.~\eqref{eq:total dist}, we get the distributions
\begin{equation}
    \frac{1}{\sigma}\frac{d\sigma}{d\cos\theta^\pm_i} =
    \frac{1}{2}\left(1 + B^\pm_i\cos\theta^\pm_i\right) ,
\label{eq:B dist}
\end{equation}
through which one can measure $B^\pm_i$ to obtain the quark and antiquark polarization components $b^\pm_i$. Converting the double differential distribution in eq.~\eqref{eq:total dist} to a distribution differential in the product $\cos\theta^+_i\cos\theta^-_j$, we obtain
\begin{equation}
    \frac{1}{\sigma}\frac{d\sigma}{d(\cos\theta^+_i\cos\theta^-_j)} = \frac{1}{2}\left(1 - C_{ij}\cos\theta^+_i\cos\theta^-_j\right) \ln\left(\frac{1}{|\cos\theta^+_i\cos\theta^-_j|}\right) ,
\label{eq:Cii dist}
\end{equation}
through which one can measure the $C_{ij}$ prefactors related to the spin correlations. We imagine using them to extract the diagonal ($c_{ii}$) components of the spin correlation matrix. For the off-diagonal components, which are useful to divide into symmetric ($c_{ij}$) and antisymmetric ($c_\ell$) parts because of their different $CP$ properties (recall table~\ref{tab:P and CP of spin}), one can derive from eq.~\eqref{eq:total dist pre simple} the distribution
\begin{equation}
    \frac{1}{\sigma} \frac{d\sigma}{dX_\pm} =
    \frac{1}{2}\left(1 - \frac{C_{ij}^\pm}{2}X_\pm\right)\cos^{-1}(|X_\pm|) \,,
\label{eq:Cij dist}
\end{equation}
where $X_\pm = \cos\theta^+_i\cos\theta^-_j \pm \cos\theta^+_j\cos\theta^-_i$. The components $c_{ij}$ (with $i \neq j$) and $c_\ell$ can be computed from $C_{ij}^\pm$ via eq.~\eqref{eq:B,Cii,Cij+-}.

\begin{figure}
\centering
\includegraphics[width=0.9\linewidth]{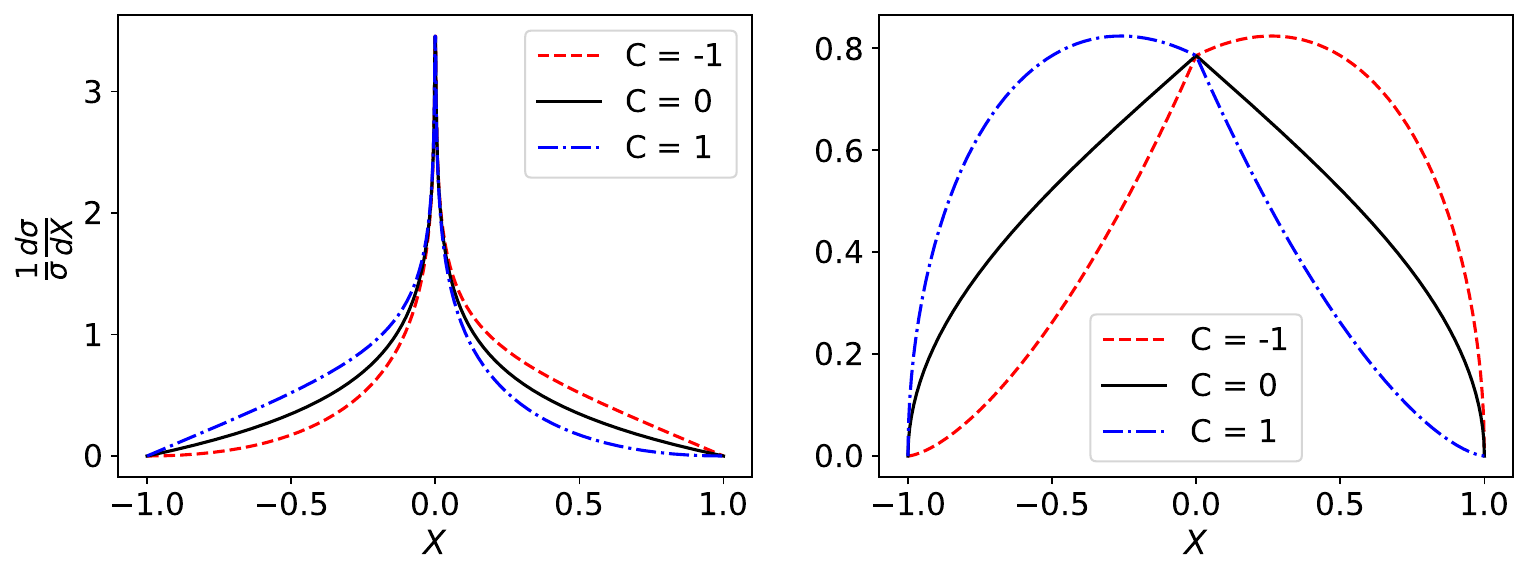}
\caption{Distributions of $X=\cos\theta^+_i\cos\theta^-_i$ (left) and $X=\cos\theta^+_i\cos\theta^-_j \pm \cos\theta^+_j\cos\theta^-_i$ (right), from which the spin correlations can be extracted. The distributions are based on eqs.~\eqref{eq:Cii dist}--\eqref{eq:Cij dist}, and are shown for three different values of the coefficient $C_{ii}$ (left) and $C_{ij}^\pm/2$ (right).}
\label{fig:spin corr theo dist}
\end{figure}

The distributions in eq.~\eqref{eq:B dist} are straight lines, while the shapes of the distributions in eqs.~\eqref{eq:Cii dist} and~\eqref{eq:Cij dist} are shown in figure~\ref{fig:spin corr theo dist}. They are plotted for $C_{ii}$ and $C_{ij}^\pm/2$ values of $\pm 1$ (which are the upper and lower bounds) and $0$.
 
\subsection{Statistical Uncertainty Evaluation}
\label{sub:uncertainty}

The statistical uncertainty of the coefficients $B^\pm_i$, $C_{ij}$, and $C_{ij}^\pm$, when they are extracted from fits of data to eqs.~\eqref{eq:B dist}, \eqref{eq:Cii dist}, and~\eqref{eq:Cij dist}, respectively, is given approximately by
\begin{equation}
    \Delta B^\pm_i \simeq \frac{\sqrt 3}{\sqrt N}\;,\quad
    \Delta C_{ij} \simeq \frac{3}{\sqrt N}\;,\quad
    \Delta C_{ij}^\pm \simeq \frac{3\sqrt2}{\sqrt N}\;,
\label{uncert-at-0}
\end{equation}
where $N$ is the number of events. These results are derived in appendix~\ref{app:uncertainty}. Using eq.~\eqref{eq:B,Cii,Cij+-} we then obtain for the statistical uncertainties of the polarization and spin correlation components
\begin{gather}
    \Delta b^\pm_i \simeq \frac{\sqrt3}{|r_i\alpha_\pm|\sqrt{fN_{\rm sig}}}
    \label{eq:b err}\;,\\
    \Delta c_{ii} \simeq \frac{3}{r_i^2|\alpha_+\alpha_-|\sqrt{fN_{\rm sig}}}\;,\\
    \Delta c_{ij(\ell)} \simeq \frac{3}{\sqrt2\,|r_i r_j\alpha_+\alpha_-|\sqrt{fN_{\rm sig}}}\;, \label{eq:c err}
\end{gather}
where $N_{\rm sig }$ is the number of signal events, and the notation $c_{ij(\ell)}$ means to refer at once to $c_{ij}$ from eq.~\eqref{c_ij} and $c_\ell$ from eq.~\eqref{c_l}. The uncertainty in eq.~\eqref{eq:b err} applies to $b_i^+$ and $b_i^-$ separately. The quantities with definite $P$ and $CP$ properties formed from the sums and differences of $b_i^+$ and $b_i^-$ (recall table~\ref{tab:P and CP of spin}) will have lower relative statistical uncertainties since both the quark and antiquark measurements will contribute.

We note that these formulas only provide rough estimates of the expected statistical uncertainties, mainly because they do not take into account effects of unfolding or nontrivial angular distributions of the background.
	
\section{Quark Polarization and Spin Correlations in QCD}
\label{sub:bb cc spin corr}
	
We used MadGraph~\cite{Alwall:2014hca} to obtain the leading-order (LO) QCD expectations for the polarization and spin correlations in $pp \to b\bar b$ and $pp \to c\bar c$.

We first validated our procedure on the process $pp \to t\bar t$, results for which are available in the literature~\cite{Severi:2022qjy}. As a technical tool, we decayed the top quark as $t\to b\ell^+v_\ell$ with MadSpin~\cite{Artoisenet:2012st} (and similarly for $\bar t$) to obtain the polarization and spin correlation information of the $t$ and $\bar t$ from the angular distributions of the leptons. We used the NNPDF2.3QED LO parton distribution functions with $\alpha_s(m_Z)=0.130$~\cite{Ball:2013hta} and the default dynamical factorization and renormalization scales defined in MadGraph. The simulation was inclusive, with no cuts applied, as relevant for the comparison with the numbers in ref.~\cite{Severi:2022qjy}. As a check, we have also run the full matrix element simulation in MadGraph, namely $pp \to b\ell^+\nu_\ell \bar b\ell^-\bar\nu_\ell$ without separating the process into production and decay, and obtained consistent results.

The symmetries of LO QCD dictate that only the components $c_{kk}$, $c_{rr}$, $c_{nn}$, and $c_{rk}$ can be nonzero~\cite{Bernreuther:2015yna}. We indeed find all the other components to be consistent with zero. For the non-vanishing components, we found good agreement with the LO results of ref.~\cite{Severi:2022qjy}.
	
To obtain the polarizations and spin correlations for $pp \to b\bar b$ and $pp \to c\bar c$, we cannot follow a procedure exactly analogous to what we did for $pp \to t\bar t$ since MadGraph does not allow decaying $b$ and $c$ quarks (which we need as a technical tool to extract the spin information of the quarks). Instead, relying on the flavor blindness of QCD, we simulated $pp \to t\bar t$ with the top-quark mass set to $m_b=4.7$~GeV or $m_c=1.275$~GeV (and its width set to a negligible value, $10^{-3}$~GeV). We have also lowered the masses of the particles that participate in the top decay $t\to b\ell^+v_\ell$, so that the decay will still happen. We changed the masses of the $W$ and $b$ to 1~GeV for the $b$ simulation and $0.5$~GeV for the $c$ simulation and set the $W$ width to $10^{-3}$~GeV. We have verified, by simulating the decay of a polarized top quark in these models using the method of ref.~\cite{BuarqueFranzosi:2019boy}, that the spin analyzing power of the lepton remains the same.

Unlike in the $t\bar t$ case, where an inclusive measurement without any cuts on the $t$ and $\bar t$ is possible (with the use of unfolding), a measurement in $b\bar b$ or $c\bar c$ will typically be limited by triggers. (Triggers will be discussed in detail in section~\ref{sub:triggering}.) As we will see in later sections, the muon-based triggers are the best for our purposes. Since we work in MadGraph, at the level of parton-level quarks, instead of applying cuts on the muons produced in the hadron decays, we will apply roughly equivalent cuts on the quarks. Using a Pythia simulation, we found that applying the Run~2 ATLAS dimuon trigger threshold (shown later in table~\ref{tab:HL-LHC Run2 compare}) to muons from $b\to c$ transitions in the $b\bar b$ case is equivalent in terms of the cross section to applying the cuts $|\eta|<2.4$ and $p_T>79$~GeV on the quarks (``jets''). In the $c\bar c$ case the same procedure leads to a $p_T>115$~GeV cut on the quarks.\footnote{The difference between $b\bar b$ and $c\bar c$ is mostly related to the fact that a $b$ hadron typically carries around $70\%$ of the $b$-jet momentum, while a $c$ hadron carries around $50\%$ of the $c$-jet momentum, so $c$ quarks need to be more energetic for their muons to pass the trigger threshold.} We will use these cuts here as an example.

\begin{table}
\centering
\hspace*{-3mm}\begin{tabular}{c|ccccc} \hline\hline
    & $t\bar t$, no cuts & $b\bar b$, no cuts & $c\bar c$, no cuts & $b\bar b$ with cuts & $c\bar c$ with cuts \\\hline
    $c_{kk}$  & \pl$0.324\pm0.006$ & \pl$0.296\pm0.004$ & \pl$0.284\pm0.004$ & $-0.987\pm0.004$    & $-0.984\pm0.006$\\ 
    $c_{rr}$  & \pl$0.009\pm0.006$ & \pl$0.004\pm0.004$ & $-0.006\pm0.004$   & $-0.603\pm0.004$    & $-0.609\pm0.006$\\ 
    $c_{nn}$  & \pl$0.333\pm0.006$ & \pl$0.299\pm0.004$ & \pl$0.298\pm0.004$ & \pl$0.591\pm0.004$  & \pl$0.603\pm0.006$\\ 
    $2c_{rk}$ & $-0.211\pm0.008$   & $-0.197\pm0.006$   & $-0.188\pm0.006$   & $-0.038\pm0.006$    & $-0.008\pm0.009$\\\hline\hline
\end{tabular}
\caption{Spin correlations in $b\bar{b}$ and $c\bar{c}$ events with and without cuts for Run~2 ($\sqrt s = 13$~TeV). For comparison, we also show the same quantities for the $t\bar t$ case. The uncertainties shown are the statistical uncertainties of our simulation.}
\label{tab:bb cc spin correlations}
\end{table}
	
The results for the non-vanishing spin correlation components are shown in table~\ref{tab:bb cc spin correlations}. We also present the inclusive values for a more meaningful comparison with the $t\bar t$ case, as a check. For the inclusive $b\bar b$ and $c\bar c$ simulations we fixed the factorization and renormalization scales to $10$~GeV, to avoid large artifacts from low energies. As can be seen in the table, the inclusive values are rather similar between $t\bar t$, $b\bar b$, and $c\bar c$. The cuts, on the other hand, take us to a completely different regime. This is understandable since the inclusive contributions are dominated by a region near the production threshold while the cuts select regions away from the threshold.

Table~\ref{tab:bb cc spin correlations hi lumi} presents the analogous results for the HL-LHC with $\sqrt s = 14$~TeV, where our effective cuts on the quarks are $|\eta|<2.5$ and $p_T>59$~GeV for $b\bar b$ and $p_T>80$~GeV for $c\bar c$. We can see similar effects from the cuts as in the Run~2 case.

To assess systematic uncertainties, we examined the effects of varying the renormalization and factorization scales up and down by a factor of $2$. While there was a significant effect on the cross sections, there were no significant effects on the spin correlation coefficients relative to the statistical uncertainties of our simulations, which are shown in tables~\ref{tab:bb cc spin correlations} and~\ref{tab:bb cc spin correlations hi lumi}.

\begin{table}
\centering
\begin{tabular}{c|cccc}\hline\hline
              & $b\bar b$, no cuts & $c\bar c$, no cuts & $b\bar b$ with cuts & $c\bar c$ with cuts \\\hline
    $c_{kk}$  & \pl$0.298\pm0.004$ & \pl$0.276\pm0.004$ & $-0.972\pm0.004$    & $-0.995\pm0.006$\\ 
    $c_{rr}$  & $-0.001\pm0.004$   & $-0.009\pm0.004$   & $-0.587\pm0.004$    & $-0.610\pm0.006$\\ 
    $c_{nn}$  & \pl$0.302\pm0.004$ & \pl$0.284\pm0.004$ & \pl$0.577\pm0.004$  & \pl$0.596\pm0.006$\\ 
    $2c_{rk}$ & $-0.187\pm0.006$   & $-0.188\pm0.006$   & $-0.054\pm0.006$    & $-0.015\pm0.009$\\\hline\hline
\end{tabular}
\caption{Spin correlations in $b\bar{b}$ and $c\bar{c}$ events with and without cuts for the HL-LHC with $\sqrt s = 14$~TeV. The uncertainties shown are the statistical uncertainties of our simulation.}
\label{tab:bb cc spin correlations hi lumi}
\end{table}

\section{Proposed Analyses and Their Prospects}
\label{Sec: bcs}
	
In this section we will consider a variety of analysis channels for measuring the polarizations or spin correlations in $pp \to q\bar q$ processes with $q = s$, $c$, or $b$, using the baryon decays that were listed in table~\ref{tab:decay scheme}.

We will need to address a variety of backgrounds. There are \emph{intrinsic backgrounds}, which arise from the same parton-level process as the signal but with a different hadron decay passing the selection. There are also \emph{extrinsic backgrounds}, which arise due to other parton-level processes that may involve the same baryon decay as the signal or another similar hadron decay. Lastly, there are \emph{combinatorial backgrounds} (which may be of an intrinsic or extrinsic origin), which are a result of random tracks forming by chance a vertex similar to that of the baryon decay of interest. While the probability of this happening will usually be low, such a background can still be significant if the total cross section of the corresponding process is large.

We will assess the feasibility of each channel in terms of the precision that can be achieved and the sample purity. We will do it with the help of MadGraph~\cite{Alwall:2014hca} and/or Pythia~\cite{Sjostrand:2014zea} simulations and reliance on elements of existing ATLAS and CMS analyses. For jet clustering, the Pythia simulations are interfaced with FastJet~\cite{Cacciari}, where we use the anti-$k_t$ algorithm with radius $R=0.4$~\cite{Cacciari:2008gp}. Apart from trigger-motivated cuts and background reduction cuts relevant in each case, we will present our results for several values of dijet invariant mass ($m_{jj}$) cuts (in cases where statistics allows that) since such a selection can enhance the sensitivity to BSM effects that become sizable only at high energies.
	
We will start by presenting our assumed datasets, based on the current and future planned LHC parameters and triggers, and then proceed to discussing the individual analysis channels, each with its own backgrounds and selection strategy.

\subsection{Benchmark Datasets}
\label{sub:triggering}
	
We will consider the currently available full Run~2 dataset as well as the HL-LHC dataset. Table~\ref{tab:HL-LHC Run2 compare} presents the main parameters defining these datasets, including the standard trigger-motivated cuts that we will be assuming. The numbers shown in the table are for the offline cuts from ATLAS~\cite{ATL-PHYS-PUB-2019-005} and the online ones for CMS~\cite{Collaboration:2759072}. We will be using the ATLAS cuts.
	
\begin{table}
\centering
\hspace*{-1mm}\begin{tabular}{l c c c c} \hline\hline
    & \multicolumn{2}{c}{ATLAS} & \multicolumn{2}{c}{CMS}\\
    & Run~2 &HL-LHC& Run~2 &HL-LHC \\\hline
    Collider energy $\sqrt s$ [TeV] &13 &14&13&14\\
    Integrated luminosity $\mathcal{L}$ [fb$^{-1}$] & 140&3000&140&3000\\
    Jet $p_T$ cut [GeV]&460&400&500&520\\
    Double muon $p_T$ cut (without isolation) [GeV] &15&10&37,~27&37,~27\\
    Single muon $p_T$ cut (with isolation) [GeV]&27&20&24&24\\
    Double electron $p_T$ cut (without isolation) [GeV]&18&10&25&25\\
    Single electron $p_T$ cut (with isolation) [GeV]&27&22&28&32 or 26\\
    Jet $|\eta|$ cut &2.4&3.8&2.4&4.0\\
    Muon $|\eta|$ cut &2.4 &2.5&2.4&2.4\\
    Electron $|\eta|$ cut &2.4&2.5&2.4&2.4\\
    \hline\hline
\end{tabular}
\caption{The collider energy, luminosity, and trigger-motivated cuts for Run~2 of the LHC and those that are planned for the HL-LHC in ATLAS~\cite{ATL-PHYS-PUB-2019-005} and CMS~\cite{Collaboration:2759072}. We will be using the ATLAS cuts.}
\label{tab:HL-LHC Run2 compare}
\end{table}

For the jet based triggers, which are relevant for the hadronic channels $c\to\Lambda_c^+ \to pK^-\pi^+$ and $s\to\Lambda \to p\pi^-$, for Run~2 we added the requirement $|\eta|<2.4$ (even though the trigger functions up to $|\eta|=2.8$~\cite{Owen:2302730}) so that the jets will be within the tracker. For the HL-LHC we require $|\eta|<3.8$.

For the semileptonic channels $b\to\Lambda_b \to X_c\mu^-\bar{\nu}_\mu$ and $c\to\Lambda_c^+ \to \Lambda\mu^+\nu_\mu$ we can use the muon triggers, whose thresholds are much softer than those of the jet triggers. Even though the muon carries only a fraction of the jet energy, the muon triggers will still provide higher statistics. Since our muons are inside jets, the triggers of interest are primarily those that do not require the muons to be isolated. That is not a problem for events with two muons since double muon triggers without isolation requirements have sufficiently low thresholds. However, in some of the analyses that we will describe (semileptonic $\Lambda_c^+$ decay on one side and hadronic on the other side, or polarization measurements without any requirement on the second jet) just a single muon will be present. Single muon triggers without isolation have the relatively high thresholds of $50$~GeV~\cite{ATLAS:2020gty,Collaboration:2759072}. We will instead be relying on the ATLAS single muon trigger (included in table~\ref{tab:HL-LHC Run2 compare}) with a loose isolation criterion, which has around $50\%$ efficiency for muons in heavy-flavor jets~\cite{ATLAS:2020gty}.

As was mentioned in section~\ref{sec:decay chains}, decays with electrons instead of muons can be considered as well (even though reconstruction of electrons inside jets usually has low efficiency or high background), and table~\ref{tab:HL-LHC Run2 compare} shows the corresponding triggers.
	
In the context of the $b\to\Lambda_b \to X_c\mu^-\bar{\nu}_\mu$ channel, we also looked at $b$-tagging triggers. In ATLAS, the single $b$-jet trigger~\cite{ATLAS:2021piz} requires $E_T>225$~GeV (which is expected to become $p_T>180$~GeV at the HL-LHC~\cite{ATL-PHYS-PUB-2019-005}), and the double $b$-jet trigger demands $E_T>150$~GeV for the leading jet and $E_T>50$~GeV for the subleading one. Similarly, CMS is expected to have a double $b$-jet trigger with a $p_T>128$~GeV cut at the HL-LHC~\cite{Collaboration:2759072}. Even though these thresholds are significantly lower than those of the generic jet triggers, we have checked that the much lower thresholds of the muon-based triggers still result in more statistics. This happens in our particular case because the $b$ jets we are interested in always contain a muon. The $b$-jet triggers can however be useful at high invariant masses, where they can recover much of the efficiency loss due to the loose isolation requirement of the low-$p_T$ single-muon triggers.

In addition to the standard trigger paths listed in table~\ref{tab:HL-LHC Run2 compare}, we will consider the CMS ``parked'' $b$-hadron dataset that was collected during part of Run~2 using the \emph{data parking} strategy with a single displaced muon trigger~\cite{CMS-DP-2019-043,Bainbridge:2020pgi,CMS:2024syx}. The muon $p_T$ threshold varied between 7 and 12~GeV depending on the luminosity, its track was required to be within $|\eta| < 1.5$, and satisfy a requirement on impact parameter significance. Despite the lower integrated luminosity of this dataset ($42$~fb$^{-1}$) and the $\eta$ restriction, the soft $p_T$ threshold allowed collecting more statistics than the standard Run~2 muon triggers.
	
\subsection{Analyses of \texorpdfstring{$pp \to s\bar{s}$}{pp->ssbar}}
	
For measuring the polarization and spin correlations in $pp\to s\bar{s}$, we consider events with $s\to\Lambda$ and $\Lambda\to p \pi^-$.
	
\subsubsection{\texorpdfstring{$\Lambda$}{Lambda} Reconstruction, Efficiency and Signal Yield}
\label{sub:Lambda reco}

The decay $\Lambda \to p \pi^-$ has a very distinct signature of a highly displaced vertex with a pair of oppositely charged tracks that reconstruct the $\Lambda$ mass if they are assigned the proton and charged pion masses. The other similar decay, $K_S \to \pi^+\pi^-$, will usually fail the $\Lambda$ mass reconstruction, and moreover can be vetoed without significant loss of signal efficiency by requiring that the two tracks do not reconstruct the $K_S$ mass when assigned the charged pion masses. These $\Lambda$ decays were reconstructed in multiple analyses in ATLAS (e.g.,~\cite{ATLAS:2011xhu,ATLAS:2012cvl,ATLAS:2014swk,ATLAS:2019pqg}) and CMS (e.g.,~\cite{CMS:2011jlm,CMS:2018wjk,CMS:2019isl,CMS:2020zzv,CMS:2021rvl}).

There is, however, an important obstacle to reconstructing the decays of energetic $\Lambda$ baryons. With increasing $p_T$ they quickly become too displaced to be successfully reconstructed within the volume of the tracker. This can result in a very significant efficiency loss. To optimize the reconstruction efficiency of highly displaced tracks, ATLAS have developed the Large Radius Tracking (LRT) algorithm~\cite{ATL-PHYS-PUB-2017-014,ATL-PHYS-PUB-2019-013,ATLAS:2023nze}. It has been used in multiple searches for long-lived BSM particles~\cite{ATLAS:2017tny,ATLAS:2019kpx,ATLAS:2019fwx,ATLAS:2019jcm,ATLAS:2020xyo,ATLAS:2020wjh,ATLAS:2021jig,ATLAS:2023oti}. The LRT algorithm looks at hits remaining after the standard reconstruction, and tries to reconstruct the remaining tracks with looser conditions on the transverse and longitudinal impact parameters. The addition of this algorithm allows keeping decent track reconstruction efficiency up to decay radii $d_T$ of about $d_T^{\rm max} \approx 300$~mm, with the mean reconstruction efficiency up to this decay radius being roughly $\epsilon_{\rm track} \approx 65\%$. It is of note that LRT was applied to only about $10\%$ of the events in Run~2, but is going to be used regularly in Run~3 and at the HL-LHC after the LRT processing time was significantly improved~\cite{ATLAS:2023nze}. Moreover, with the ATLAS tracker upgrade planned for the HL-LHC, it will be possible to address even larger decay radii, with average reconstruction efficiency of roughly $\epsilon_{\rm track} \approx 80\%$ up to $d_T^{\rm max} \approx 400$~mm~\cite{Strebler:2022nkh}. We will be using the efficiency numbers with the LRT algorithm to estimate the $\Lambda$ reconstruction efficiency, after accounting for the probability for the $\Lambda$ to decay within the ranges mentioned above.

The average $\Lambda$ decay radius in $pp \to s\bar s$ events can be estimated as
\begin{equation}
    \langle d_T \rangle = \langle \gamma \beta_T \rangle\, c\tau
    = \frac{\langle p_{T,\Lambda}\rangle\, c\tau}{m_\Lambda}
    = \frac{\langle z\, p_{T,{\rm jet}}\rangle\, c\tau}{m_\Lambda} \;,
\end{equation}
where $c\tau \approx 7.9$~cm is the $\Lambda$ lifetime and $z$ is the $\Lambda$ momentum fraction from eq.~\eqref{eq:z definition}. Since the cross section decreases fast as a function of $p_{T,{\rm jet}}$ and $z$, we can obtain rough estimates for $\langle d_T \rangle$ by taking $p_{T,{\rm jet}}$ to be the trigger-motivated jet $p_T$ threshold from table~\ref{tab:HL-LHC Run2 compare}, and $z = 0.3$, which is the lowest value of $z$ we will be willing to use since much softer $\Lambda$ baryons are not correlated with the original strange-quark polarization, as was discussed in section~\ref{sec: polarization retention in baryons}. This gives $\langle d_T \rangle \sim 9.8$~m for Run~2 and $8.5$~m for the HL-LHC. Since $\langle d_T \rangle \gg d_T^{\rm max}$, the probability for the $\Lambda$ to decay sufficiently early is $\epsilon_{d_T} \simeq d_T^{\rm max}/\langle d_T \rangle \approx 3.1\%$ for Run~2 and $4.7\%$ for the HL-LHC. More accurate numbers that we obtained from a Pythia simulation (which accounts for the full jet $p_T$ and $z$ distributions above their corresponding thresholds) are $\epsilon_{d_T} \approx 2.2\%$ for Run~2 and 3.4\% for the HL-LHC. For background jets (which are a mixture of all $pp \to jj$ processes apart from $pp \to s\bar s$) the numbers are $2.4\%$ and $3.7\%$, respectively.

The full reconstruction efficiency for the $\Lambda \to p\pi^-$ decay in signal jets is $\epsilon_{\Lambda} = \epsilon_{d_T}\,\epsilon_{\rm track}^2 \approx 0.93\%$ for Run~2 and $2.2\%$ for the HL-LHC. For spin correlation measurements we need the efficiency for reconstructing both a $\Lambda$ and a $\bar\Lambda$. Despite the correlation between the $p_T$ of the two jets, we can still obtain a rough estimate by simply squaring the efficiency, $\epsilon_{\Lambda\bar\Lambda} \simeq \epsilon_\Lambda^2$, because the jet $p_T$ values are distributed mainly near the threshold. This gives $\epsilon_{\Lambda\bar\Lambda} \approx 9 \times 10^{-5}$ for Run~2 and $5 \times 10^{-4}$ for the HL-LHC.

\begin{table}
\centering
\begin{tabular}{ccc}\hline\hline
    \multicolumn{3}{c}{Run 2} \\\hline
     $\sigma_{s\bar{s}}$ [pb] & $N_{s}$ & $N_{s\bar{s}}$ \\\hline
     3.1 & 54 & 0.0074 \\\hline\hline
\end{tabular}
\begin{tabular}{|ccc}\hline\hline
    \multicolumn{3}{|c}{HL-LHC} \\\hline
    $\sigma_{s\bar{s}}$ [pb] & $N_s$ & $N_{s\bar{s}}$ \\\hline
    8.3 & $8.1 \times 10^{3}$ & 1.7 \\\hline\hline
\end{tabular}
\caption{Cross sections (with trigger-motivated cuts) and signal event counts (after the full selection) for measurements of $s$-quark polarization ($N_s$) and spin correlations ($N_{s\bar s}$).}
\label{fig:s num with pt frag}
\end{table}

Table~\ref{fig:s num with pt frag} shows the numbers of expected signal events in Run~2 and at the HL-LHC. We show both the number of $s$ jets available for the polarization measurement ($N_s$) and the number of $s\bar s$ pairs available for the spin correlation measurement ($N_{s\bar s}$) computed as
\begin{align}
    N_{s} &= \mathcal{L}\,\sigma_{s\bar{s}}\,f(s\to\Lambda,~z>0.3)\BR(\Lambda\to p\pi^-)\,\epsilon_{\Lambda} \,,\\
    N_{s\bar{s}} &= \mathcal{L}\,\sigma_{s\bar{s}}\,f^2(s\to\Lambda,~z>0.3)\BR^2(\Lambda\to p\pi^-)\,\epsilon_{\Lambda\bar\Lambda} \,,
\end{align}
where the cross sections $\sigma_{s\bar s}$ (with the trigger-motivated cut on the jets) were computed in MadGraph. We see from table~\ref{fig:s num with pt frag} that the number of events available for spin correlation measurements is going to be too low even at the HL-LHC. We will therefore proceed with investigating the prospects of polarization measurements only.

\subsubsection{Background}
	
The dominant background is due to $\Lambda$ baryons produced in dijet processes other than $pp\to s\bar s$. While the number of $\Lambda$ baryons produced in most of these processes falls with $z$ faster than in the signal, their total cross section is large. As a result, their contribution ends up being significant.

\begin{figure}
\centering
\includegraphics[width=0.49\linewidth,trim= 0 0 0 1cm,clip]{"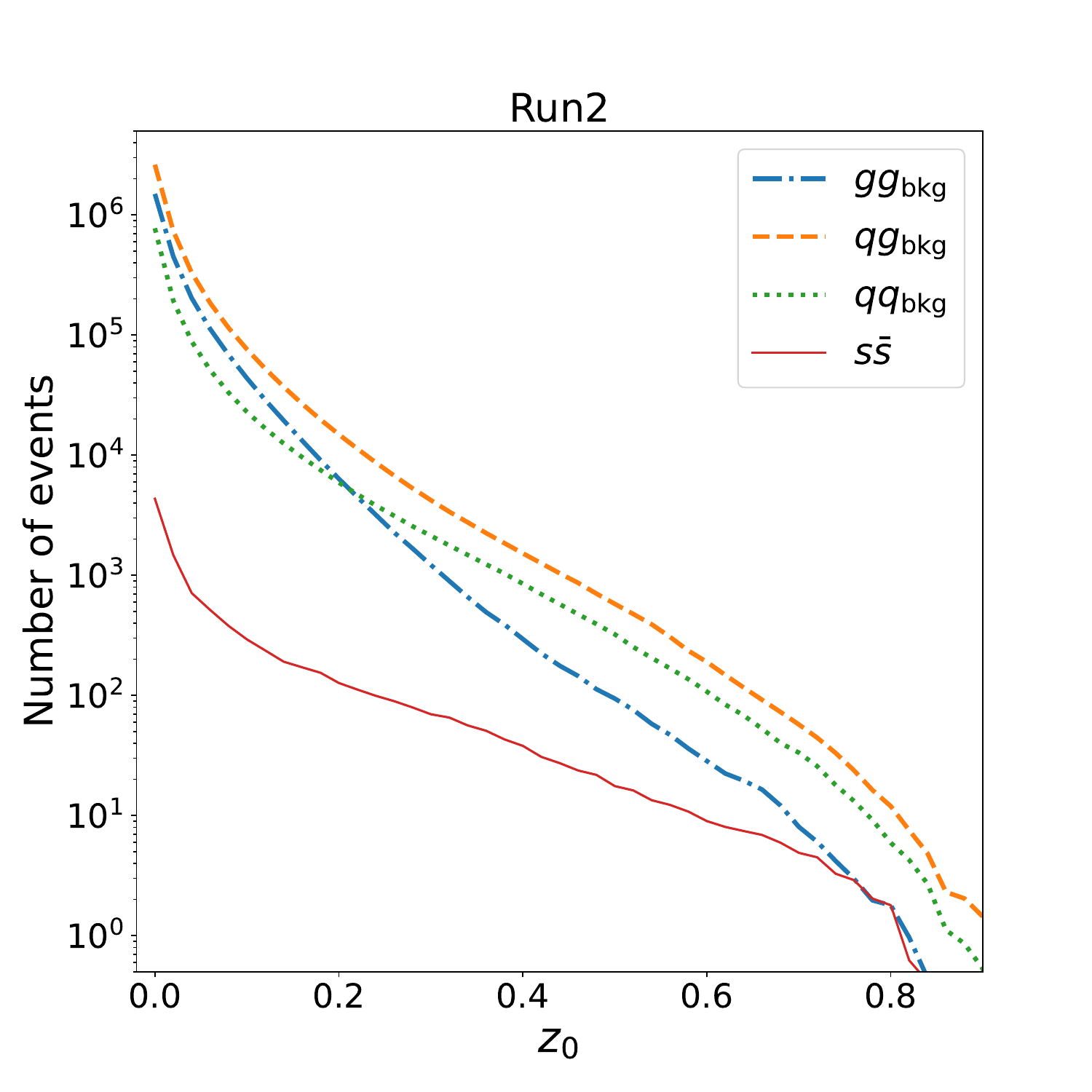"}
\includegraphics[width=0.49\linewidth,trim= 0 0 0 1cm,clip]{"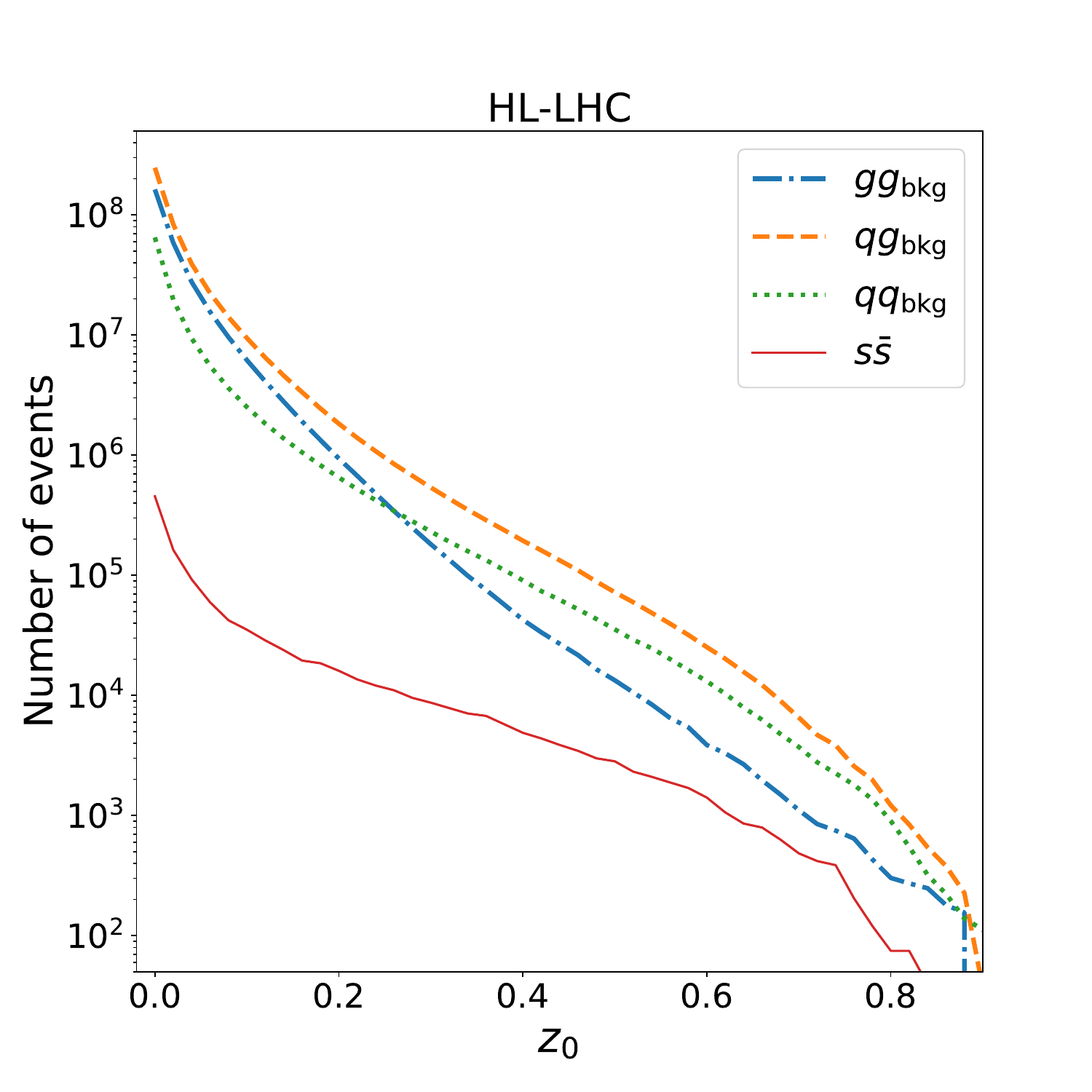"}
\caption{Expected number of events with reconstructed $\Lambda$ baryons satisfying $z > z_0$ for the signal originating from $s\bar{s}$ events and backgrounds with the final state partons specified in the legend, for Run~2 (left) and the HL-LHC (right).}
\label{fig:bg&signal_Pythia single}
\end{figure}

Figure~\ref{fig:bg&signal_Pythia single} shows the results of a Pythia simulation for the signal and backgrounds, where the backgrounds are split into three categories according to the produced partons: $gg$, $qg$, and $qq$ (where $q$ represents all flavors of quarks and antiquarks). The two leading jets in each event are considered as $s$ jet candidates, except for $s\bar s$ events, where we matched the jets to partons to count only $\Lambda$ baryons from $s$ (but not $\bar s$) as signal. The numbers are presented as a function of $z_0$, the cut on the momentum fraction $z$ carried by the $\Lambda$. From these plots we can calculate the sample purity. For $z_0 = 0.3$, the purity is $f \approx 0.9\%$ for both Run~2 and the HL-LHC. 

\subsubsection{Measurement prospects}
	
With the signal event counts and purities, we can use eq.~\eqref{eq:b err} to compute the expected precision of the polarization measurements. We find the expected statistical uncertainty on the polarization components $b^\pm_i$ at the HL-LHC to be given by $r_i\Delta b^\pm_i \approx 0.27$. We show the result for the product $r_i\Delta b^\pm_i$ to provide numbers that are independent of the systematic uncertainty of the polarization retention factors $r_i$. We remind the reader, however, that $r_L \sim 0.6$, as mentioned in section~\ref{sec: polarization retention in baryons}. The physical range of $b^\pm_i$ values is $[-1,1]$. Only the HL-LHC results were discussed here since the Run~2 numbers are far from being promising.
 
We conclude that with the standard triggers we assumed, $s\bar s$ spin correlations cannot be measured, while polarization measurements might be possible at the HL-LHC, although the statistical uncertainty of the result is expected to be high, and the low purity of the sample will make the measurements difficult.

\subsection{Analyses of \texorpdfstring{$pp \to c\bar{c}$}{pp->ccbar}}
	
For $c\bar c$ polarization and spin correlation measurements, we will consider in turn three possible analysis channels in terms of the $\Lambda_c^+$ decays: the \emph{hadronic channel} where $\Lambda_c^+ \to pK^-\pi^+$, the \emph{semileptonic channel} where $\Lambda_c^+ \to \Lambda\mu^+\nu_\mu$, and the \emph{mixed channel} with the hadronic decay in one jet and the semileptonic decay in the other.
	
\subsubsection{Hadronic Channel}
\label{sub:c-hadronic}

The signature of the $\Lambda_c^+\to pK^-\pi^+$ decay is one negatively and two positively charged tracks coming from a common vertex. They should also reconstruct the $\Lambda_c^+$ mass for an assignment of the proton and $\pi^+$ masses to the positively charged tracks and the $K^-$ mass to the negatively charged one. Such decays were reconstructed by CMS in refs.~\cite{CMS:2019uws,CMS:2023frs}.

There are intrinsic backgrounds from various other decays of charmed hadrons, with examples shown in table~\ref{tab:c had bg}. Some approaches for reducing them were discussed in ref.~\cite{Galanti:2015pqa}. Extrinsic backgrounds from $pp\to b\bar b$ should be considered as well. They include $\Lambda_c^+$ baryons produced in $\Lambda_b$ or $B$-meson decays as well as other decays that produce vertices that pass the selection. These contributions can be suppressed significantly using the large displacement of the $b$-hadron decays ($c\tau \approx 0.45$~mm) relative to the $\Lambda_c^+$ ($c\tau \approx 0.06$~mm). There is also the combinatorial background of three general tracks forming by chance a fake $\Lambda_c^+$-like vertex.

\begin{table}
\centering
\begin{tabular}{cc|lc}\hline\hline
\multicolumn{2}{c|}{Fragmentation Fraction [\%]} & Decay Scheme & Branching Ratio [\%]\\
\hline
    \multirow{4}{*}{$c\to\Lambda_c^+$} & \multirow{4}{*}{6.4}
     & $\Lambda_c^+\to pK^-\pi^+$ (signal) & 6.3 \\
    && $\Lambda_c^+\to pK^-\pi^+\pi^0$ & 4.4\\
    && $\Lambda_c^+\to \Sigma^+\pi^-\pi^+$ & 4.5\\
    && $\Lambda_c^+\to \pi^+\pi^-\pi^+\Lambda$ & 3.6\\
\hline
    \multirow{4}{*}{$c\to D^+$} & \multirow{4}{*}{22.7}
     & $D^+\to\pi^+K^-\pi^+$ & 9.4\\
    && $D^+\to\pi^+K^-\pi^+\pi^0$ & 6.2\\
    && $D^+\to\pi^+\pi^-\pi^+K_S^0$ & 3.1\\
    && $D^+\to\pi^+\pi^-\pi^+\pi^0$ & 1.2\\
\hline
    \multirow{2}{*}{$c\to D^0$} & \multirow{2}{*}{61.8}
     & $D^0\to\pi^+K^-\pi^+\pi^-$ & 8.2\\
    && $D^0\to\pi^+K^-\pi^+\pi^-\pi^0$ & 4.3\\
\hline
    \multirow{3}{*}{$c\to D_s^+$} & \multirow{3}{*}{8.2}
     & $D_s^+\to K^+K^-\pi^+$ & 5.4\\
    && $D_s^+\to K^+K^-\pi^+\pi^0$ & 5.5\\
    && $D_s^+\to \pi^+K^-\pi^+K^0_S$ & 1.5\\
\hline\hline
\end{tabular}
\caption{The $\Lambda_c^+\to pK^-\pi^+$ signal and some of its intrinsic backgrounds. The numbers are from refs.~\cite{Lisovyi:2015uqa,Workman:2022ynf}. The background decays will usually fail to reconstruct the $\Lambda_c^+$ mass---a requirement that will strongly suppress them.}
\label{tab:c had bg}
\end{table}

Simulating the backgrounds for the $\Lambda_c^+\to pK^-\pi^+$ decay with publicly available tools is nontrivial for us due to several reasons. First, the displacement of the $\Lambda_c^+$ vertex is small, so a careful simulation of the tracking resolution, as well as the vertexing algorithm, are needed to assess the impact of any displacement-related cuts. Second, simulating combinatorial background requires good control of tails of distributions of high cross section processes.

We can however get an idea about the size of the expected background and the signal efficiency from the CMS analysis in ref.~\cite{CMS:2023frs}. CMS used several variables to select $\Lambda_c^+\to pK^-\pi^+$ candidates: the $\chi^2$ of the vertex fit to the three tracks, the angle between the $\Lambda_c^+$ candidate momentum and the vector connecting the production and decay vertices in the transverse plane, the transverse separation significance between the two vertices, and the $\Lambda_c^+\,$ $p_T$ fractions carried by the kaon and by the proton. A distribution of the $pK^-\pi^+$ invariant mass was then constructed, with the $\Lambda_c^+$ contribution appearing as a narrow peak. The number of events in the peak is about $7.9\%$ of the total number of events in the peak region for $\Lambda_c^+$ candidates with $20 < p_T < 30$~GeV. CMS estimate the prompt $\Lambda_c^+$ fraction in the peak to be between $0.79$ to $0.88$. We will therefore take our rough estimate for the sample purity (when a single jet is considered) to be $f \approx 6.9\%$. CMS also reports the signal efficiency to be roughly $\epsilon_{\Lambda_c^+} \approx 25\%$, which we will also assume. It should be noted that the invariant mass resolution will be improved by $20$--$50\%$ at the HL-LHC~\cite{ATL-PHYS-PUB-2018-041,ATL-PHYS-PUB-2018-032,ATL-PHYS-PUB-2018-035,CMS-PAS-FTR-18-013,CMS-PAS-FTR-18-033} so the purity will improve accordingly. We shall be optimistic and reduce the background under the peak by a factor of two (i.e., approximately double the purity) in our estimates for the HL-LHC. We note that the upgraded tracking detectors will likely improve the efficacy of the other cuts used in the selection as well.

The sidebands of the $pK^-\pi^+$ invariant mass distribution can be used to measure the angular distributions of the background. They will need to be subtracted to obtain the polarization information of the signal from the events in the peak region.

\begin{table}
\centering
\begin{tabular}{c|cccc}\hline\hline
    \textbf{\bm{$c\bar c$}, hadronic} & \multicolumn{4}{c}{Run 2} \\\hline
    $m_{jj}$ cut [GeV] & $\sigma_{c\bar{c}}$ [pb] & $N_c$ & $r_i\Delta b^\pm_i$ & $N_{c\bar{c}}$ \\\hline
    \mbox{no cut} & 3.0  & 420 & 0.48 & 0.42 \\ 
             1000 & 2.6  & 360 & & \\ 
             1500 & 0.57 &  80 & & \\ 
             2000 & 0.13 &  18 & & \\ \hline\hline
\end{tabular}
\begin{tabular}{|cccc}\hline\hline
    \multicolumn{4}{|c}{HL-LHC} \\\hline
    $\sigma_{c\bar{c}}$ [pb] & $N_c$ & $r_i\Delta b^\pm_i$ & $N_{c\bar{c}}$ \\\hline
    8.0  & $2.4\times 10^{4}$ & 0.045 & 24 \\ 
    4.7  & $1.4\times 10^{4}$ & 0.059 & 14 \\ 
    0.93 &               2800 & 0.13  & 2.8 \\ 
    0.21 &                650 & 0.28  & \\ 
\hline\hline
\end{tabular}
\label{c num with pt frag: a}
\caption{Cross sections (with trigger-motivated cuts) and signal event counts (after the full selection) for measurements of $c$-quark polarization ($N_c$) and spin correlations ($N_{c\bar c}$) in the hadronic channel. The expected statistical uncertainties for the polarization measurements are shown as well. The sample purity is $6.9\%$ for Run~2, and double that for the HL-LHC.}
\label{fig:c num with pt frag}
\end{table}

To calculate the expected number of signal events, we computed the $pp\to c\bar c$ cross sections $\sigma_{c\bar c}$ with the jet trigger cuts from table~\ref{tab:HL-LHC Run2 compare} using a MadGraph simulation. In addition, for the $c$ polarization measurements we require the $c$ jet to contain a $\Lambda_c^+\to p K^-\pi^+$ decay. For the spin correlations we require this $\Lambda_c^+$ decay in one jet and the analogous $\bar{\Lambda}_c^-$ decay in the other. The expected numbers of events for measurements of polarization ($N_c$) and spin correlations ($N_{c\bar c}$), calculated as
\begin{align}
    N_{c} &= \mathcal{L}\,\sigma_{c\bar{c}}\,f(c\to\Lambda_c^+) \BR(\Lambda_c^+\to p K^-\pi^+)\,\epsilon_{\Lambda_c^+} \,,\\
    N_{c\bar{c}} &= \mathcal{L}\,\sigma_{c\bar{c}}\,f^2(c\to\Lambda_c^+) \BR^2(\Lambda_c^+\to p K^-\pi^+)\,\epsilon_{\Lambda_c^+}^2 \,,
\end{align}
are shown in table~\ref{fig:c num with pt frag} for Run~2 and the HL-LHC. We also provide results that correspond to different cuts on the dijet invariant mass, $m_{jj}$, as such cuts can enhance the sensitivity to new physics contributions. As can be seen in the table, the number of events available for the spin correlation measurements in this channel is small even for the HL-LHC (considering also the relatively low purity), so we proceed with the analysis of polarization measurements only. The table shows the expected statistical uncertainties for the polarization measurements (multiplied by the polarization retention factors) $r_i\Delta b^\pm_i$ based on eq.~\eqref{eq:b err}. The prospects are borderline for Run~2 but seem good for the HL-LHC.

\subsubsection{Semileptonic Channel}
\label{sub:c semi analysis}
	
A potentially more promising avenue is the semileptonic decay $\Lambda_c^+\to\Lambda\mu^+\nu_\mu$ with $\Lambda\to p\pi^-$. While the branching ratio of this decay chain ($3.5\%$ for $\Lambda_c^+\to\Lambda\mu^+\nu_\mu$, $64\%$ for $\Lambda\to p\pi^-$~\cite{Workman:2022ynf}) is lower than that of the hadronic decay, the muon triggers have very low $p_T$ thresholds (see table~\ref{tab:HL-LHC Run2 compare}) so there is a potential for getting more data. We do not consider a selection without a $\Lambda$ decay reconstructed in the tracker because it will be difficult to extract the $\Lambda_c^+$ polarization information if the muon will be the only product associated with the $\Lambda_c^+$ decay. Besides that, the $\Lambda$ requirement strongly suppresses the intrinsic background due to semileptonic $D$-meson decays since they are kinematically forbidden from producing a $\Lambda$ (which would have to be produced together with an antibaryon, for baryon number conservation, while no sufficiently light antibaryons exist).

This channel also has disadvantages, such as the low reconstruction efficiency of $\Lambda$ decays, and the shortness of the $\Lambda_c^+$ lifetime ($c\tau \approx 0.06$~mm) which leads to large uncertainties in its flight direction reconstruction, which is needed for the neutrino reconstruction. Although these challenges exist we will analyze the potential of this channel.

Our selection in this channel requires the jet to contain a muon (as done sometimes for charm jet tagging~\cite{ATLAS:2015thz,CMS:2017wtu,ATLAS-CONF-2018-055,CMS:2021scf,CMS:2023aim}) as well as a reconstructed $\Lambda \to p \pi^-$ decay. To ensure that the $\Lambda$ originates from a $\Lambda_c^+$ decay, one should demand the $\Lambda$ trajectory (inferred from the momenta of its decay products) to form a displaced vertex with the muon. Events in which the $\Lambda$ trajectory is consistent with both a common vertex with the muon and the primary vertex, can still be accepted if the $\Lambda$ carries a significant fraction (e.g., above $20\%$) of the jet momentum, since $\Lambda$ baryons produced in parton showering and hadronization will usually be soft. We expect this requirement to have high signal efficiency and significant background suppression, but estimating the efficiency and purity quantitatively requires a detailed simulation which is beyond the scope of the current work.

There is an extrinsic background from $b\bar b$ production. We can estimate the size of this background by starting with the inclusive branching ratio for $b$ jets to contain a $\Lambda$ or $\bar\Lambda$ baryon, which was measured to be~\cite{Workman:2022ynf}
\begin{equation}
    \BR\left(b \to \mbox{($b$-hadron)} \to \Lambda/\bar\Lambda + X\right) \approx (5.9 \pm 0.6)\% \,.
    \label{eq:b2Lambda}
\end{equation}
Accounting for us requiring a $\Lambda$ (not a $\bar\Lambda$) while on the other hand collecting background from both the $b$ and $\bar b$ jet, we are left with the same number. We will further assume that the probability of having a $\mu^+$ in association with the $\Lambda$ is roughly $10\%$.\footnote{This assumption is based on the fact that the off-shell $W$ boson that produces the muon can also produce other flavors of leptons and quarks, to the extent allowed by phase space. While the most common decay chain $b \to c \to s$ (for example) involves two $W$ bosons, they produce muons of opposite charges, only one of which is relevant to our signal.} The corresponding numbers are compared with the signal in table~\ref{tab:br and frag for eff and pure c}.

\begin{table}
\centering
\begin{tabular}{l|cccc}\hline\hline
Fragmentation and Decay & FF [\%] & BR [\%] & FF$\times$BR [\%] & Lifetime [s]\\\hline
Signal: $c\to \Lambda_c^+\to\Lambda\mu^+\nu_\mu$ & 6.4 & 3.5 & 0.22 & $2.0 \cdot 10^{-13}$\\\hline
Background: &&&\\
$b$+$\bar{b} \to \mbox{($b$ and $\bar b$ hadrons)} \to \Lambda\mu^+\nu_\mu X$ & & & 0.59 & $1.6 \cdot 10^{-12}$ \\
\emph{based on:} $b \to \mbox{($b$-hadron)} \to \Lambda/\bar\Lambda + X$ & & & \emph{5.9} & \\\hline\hline
\end{tabular}
\caption{Properties of the semileptonic $\Lambda_c^+$ signal and the $b\bar b$ background: fragmentation fractions (FF)~\cite{Lisovyi:2015uqa}, branching ratios (BR)~\cite{Workman:2022ynf} and lifetimes~\cite{Workman:2022ynf}. The $c\bar c$ and $b\bar b$ production cross sections are similar for given cuts on the jets. The number shown for the background is already summed over the two jets in the event and involves the assumption that roughly $10\%$ of the events in the $\Lambda X$ sample contain a $\mu^+$. The numbers shown are before any discriminating cuts.}
\label{tab:br and frag for eff and pure c}
\end{table}

There are a number of ways to reduce the $b\bar b$ background significantly. One can immediately veto events in which both the secondary vertex from the $b$-hadron decay and the tertiary vertex from the subsequent charmed hadron decay can be distinguished. Moreover, one can use the order-of-magnitude difference between the $\Lambda_c^+$ and $b$-hadron lifetimes (see table~\ref{tab:br and frag for eff and pure c}) and apply an upper bound on the transverse displacement $d_T$ of the secondary vertex. For example, requiring $d_T < 0.77~\mbox{mm}\times(p_T^{\rm jet}/115~\mbox{GeV})$ has an efficiency of $40\%$ for the signal and $10\%$ for the background, giving $60\%$ purity.\footnote{To obtain these numbers we assumed that the jet $p_T$ fraction carried by the hadron is $0.5$ for the $\Lambda_c^+$ and $0.7$ for the $b$ hadrons.} Various additional discriminants between $b$ and $c$ jets exist and are used in $c$ tagging algorithms. For a $c$-jet efficiency of $40\%$, a $b$-jet efficiency as low as $6\%$ is achieved in both ATLAS~\cite{ATLAS:2022qxm} and CMS~\cite{CMS:2021scf}, which would lead to $71\%$ purity in our case. The performance of charm tagging should be even better when applied to the $\Lambda_c^+$ sample than to an inclusive sample of charmed hadrons because the lifetimes of the more common charmed hadrons are closer to the $b$-hadron lifetimes. In addition, the properties of our decay of interest can suppress the background further. In particular, one can require that the displaced vertex formed by the muon track and the inferred $\Lambda$ trajectory (if it is distinct from the primary vertex) should not contain any additional tracks. Based on these arguments, we will assume a charm tagging efficiency of $\epsilon_c \approx 40\%$ and neglect the remaining background.

\begin{table}
\centering 
\begin{tabular}{c|ccc}\hline\hline
    \textbf{$\bm{c\bar c}$, semileptonic} & \multicolumn{3}{c}{polarization} \\\hline 
    $m_{jj}$ cut~[GeV]& $\sigma_{c\bar{c}}$~[pb]& $N_{c}$ & $r_i\Delta b^\pm_i$ \\\hline
    \mbox{no cut} & 2500 & $8.3\times 10^{3}$ & 0.019 \\ 
              100 & 2200 & $7.2\times 10^{3}$ & 0.020 \\ 
              300 &  350 & $1.2\times 10^{3}$ & 0.051 \\ 
              500 &   66 &                160 & 0.14 \\ 
              750 &   14 &                 22 & 0.37 \\     
     \hline\hline
\end{tabular}
\begin{tabular}{|cc}\hline\hline
    \multicolumn{2}{|c}{spin correlations} \\\hline
    $\sigma_{c\bar{c}}$~[pb]& $N_{c\bar{c}}$ \\\hline
    2300 & 2.9 \\ 
    2000 & 2.4 \\ 
     250 &  \\ 
      48 &  \\ 
     9.3 &  \\ 
\hline\hline
\end{tabular}
\caption{Run~2 cross sections (with trigger-motivated cuts) and signal event counts (after the full selection) for measurements of $c$-quark polarization and spin correlations in the semileptonic channel. The expected statistical uncertainties for the polarization measurements are shown as well.}
\label{tab:muon c number of evt}
\end{table}

Table~\ref{tab:muon c number of evt} shows the Run~2 cross sections and numbers of events available for measurements of polarization ($N_c$) and spin correlations ($N_{c\bar c}$). The cross sections are based on the single muon trigger acceptance for the polarization and the double muon trigger for spin correlations. They were obtained using a Pythia simulation of $pp\to c\bar{c}$, where we allowed the charmed hadrons to decay only to final states with a muon and applied the trigger cuts from table~\ref{tab:HL-LHC Run2 compare} to muons from charmed hadrons inside the two leading jets. These cross sections do not include branching ratios. The numbers of events were calculated as
\begin{align}
    N_{c} &= \mathcal{L}\,\sigma_{c\bar{c}}\,f(c\to\Lambda_c^+) \BR(\Lambda_c^+\to \Lambda \mu^+ \nu_\mu) \BR(\Lambda\to p \pi^-)\,\epsilon_\mu\,\epsilon_c\,\epsilon_\Lambda \,,\\
    N_{c\bar{c}} &= \mathcal{L}\,\sigma_{c\bar{c}}\,f^2(c\to\Lambda_c^+) \BR^2(\Lambda_c^+\to \Lambda \mu^+ \nu_\mu) \BR^2(\Lambda\to p \pi^-)\,\epsilon_{c,2}\,\epsilon^2_\Lambda \,,
\end{align}
where $\epsilon_\mu \approx 50\%$ is the efficiency for the muon to pass the isolation requirement of the single muon trigger~\cite{ATLAS:2020gty}, $\epsilon_{c,2} \equiv 2\epsilon_c - \epsilon_c^2$ is the efficiency for any of the two jets to pass charm tagging, and $\epsilon_\Lambda$ includes the decay radius and reconstruction efficiencies of the $\Lambda$, where we again rely on the LRT algorithm~\cite{ATL-PHYS-PUB-2017-014,ATL-PHYS-PUB-2019-013,ATLAS:2023nze}. As discussed in section~\ref{sub:Lambda reco}, for Run~2, in the range of $\Lambda$ decay radii $d_T$ up to $d_T^{\rm max} \approx 300$~mm, the average reconstruction efficiency for each track is expected to be $\epsilon_{\rm track} \approx 65\%$, and for the HL-LHC, in the range of up to $d_T^{\rm max} \approx 400$~mm, it will be $\epsilon_{\rm track} \approx 80\%$. Averaging over the whole range is justified as the mean decay radius is $\langle d_T\rangle \gg d_T^{\rm max}$. We estimate $\langle d_T\rangle$ with the assumption that $p_T^{\Lambda} \sim (0.5/3)\,p_T^{\rm jet}$, where the factor of $0.5$ is the typical $c$-jet momentum fraction carried by the $\Lambda_c^+$ and the division by $3$ roughly accounts for the $\Lambda$ being one out of three decay products. For the jet $p_T$, we used the rough estimate $p_T^{\rm jet} \sim \max(\bar p_T^{\rm jet},m_{jj}^{\rm cut}/3)$ where $\bar p_T^{\rm jet} \approx 115$~GeV is the cut on jets that is equivalent to the cuts of both the single and double muon triggers of Run~2, like we already discussed in a slightly different context in section~\ref{sub:bb cc spin corr}. For the HL-LHC, the equivalent jet cuts are $85$~GeV and $80$~GeV for the single and double muon triggers, respectively. We use $m_{jj}^{\rm cut}/3$ since in the dijet CM frame, each jet has a momentum of $|\bm{p}| \simeq m_{jj}/2$ and its transverse component is smaller, around $p_T \sim m_{jj}/3$ on average.

Table~\ref{tab:muon c number of evt} also shows the expected statistical uncertainty of the polarization measurements, while the number of events for spin correlation measurements in Run~2 is too small to be useful. Table~\ref{tab:muon c number of evt high-lumi} presents the analogous results for the HL-LHC, where spin correlation measurements become feasible. We see that the semileptonic channel is superior to the hadronic channel, except for polarization measurements at high $m_{jj}$ at the HL-LHC, where the two channels are comparable. 

\begin{table}
\centering 
\begin{tabular}{c|ccc}\hline\hline
    \textbf{$\bm{c\bar c}$, semilep.} & \multicolumn{3}{c}{polarization} \\\hline 
    $m_{jj}$ cut~[GeV]& $\sigma_{c\bar{c}}$~[pb]& $N_{c}$ & $r_i\Delta b^\pm_i$ \\\hline
    \mbox{no cut} & 9400 & $1.7\times 10^{6}$ & 0.001 \\ 
              100 & 6700 & $1.2\times 10^{6}$ & 0.002 \\ 
              300 &  620 & $9.8\times 10^{4}$ & 0.005 \\ 
              500 &  110 & $1.1\times 10^{4}$ & 0.017 \\ 
              750 &   23 & $1.6\times 10^{3}$ & 0.043 \\ 
             1000 &  5.4 &                280 & 0.10 \\ 
             1500 & 0.62 &                 22 & 0.37  \\  
    \hline\hline
\end{tabular}
\begin{tabular}{|cccc}\hline\hline
    \multicolumn{4}{|c}{spin correlations} \\\hline
    $\sigma_{c\bar{c}}$~[pb] & $N_{c\bar{c}}$ & $r_i^2\Delta c_{ii}$ & $r_ir_j\Delta c_{ij(\ell)}$ \\\hline
    13000 & $2.4\times10^3$ &  0.060 & 0.042 \\ 
     7700 & $1.5\times10^3$ &  0.078 & 0.054 \\ 
      540 &  70  &  0.35    & 0.25  \\ 
       89 &  4.8   &    \\ 
       16 &  \\ 
      4.1 &  \\ 
      0.47 &  \\ 
    \hline\hline
\end{tabular}
\caption{HL-LHC cross sections (with trigger-motivated cuts) and signal event counts (after the full selection) for measurements of $c$-quark polarization and spin correlations in the semileptonic channel. The expected statistical uncertainties are shown as well. The shorthand $\Delta c_{ij(\ell)}$ denotes the uncertainties in $c_{ij}$ and $c_\ell$, which are of the same size.}
\label{tab:muon c number of evt high-lumi}
\end{table}

\subsubsection{Mixed Channel}
	
We can also look at a mixed channel, where the decay in one of the jets is semileptonic and in the other hadronic. This channel combines the ability to trigger on a muon (with a low $p_T$ threshold) on one side with the higher BR and a cleaner decay (without a neutrino) on the other side.

\begin{table}
\centering
\begin{tabular}{ccc}\hline\hline
    $h_c$ & $f(c\to h_c)$ [\%] & $\BR(h_c\to\mu^+X)$ [\%]\\\hline
    $\Lambda_c^{+}$ &  $6.4$ &  $3.5$ \\
    $D^+$           & $22.7$ & $17.6$ \\
    $D^0$           & $61.8$ &  $6.8$ \\
    $D_s^+$         &  $8.2$ & $\approx 6$ \\
\hline\hline
\end{tabular}
\caption{The fragmentation fractions~\cite{Lisovyi:2015uqa} and branching ratios~\cite{Workman:2022ynf} of inclusive semi-muonic decays of the common charmed hadrons.}
\label{tab: c FF BR-semilep}
\end{table}

We propose using the hadronic side of the event for polarization measurements due to the ability to fully reconstruct the decay kinematics, without the need to account for a neutrino. Moreover, one can then enjoy the \emph{inclusive} semimuonic decays of \emph{all} charmed hadrons (see table~\ref{tab: c FF BR-semilep}) in the second jet to trigger the event. The expected number of signal events in the sample is then
\begin{align}
    N_c &= \mathcal{L}\,\sigma_{c\bar{c}}\,f(c\to\Lambda_c^+) \BR(\Lambda_c^+\to pK^-\pi^+)\,\epsilon_{\Lambda_c^+ \to p K^- \pi^+} \nonumber\\
    &\quad \times \sum_{h_c=\Lambda_c^+,D^+,D^0,D_s^+} f(c\to h_c) \BR(h_c\to\mu^+X)\,\epsilon_\mu \,,
\end{align}
where we assume an efficiency of $\epsilon_{\Lambda_c^+ \to p K^- \pi^+} \approx 25\%$ for the hadronic decay selection~\cite{CMS:2023frs} and $\epsilon_\mu \approx 50\%$ for passing the isolation requirement of the single muon trigger~\cite{ATLAS:2020gty}. The resulting numbers of events are given in table~\ref{tab:mix c B}.

\begin{table}
\centering
\begin{tabular}{c|ccc|ccc}\hline\hline
    \multicolumn{7}{c}{\textbf{$\bm{c\bar c}$, mixed channel, polarization}} \\\hline
    $m_{jj}$ cut & \multicolumn{3}{c|}{Run 2} & \multicolumn{3}{c}{HL-LHC} \\\cline{2-7}
    \,[GeV]\, & $\sigma_{c\bar{c}}$~[pb]&$N_c$&$r_i\Delta b^\pm_i$ & $\sigma_{c\bar{c}}$~[pb]&$N_c$&$r_i\Delta b^\pm_i$ \\\hline
    \mbox{no cut} & 2500 & $1.6\times10^4$ & $0.021-0.080$  & 9400 & $1.3\times10^6$ & $0.002-0.006$ \\ 
              100 & 2200 & $1.3\times10^4$ & $0.022-0.085$  & 6700 & $9.0\times10^5$ & $0.003-0.007$ \\ 
              300 &  350 &            2600 & $0.055-0.21$   & 620 & $8.4\times10^4$ & $0.009-0.024$ \\ 
              500 &   66 &             520 & $0.13-0.49$    & 110 & $1.4\times10^4$ & $0.021-0.059$ \\ 
              750 &   14 &              86 & $\gtrsim 0.28$ & 23 & $3.1\times10^3$ & $0.047-0.13$ \\ 
             1000 &  3.5 &              22 &                & 5.4 &             730 & $0.097-0.26$ \\
             1500 &  0.40 &              2.5&                & 0.62 &              84 & $\gtrsim 0.29$ \\
             \hline\hline
\end{tabular}
\caption{Cross sections, expected numbers of signal events and expected statistical uncertainties for the $c$-quark polarization measurements using the hadronic decay in the mixed channel of $c\bar{c}$. We show a range of values for the statistical uncertainties, corresponding to purities between $100\%$ and $6.9\%$ for Run~2 and $13.8\%$ for the HL-LHC (see text).}
\label{tab:mix c B}
\end{table}

The requirement of the muon in the second jet is also expected to remove part of the background that is observed under the $\Lambda_c^+ \to p K^- \pi^+$ peak in the CMS measurement~\cite{CMS:2023frs}. While it will not eliminate background coming from $c\bar c$ or $b\bar b$, the muon requirement will strongly suppress the combinatorial background from high cross section dijet final states without heavy flavors (e.g., $gg$). Since we do not know the composition of the background observed by CMS, we present in table~\ref{tab:mix c B} a range of values for the statistical uncertainties, for purities varying between $100\%$ and the hadronic decay purity of $6.9\%$ for Run~2 (as in the CMS measurement) and $13.8\%$ for the HL-LHC (recall the discussion in section~\ref{sub:c-hadronic}). The results in table~\ref{tab:mix c B} are comparable to those of the semileptonic channel.
 
For spin correlation measurements, the expected number of signal events is
\begin{align}
    N_{c\bar c} &= \mathcal{L}\,\sigma_{c\bar{c}}\,f^2(c\to\Lambda_c^+) \BR(\Lambda_c^+\to pK^-\pi^+) \BR(\Lambda_c^+\to\Lambda\mu^+\nu_\mu) \BR(\Lambda\to p \pi^-) \nonumber\\
    &\quad \times \epsilon_{\Lambda_c^+ \to p K^- \pi^+}\, \epsilon_\Lambda\,\epsilon_\mu\,\epsilon_c \,,
\end{align}
where $\epsilon_{\Lambda_c^+ \to p K^- \pi^+} \approx 25\%$ is the hadronic decay efficiency~\cite{CMS:2023frs}, $\epsilon_\Lambda$ describes the $\Lambda$ decay acceptance and reconstruction efficiency (estimated as in section~\ref{sub:c semi analysis}), $\epsilon_\mu \approx 50\%$ is the efficiency loss due to the isolation requirement of the single muon trigger~\cite{ATLAS:2020gty}, and $\epsilon_c$ is the charm tagging efficiency discussed in section~\ref{sub:c semi analysis}. Table~\ref{tab:mix c num evt} shows the expected numbers of events for Run~2 and the HL-LHC. While the same single-muon trigger is used here as in the polarization measurements, the cross sections are higher by a factor of $2$ because either a $\mu^+$ from the $c$ jet or a $\mu^-$ from the $\bar c$ jet can trigger the event. This also affects the equivalent $p_T^{\rm jet}$ values we use to estimate the $\Lambda$ displacement acceptance, which become $\bar{p}_T^{\rm jet}=97$~GeV for Run~2 and $\bar{p}_T^{\rm jet}=72$~GeV for the HL-LHC.

\begin{table}
\centering
\begin{tabular}{c|cc|cccc}\hline\hline
    \multicolumn{7}{c}{\textbf{$\bm{c\bar c}$, mixed channel, spin correlations}} \\\hline
    $m_{jj}$ cut & \multicolumn{2}{c|}{Run 2} & \multicolumn{4}{c}{HL-LHC} \\\cline{2-7}
    \,[GeV]\, & $\sigma_{c\bar{c}}$ [pb] & $N_{c\bar{c}}$ & $\sigma_{c\bar{c}}$ [pb] & $N_{c\bar{c}}$ & $r_i^2\Delta c_{ii}$ & $r_ir_j\Delta c_{ij(\ell)}$ \\\hline
    \mbox{no cut} & 5000 & 19 & 19000 & $3.9\times 10^{3}$ & $0.072-0.19$   & $0.050-0.13$  \\ 
              100 & 4300 & 16 & 13000 & $2.7\times 10^{3}$ & $0.085-0.23$   & $0.060-0.16$  \\ 
              300 &  710 &  2.7 &  1200 &                200 & $\gtrsim 0.32$ & $\gtrsim 0.22$ \\
	       \hline\hline
\end{tabular}
\caption{Cross sections, expected numbers of signal events and expected statistical uncertainties for the $c\bar c$ spin correlation measurements in the mixed channel. We show a range of values for the statistical uncertainties, corresponding to purities between $100\%$ and $13.8\%$ (see text).}
\label{tab:mix c num evt}
\end{table} 

While the semileptonic decay selection is almost background-free (as discussed in section~\ref{sub:c semi analysis}), various processes can mimic the hadronic decay. If the background observed under the hadronic $\Lambda_c^+$ peak in the inclusive CMS sample~\cite{CMS:2023frs} is primarily intrinsic (i.e., from $c\bar c$), it will not be affected by the semileptonic selection on the other side of the event and our rough purity estimates will be $6.9\%$ in Run~2 and $13.8\%$ at the HL-LHC. If, on the other hand, the background is mostly extrinsic and comes from a process such as $pp\to gg$, where the jets rarely contain a muon and a $\Lambda$, the semileptonic selection will eliminate most of it and the sample will be almost background-free. For Run~2, the number of events is too small for a meaningful measurement regardless of the background. For the HL-LHC, we show in table~\ref{tab:mix c num evt} a range of values for the expected precision corresponding to the above range of possible purity values. The expected precision is comparable to that of the semileptonic channel. 

\subsection{Analyses of \texorpdfstring{$pp \to b\bar{b}$}{pp->bbbar}}
\label{sub:num event bb}

We propose using the $b\to\Lambda_b \to X_c\mu^-\bar{\nu}_\mu$ process, with the muon-based triggers from table~\ref{tab:HL-LHC Run2 compare}, to measure the polarization and spin correlations in $b\bar b$ events.

As introduced in section~\ref{sec:decay chains}, we will consider three types of selection: Inclusive, Semi-inclusive, and Exclusive. In the Inclusive Selection, no attempt is made to reduce the intrinsic background due to semileptonic $B$-meson decays. Such a selection can still be competitive because of its high signal efficiency. In the Semi-Inclusive selection, we require in addition to the muon the reconstruction of a $\Lambda$ baryon (via its decay $\Lambda \to p\pi^-$) originating from the vicinity of the displaced vertex. This reduces the $B$-meson background. The last selection type is the Exclusive Selection. In this selection, in addition to the muon, we require a full reconstruction of one of $\Lambda_c^+$ decays by a set of tracks consistent with originating from a common vertex. This significantly suppresses the $B$-meson background too. Another advantage of the complete reconstruction of a $\Lambda_c^+$ decay is that the kinematics of the $\Lambda_b$ decay as a whole can then be reconstructed more accurately. In table~\ref{fig:decay_scheme} we list all the decay channels relevant to the signal in each selection type.\footnote{In the Exclusive Selection, we include decays with the $\Sigma^\pm$ baryons in the final state. While these particles will often decay before passing through the entire tracker, reconstruction of such short tracks is possible~\cite{ATLAS:2017oal,ATLAS:2022rme,CMS:2019ybf,CMS:2020atg,CMS:2023mny}.}

\begin{table}
\begin{center}
\begin{tabular}{c|c|c}\hline\hline
    Selection & Decay Modes & Branching Ratio \\\hline
    Inclusive & $\Lambda_b\to X_c\mu^-\bar\nu_\mu$ & 11\%\\\hline
    \multirow{2}{*}{Semi-inclusive}
        & $\Lambda_c^+\to \Lambda X$ & 38\%\\
        & $\Lambda\to p\pi^-$ & 64\%\\\hline
    \multirow{8}{*}{Exclusive}
        & $\Lambda_c^+\to p K^- \pi^+$ & 6.3\% \\
	& $\Lambda_c^+\to\Lambda\pi^+ \to p \pi^- \pi^+$ & 0.8\% \\
	& $\Lambda_c^+\to pK_S\to p \pi^- \pi^+$ & 1.1\% \\
	& $\Lambda_c^+\to \Lambda \pi^+\pi^+\pi^-\to p\pi^+\pi^+\pi^-\pi^-$ & 2.3\% \\
	& $\Lambda_c^+\to p K_S \pi^+ \pi^- \to p \pi^+ \pi^+ \pi^- \pi^-$ & 1.1\% \\
	& $\Lambda_c^+\to \Sigma^+ \pi^+ \pi^-$ & 4.5\%\\
	& $\Lambda_c^+\to \Sigma^- \pi^+ \pi^+$ & 1.9\%\\\cline{2-3}
        & total & 18\% \\\hline\hline
\end{tabular}
\end{center}
\caption{Decay modes relevant to the three $\Lambda_b$ selections and their branching ratios~\cite{Workman:2022ynf}. The Semi-inclusive and Exclusive selections are done on top of the Inclusive selection.}
\label{fig:decay_scheme}
\end{table}

\subsubsection{Inclusive Selection}
\label{sub:inclusive}

The main requirement of this selection is the presence of a muon in the jet, similar to \emph{soft muon $b$ tagging}~\cite{CMS:2012feb,ATLAS:2015thz,ATLAS:2016lxn,CMS:2017wtu,ATLAS:2021piz,ATLAS:2022jbw}.

To suppress extrinsic backgrounds with prompt or mildly displaced (in particular, $c\bar c$) muons, we assume applying a $b$ tagging algorithm that may use the muon impact parameter significance, or $p_T^{\rm rel}$ (the component of muon momentum transverse to the jet axis)~\cite{ATLAS:2015thz}, or other variables. For our estimates, we will assume the $b$ tagging efficiency to be $\epsilon_b \approx 80\%$ and allow ourselves to neglect the above backgrounds.

Another source of extrinsic background is $pp \to t\bar{t} \to b\bar{b}X$. This background will be small since the $t\bar t$ cross section is smaller than the $b\bar b$ cross section for any fixed $b\bar b$ invariant mass. It can be suppressed further by vetoing the presence of additional objects, such as isolated leptons or multiple energetic jets.

\begin{figure}
\centering
\includegraphics[width=0.51\linewidth, trim =0cm 0cm 0cm 0cm,clip]{"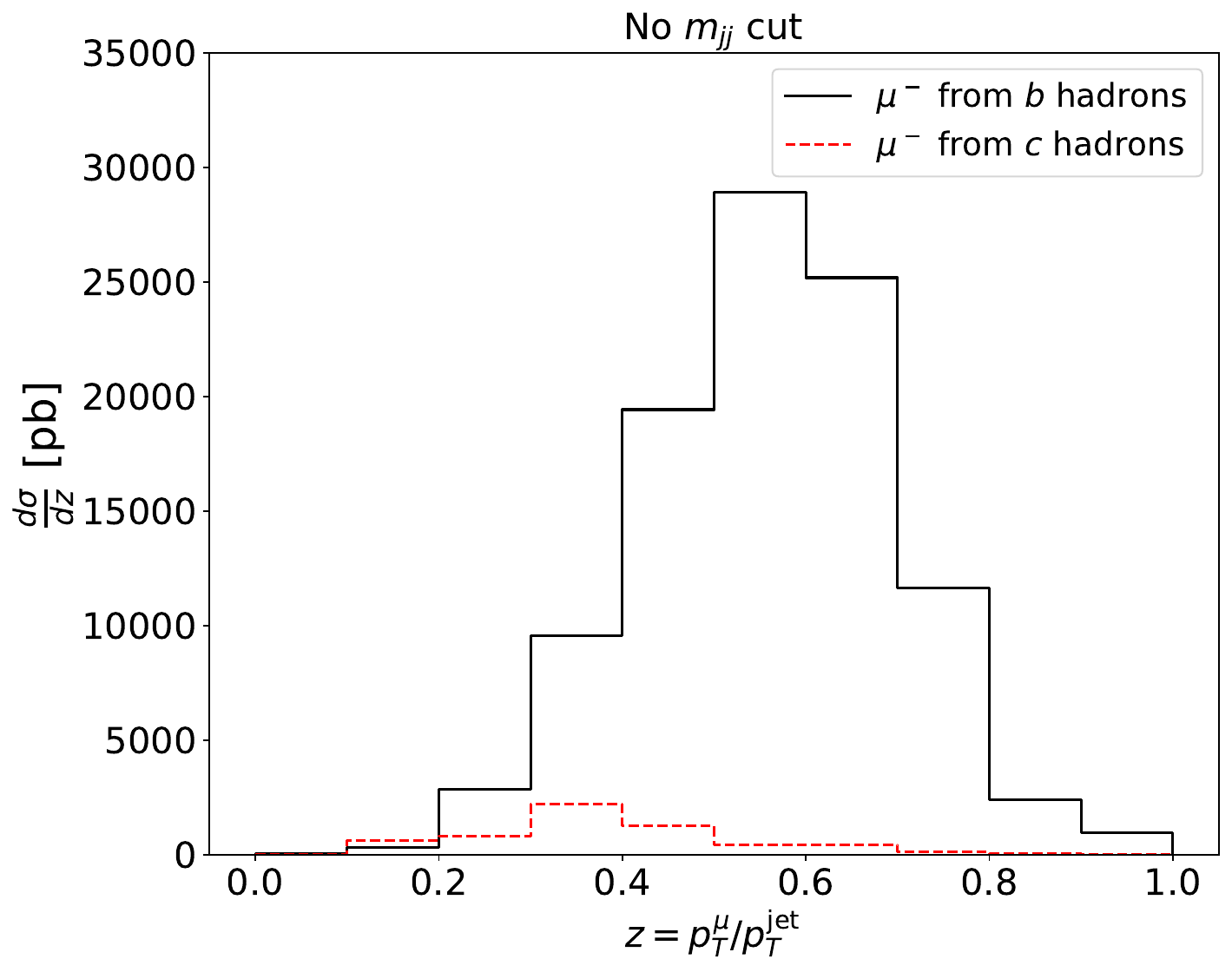"}
\includegraphics[width=0.485\linewidth, trim =0cm 0cm 0cm 0cm,clip]{"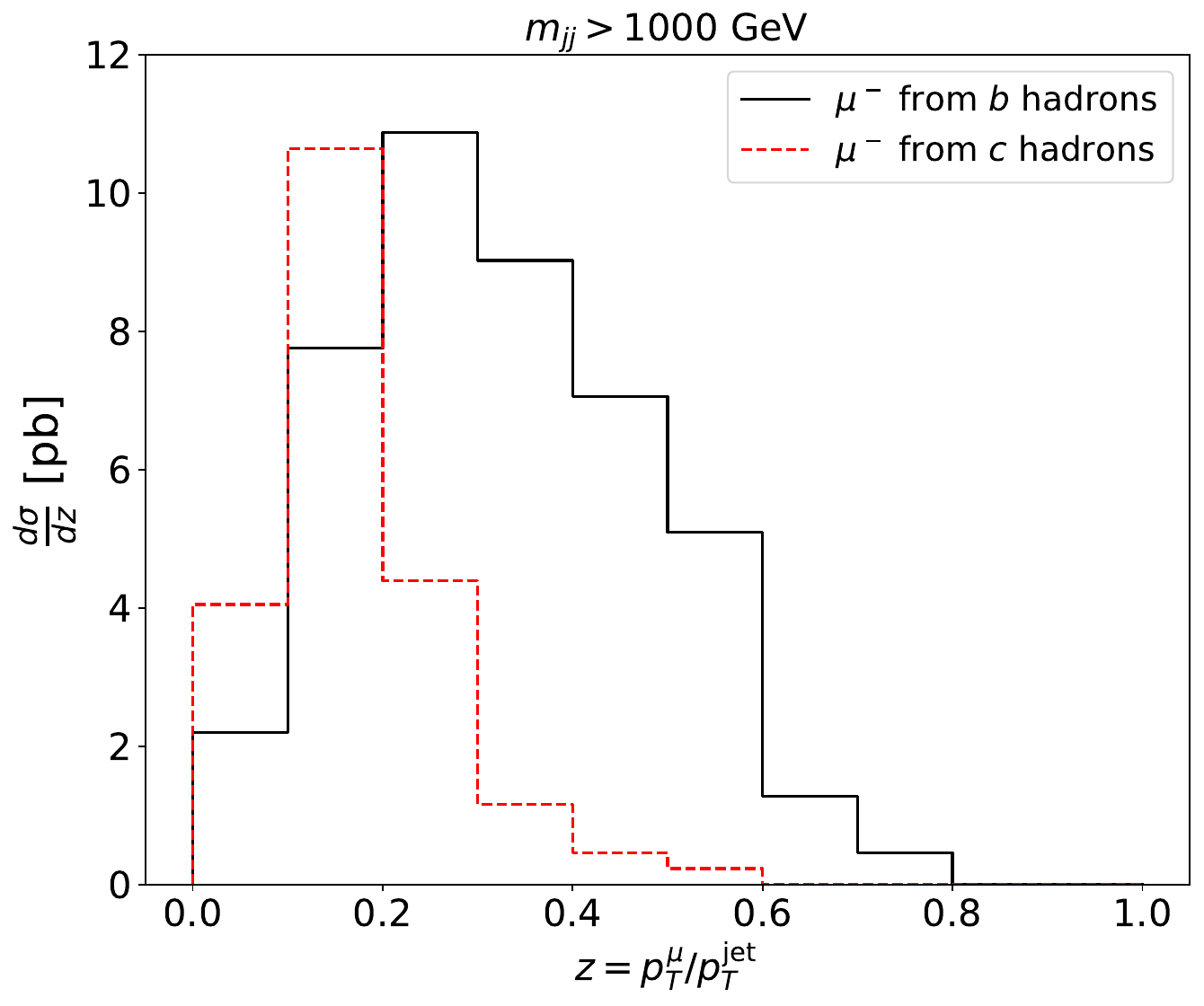"}
\caption{The momentum fraction $z$ of the muons relative to their respective jets. The graphs compare the distributions for muons originating directly from $b$ hadrons with those from $c$ hadrons in $b$ jets, with no invariant mass cut (left) and with $m_{jj}>1000$~GeV (right). (The plot normalizations do not include the branching ratios.)}
\label{fig:Z dist}
\end{figure}
	
The intrinsic background is the more prominent one. An important contribution is made by the semileptonic $B$-meson decays via the same underlying process as in the signal, $b\to c\mu^-\bar\nu_\mu$, but we do not attempt to reduce it in the inclusive selection as we want to keep the signal efficiency high. Another source of intrinsic background is the decay chain $b\to cf\bar f'$, $c\to s \mu^+ \nu_\mu$. The fact that the muon in this case has the wrong charge cannot be used to eliminate this background because in the inclusive selection we do not know which jet is coming from the $b$ and which from the $\bar b$. There is a different way to reduce this background significantly. The muons emitted in $b\to c$ transitions (as in our signal) are usually more energetic than those emitted in the $c\to s$ transitions in the $b\to c\to s$ chain (which are background), so it can be useful to look into the ratio between the muon and jet $p_T$, $z=p_T^\mu/p_T^{\rm jet}$. Figure~\ref{fig:Z dist} shows the $z$ distributions for muons from both $b$ and $c$ hadrons in $pp\to b\bar{b}$ events with $\sqrt{s}=13$~TeV simulated in Pythia with the single-muon trigger cuts. In a sample without any further cuts, the contribution of muons from $c$-hadron decays is small to begin with, due to their lower efficiency to pass the trigger $p_T$ cut. For a dijet invariant mass cut of $m_{jj}>1000$~GeV, their relative contribution is significant, but concentrated at lower values of $z$ than the muons from $b$ hadrons. Based on these results, we will apply a cut of $z > 0.2$ in all cases. In the example with the $m_{jj} > 1000$~GeV cut, this $z$ cut results in decent efficiency and purity ($\epsilon_z \approx 77\%$, $f_z \approx 84\%$), while the impact on the no-cut case is small ($\epsilon_z \approx 99\%$, $f_z \approx 94\%$). From these results, we find that the $c$ hadrons background is small and we will neglect it.

The numbers of events available for measurements of polarization and spin correlations, respectively, are given by
\begin{align}
    N_{b}^i &= \mathcal{L}\,\sigma_{b\bar{b}}\,f(b\to\Lambda_b) \BR(\Lambda_b\to X_c\mu^-\bar\nu_\mu)\, \epsilon_b\, \epsilon_z\, \epsilon_\mu \,,\\
    N_{b\bar{b}}^{ii} &= \mathcal{L}\,\sigma_{b\bar{b}}\,f^2(b\to\Lambda_b) \BR^2(\Lambda_b\to X_c\mu^-\bar\nu_\mu)\, \epsilon_{b,2}\, \epsilon_{z,2} \,,
\end{align}
where $\epsilon_\mu \approx 50\%$ is the efficiency for the muon to pass the isolation requirement of the single muon trigger~\cite{ATLAS:2020gty}, $\epsilon_{b,2} \equiv 2\epsilon_b - \epsilon_b^2$ is the efficiency for any of the two jets to pass the $b$ tagging condition, and $\epsilon_{z,2} \equiv 2\epsilon_z - \epsilon_z^2$ is the efficiency for any of the two muons to pass the momentum fraction cut.\footnote{Once one of the muons is classified as signal, the second muon can be classified as signal or background based on its charge.} The cross sections were obtained using a Pythia simulation of $pp\to b\bar{b}$, where we allowed the bottom hadrons to decay only to final states with a muon (not including cases in which the muon comes from a charmed hadron) and applied the single or double muon trigger cuts from table~\ref{tab:HL-LHC Run2 compare} to muons from bottom hadrons inside the two leading jets. These cross sections do not include branching ratios. The resulting numbers of events are shown in tables~\ref{fig:b num with pt frag} and \ref{fig:b num with pt frag hi lumi} for Run~2 and the HL-LHC, respectively, along with the numbers obtained for the other selections. While here we considered the standard muon-based triggers, table~\ref{fig:b num with pt frag} also presents numbers corresponding to the CMS parked dataset. It will be discussed separately in section~\ref{sub:parked data}.

\begin{table}
\centering
\begin{subtable}[b]{1\linewidth}
\centering
\begin{tabular}{c|cccc}\hline\hline
    $m_{jj}$ cut [GeV] & $\sigma_{b\bar{b}}$ [pb]& $N^i_b$& $N^s_b$& $N^e_b$\\\hline
    \mbox{no cut} & 9400 & $4.0\times 10^{6}$ & $2.0\times 10^{5}$ & $3.6\times 10^{5}$\\ 
    100 & 7200 & $3.1\times 10^{6}$ & $1.5\times 10^{5}$ & $2.8\times 10^{5}$\\ 
    300 & 560 & $2.3\times 10^{5}$ & $1.0\times 10^{4}$ & $2.1\times 10^{4}$\\ 
    500 & 82 & $3.1\times 10^{4}$ & $1.0\times10^3$ & $2.8\times 10^{3}$\\ 
    750 & 14 & $4.8\times 10^{3}$ & 120 & 430\\ 
    1000 & 3.4 & $1.1\times 10^{3}$ & 22 & 100\\ 
    1500 & 0.34 & 100 &   & 9.2\\ 
    \hline
    parked data & &$3.5\times 10^8$ & $3.6\times10^7$ & $3.1\times10^7$\\
    \hline
    purity $f$ [\%] & & $7.4$ & $57$ & $66$ \\
    \hline\hline
\end{tabular}
\caption{Polarization}
\label{b with pt frag: a}
\end{subtable}
\newline\newline
\begin{subtable}[b]{1\linewidth}
\centering
\begin{tabular}{c|ccccccc}\hline\hline
    $m_{jj}$ cut & $\sigma_{b\bar{b}}$ & \multirow{2}{*}{$N^{ii}_{b\bar b}$} & \multirow{2}{*}{$N^{ss}_{b\bar b}$} & \multirow{2}{*}{$N^{ee}_{b\bar b}$} & \multirow{2}{*}{$N^{is}_{b\bar b}$} & \multirow{2}{*}{$N^{ie}_{b\bar b}$} & \multirow{2}{*}{$N^{se}_{b\bar b}$} \\
    \,[GeV]\, & \,[pb]\, & \\\hline
    \mbox{no cut} & 10000 & $8.0\times 10^{4}$ & 200 & 640 & $8.1\times 10^{3}$ & $1.4\times 10^{4}$ & 730\\ 
    100 & 5900 & $4.7\times 10^{4}$ & 121 & 380 & $4.8\times 10^{3}$ & $8.5\times 10^{3}$ & 430\\ 
    300 & 340 & $2.7\times 10^{3}$ & 5.0 & 21 & 230 & 490 & 20\\ 
    500 & 46 & 360 &   & 2.9 & 20 & 65 & 1.8\\ 
    \hline
    parked data & & $1.1\times10^6$ & $1.1\times10^4$ & 8700 & $2.2\times10^5$ & $1.9\times10^5$ & $2.0\times10^4$\\
    \hline
    purity $f$ [\%]&&$0.55$&$32$&$44$&$4.2$&$4.9$&$38$\\
    \hline\hline
    \end{tabular}
\caption{Spin correlations}
\label{b with pt frag: b}
\end{subtable}
\caption{Run~2 cross sections (with trigger-motivated cuts) and signal event counts (after the full selection) for measurements of (a) $b$-quark polarization, and (b) $b\bar b$ spin correlations in the inclusive, semi-inclusive, and exclusive selection channels. The last row shows the expected sample purity.}
\label{fig:b num with pt frag}
\end{table}

\begin{table}
\centering
\begin{subtable}[b]{1\linewidth}
\centering
\begin{tabular}{c|cccc}\hline\hline
    $m_{jj}$ cut [GeV] & $\sigma_{b\bar{b}}$ [pb]& $N^i_b$& $N^s_b$& $N^e_b$\\\hline
    \mbox{no cut} &33000 & $3.0\times 10^{8}$ & $3.3\times 10^{7}$ & $2.7\times 10^{7}$\\ 
    100 & 18300 & $1.7\times 10^{8}$ & $1.8\times 10^{7}$ & $1.5\times 10^{7}$\\ 
    300 & 990 & $8.2\times 10^{6}$ & $7.3\times 10^{5}$ & $7.4\times 10^{5}$\\ 
    500 & 126 & $9.3\times 10^{5}$ & $6.3\times 10^{4}$ & $8.4\times 10^{4}$\\ 
    750 & 21 & $1.5\times 10^{5}$ & $7.6\times 10^{3}$ & $1.4\times 10^{4}$\\ 
    1000 & 5.6 & $3.7\times 10^{4}$ & $1.6\times 10^{3}$ & $3.3\times 10^{3}$\\ 
    1500 & 0.49 & $2.9\times 10^{3}$ & 94 & 260\\ 
    2000 & 0.088 & 490 & 12 & 44\\ 
    \hline
    purity $f$ [\%]&&$7.4$&$57$&$66$\\
    \hline\hline
\end{tabular}
\caption{Polarization}
\label{b with pt frag hi lumi: a}
\end{subtable}
\newline\newline
\begin{subtable}[b]{1\linewidth}
\centering
\begin{tabular}{c|ccccccc}\hline\hline
    $m_{jj}$ cut & $\sigma_{b\bar{b}}$ & \multirow{2}{*}{$N^{ii}_{b\bar b}$} & \multirow{2}{*}{$N^{ss}_{b\bar b}$} & \multirow{2}{*}{$N^{ee}_{b\bar b}$} & \multirow{2}{*}{$N^{is}_{b\bar b}$} & \multirow{2}{*}{$N^{ie}_{b\bar b}$} & \multirow{2}{*}{$N^{se}_{b\bar b}$} \\
    \,[GeV]\, & \,[pb]\, \\\hline
    \mbox{no cut} & 39000 & $6.7\times 10^{6}$ & $8.1\times 10^{4}$ & $5.4\times 10^{4}$ & $1.5\times 10^{6}$ & $1.2\times 10^{6}$ & $1.3\times 10^{5}$\\ 
    100 & 15000 & $2.6\times 10^{6}$ & $3.1\times 10^{4}$ & $2.1\times 10^{4}$ & $5.7\times 10^{5}$ & $4.7\times 10^{5}$ & $5.1\times 10^{4}$\\ 
    300 & 570 & $9.6\times 10^{4}$ & 610 & 780 & $1.5\times 10^{4}$ & $1.7\times 10^{4}$ & $1.4\times 10^{3}$\\ 
    500 & 74 & $1.2\times 10^{4}$ & 35 & 98 & $1.3\times 10^{3}$ & $2.2\times 10^{3}$ & 120\\ 
    750 & 12 & $2.0\times 10^{3}$ & 3.0 & 16 & 150 & 360 & 13\\ 
    1000 & 2.9 & 460 &   & 3.7 & 27 & 82 & 2.5\\ 
    \hline
    purity $f$ [\%]&&$0.55$&$32$&$44$&$4.2$&$4.9$&$38$\\
    \hline\hline
\end{tabular}
\caption{Spin correlations}
\label{b with pt frag hi lumi: b}
\end{subtable}
\caption{HL-LHC cross sections (with trigger-motivated cuts) and signal event counts (after the full selection) for measurements of (a) $b$-quark polarization, and (b) $b\bar b$ spin correlations in the inclusive, semi-inclusive, and exclusive selection channels. The last row shows the expected sample purity.}
\label{fig:b num with pt frag hi lumi}
\end{table}

\begin{table}
\centering
\begin{tabular}{ccc}\hline\hline
    Process & FF [\%] & BR [\%] \\\hline
    $b \to \Lambda_b \to \mu^-\bar\nu_\mu X$   &  7.0 &  11  \\\hline
    $b \to \bar{B}^0 \to \mu^-\bar\nu_\mu X$   & 40.1 & 10.3 \\
    $b \to B^- \to \mu^-\bar\nu_\mu X$         & 40.1 & 11.0 \\
    $b \to \bar{B}^0_s \to \mu^-\bar\nu_\mu X$ & 10.4 & 10.2 \\
    \hline\hline
\end{tabular}
\caption{The fragmentation fractions (FF)~\cite{Galanti:2015pqa,HFLAV:2014fzu} and branching ratios (BR)~\cite{Workman:2022ynf} of the signal ($\Lambda_b$) and the intrinsic background in the Inclusive Selection.}
\label{tab: b FF and BR}
\end{table}

By accounting for the fragmentation fractions and semi-muonic decay branching ratios of the $\bar B^0$, $B^-$, and $\bar B_s^0$ mesons (see table~\ref{tab: b FF and BR}), we estimate the sample purity to be $f \approx 7.4\%$ for a single jet. For spin correlation measurements, which involve two jets, this number is squared, giving the purity of $f \approx 0.55\%$. Such a low purity can be problematic as it means that even a small mismodelling of the background can ruin the measurement. The next two selection strategies that we discuss offer much higher purities, at the expense of statistics.

\subsubsection{Exclusive Selection}
\label{sub:exclusive}

The exclusive selection starts with the inclusive selection and requires, in addition, a fully reconstructed $\Lambda_c^+$ decay, to eliminate most of the background due to the semileptonic $B$-meson decays, which usually produce charmed mesons rather than baryons.

Some of the $\Lambda_c^+$ decay channels listed in table~\ref{fig:decay_scheme} come with an efficiency price: $K_S$ and $\Lambda$ can decay too far to be reconstructed, $\Sigma^\pm$ can produce a kinked track, and the channels with five tracks need each track to be successfully reconstructed. To account for this, as a rough estimate, we will assume a reconstruction efficiency of $\epsilon_{\rm reco} \approx 50\%$ for the $\Lambda_c^+$ decays.

The numbers of events available for measurements of polarization and spin correlations, respectively, using the exclusive channel are given by
\begin{align}
    N_{b}^e &= \mathcal{L}\,\sigma_{b\bar{b}}\,f(b\to\Lambda_b) \BR(\Lambda_b\to X_c\mu^-\bar\nu_\mu) \BR(\Lambda_c^+ \to {\rm reco.})\, \epsilon_b\, \epsilon_z\, \epsilon_{\rm reco}\, \epsilon_\mu \,,\\
    N_{b\bar{b}}^{ee} &= \mathcal{L}\,\sigma_{b\bar{b}}\,f^2(b\to\Lambda_b) \BR^2(\Lambda_b\to X_c\mu^-\bar\nu_\mu) \BR^2(\Lambda_c^+ \to {\rm reco.})\, \epsilon_{b,2}\, \epsilon_{z,2}\, \epsilon_{\rm reco}^2 \,,
\end{align}
where $\BR(\Lambda_c^+ \to {\rm reco.}) \approx 18\%$ is the branching fraction of the reconstructible decay modes of the $\Lambda_c^+$ from table~\ref{fig:decay_scheme} and $\epsilon_\mu \approx 50\%$ is the efficiency for the muon to pass the isolation requirement of the single muon trigger~\cite{ATLAS:2020gty}. The resulting numbers of events are included in tables~\ref{fig:b num with pt frag} and \ref{fig:b num with pt frag hi lumi}.

\begin{table}
\centering
\begin{tabular}{ccc}\hline\hline
    Process & FF [\%] & BR [\%] \\\hline
    $b \to \Lambda_b \to \Lambda_c^+X$   &  7.0 & $\approx 100$ \\\hline
    $b \to B^- \to \Lambda_c^+X$         & 40.1 & $2.8^{+1.1}_{-0.9}$ \\
    $b \to \bar{B}^0 \to \Lambda_c^+X$   & 40.1 & $5.0^{+2.1}_{-1.5}$ \\
    $b \to \bar{B}_s^0 \to \Lambda_c^+X$ & 10.4 & unknown \\
    $\bar{b} \to B^+ \to \Lambda_c^+X$   & 40.1 & $2.1^{+0.9}_{-0.6}$ \\
    $\bar{b} \to B^0 \to \Lambda_c^+X$   & 40.1 & $<3.1$ \\
    $\bar{b} \to B_s^0 \to \Lambda_c^+X$ & 10.4 & unknown \\
    \hline\hline
\end{tabular}
\caption{The fragmentation fractions (FF)~\cite{Galanti:2015pqa,HFLAV:2014fzu} and branching ratios (BR)~\cite{Workman:2022ynf} for $\Lambda_c^+$ production from the $\Lambda_b$ (signal) and $B$ mesons (background).}
\label{tab:bbbar-exclusive}
\end{table}

A remaining background in this channel is the semileptonic decays of $B$ mesons to final states with a $\Lambda_c^+$ baryon. While the branching ratios of the meson decays to baryons are small, their contribution is enhanced by their large fragmentation fractions compared with that of the $\Lambda_b$, as shown in table~\ref{tab:bbbar-exclusive}. The branching ratios for $B$-meson decays involving a $\mu^-$ in addition to the $\Lambda_c^+$ have not been measured, but such decays definitely exist. In the case of the $b$ initial state, the muon can be produced in the $b \to (c\to\Lambda_c^+)$ transition, while in the case of the $\bar b$ initial state it is produced following the $\bar b \to \bar c\,(c\to\Lambda_c^+)\,\bar s$ chain, in the subsequent $\bar c \to \bar s$ transition. Events with muons from the latter source will largely fail the cut on the muon momentum fraction $z$ discussed in section~\ref{sub:inclusive}, so only the $B^-$, $\bar B^0$ and $\bar B_s^0$ decays from table~\ref{tab:bbbar-exclusive} actually contribute to the background. Summing up their contributions (while assigning $\bar B_s^0$ the average of the $B^-$ and $\bar B^0$ BRs) and assuming the probability of having a muon to be similar to the $\Lambda_b$ case, where the muon is also produced in a $b \to c$ transition, we obtain a sample purity of $f = 66\%$. Since the muons and neutrinos in these background processes are produced in the decays of spinless mesons, their impact on the angular distributions will be trivial.

\subsubsection{Semi-inclusive Selection}

The semi-inclusive selection starts with the inclusive selection and requires, in addition, the presence of a reconstructed $\Lambda$ baryon. This requirement eliminates most of the contributions from $B$-meson decays while accepting a large fraction of $\Lambda_b$ decays since $\Lambda_b$ decays usually proceed through a $\Lambda_c^+$, which in turn produces a $\Lambda$ in about $38\%$ of the cases~\cite{Workman:2022ynf}.

The numbers of events in this channel, for measurements of polarization and spin correlations, respectively, are given by
\begin{align}
    N_{b}^s &= \mathcal{L}\,\sigma_{b\bar{b}}\,f(b\to\Lambda_b) \BR(\Lambda_b\to X_c\mu^-\bar\nu_\mu) \BR(\Lambda_c^+ \to \Lambda X) \BR(\Lambda \to p\pi^-)\, \epsilon_b\, \epsilon_z\, \epsilon_\Lambda\, \epsilon_\mu \,,\\
    N_{b\bar{b}}^{ss} &= \mathcal{L}\,\sigma_{b\bar{b}}\,f^2(b\to\Lambda_b) \BR^2(\Lambda_b\to X_c\mu^-\bar\nu_\mu) \BR^2(\Lambda_c^+ \to \Lambda X) \BR^2(\Lambda \to p\pi^-)\, \epsilon_{b,2}\, \epsilon_{z,2}\, \epsilon_\Lambda^2 \,,
\end{align}
where $\epsilon_\mu \approx 50\%$ is the efficiency for the muon to pass the isolation requirement of the single muon trigger~\cite{ATLAS:2020gty} and $\epsilon_\Lambda$ is the $\Lambda$ decay reconstruction efficiency, where we base our estimates on the ATLAS reconstruction with the LRT algorithm discussed in section~\ref{sub:Lambda reco}. The mean track reconstruction efficiency is $\sim 65\%$ ($\sim 80\%$) in the range of decay radii up to 300~mm (400~mm) for Run~2 (HL-LHC). The typical $b$-jet $p_T$ corresponding to the muons in the jets passing the double-muon trigger is $\bar p_T^{\rm jet} \approx 79$~GeV ($59$~GeV) for Run~2 (HL-LHC). For the single-muon trigger it is $81$~GeV ($62$~GeV) for Run~2 (HL-LHC). By an analytical calculation we find that with\footnote{The factor of $0.7$ is the typical $b$-jet momentum fraction carried by a $b$ hadron, and the $1/3^2$ factor is a rough estimate of the momentum splitting in a typical semileptonic $\Lambda_b$ decay and the subsequent $\Lambda_c^+$ decay.} $p_T^{\Lambda} \sim (0.7/3^2)\,p_T^{\rm jet}$, 54\% of the $\Lambda$ satisfy the decay radius condition for the Run~2 (with $p_T^{\rm jet}=79$~GeV) and 76\% satisfy it for the HL-LHC (with $p_T^{\rm jet}=59$~GeV). This efficiency will be lowered significantly when we apply high invariant mass cuts. We account for that by taking $p_T^{\rm jet} \sim \max(\bar p_T^{\rm jet},m_{jj}^{\rm cut}/3)$ in our estimate of the typical $p_T^\Lambda$. The resulting numbers of events are included in tables~\ref{fig:b num with pt frag} and \ref{fig:b num with pt frag hi lumi}.

The background in this channel can be estimated using the inclusive measurement of $\Lambda$ baryon production in $b$-hadron decays, after subtracting the signal contribution. We have already discussed this measurement, eq.~\eqref{eq:b2Lambda}, in section~\ref{sub:c semi analysis} in the context of the background for the semileptonic $c\bar c$ channel. Following similar arguments, including an assumption of a $10\%$ probability of having a $\mu^-$ alongside the $\Lambda$ (as a rough estimate), we obtain the number \emph{``Total''} in the second row in table~\ref{tab:bbbar-semi}. Part of it is the signal contribution, which is shown in the first row of the table. After subtracting it, we are left with the background, which is shown in the third row. The last row shows an estimate for the part of the background coming from processes involving the $\Lambda_c^+$ that are listed in table~\ref{tab:bbbar-exclusive}.\footnote{For the $B^0$ BR, we took its upper bound. For the $B_s^0$ and $\bar B_s^0$, we took the BR to be the average of the other BRs.} This contribution seems to account for the majority of the background (although one should keep in mind the large uncertainties in the input data and the assumptions made). As we discussed in section~\ref{sub:exclusive}, only the $b$-initiated processes from table~\ref{tab:bbbar-exclusive}, which account for roughly $60\%$ of the background contributions in that table, will pass the cut on the muon $p_T$ fraction $z$. After subtracting that contribution we obtain a sample purity of $f \approx 57\%$.

\begin{table}
\centering
\begin{tabular}{l|c|ccc}\hline\hline
    & Process & FF [\%] & BR [\%] & FF$\times$BR [\%] \\\hline
    \emph{Signal} & $b\to \Lambda_b \to \Lambda\mu^-\bar\nu_\mu X$ & 7.0 & 3.8 & 0.29 \\
    \emph{Total}  & $b$+$\bar{b} \to \mbox{(hadrons)} \to \Lambda\mu^-\bar\nu_\mu X$ & & & 0.59 \\
    \emph{Background} & & & & 0.30 \\
    \qquad via $\Lambda_c^+$ & $b$+$\bar{b} \to \mbox{(mesons)} \to (\Lambda_c^+\to\Lambda X)\mu^-\bar\nu_\mu X$ & & & 0.20 \\
    \hline\hline
\end{tabular}
\caption{Properties of the $b\bar b$ signal and background in the semi-inclusive channel: fragmentation fractions (FF) and branching ratios (BR). The number shown for \emph{Total} is already summed over the two jets in the event and involves the assumption that roughly $10\%$ of the events in the $\Lambda X$ sample contain a $\mu^-$. The \emph{Background} number is obtained as the difference between \emph{Total} and \emph{Signal}. The last row shows the part of the background that comes from the processes involving the $\Lambda_c^+$ shown in table~\ref{tab:bbbar-exclusive}.}
\label{tab:bbbar-semi}
\end{table}

\subsubsection{Mixed Selections}

For spin correlation measurements, one can also consider one type of selection (inclusive, semi-inclusive, or exclusive) for one of the jets and another type for another. The numbers of events for these mixed selections are given by
\begin{align}
    N_{b\bar{b}}^{is} &= 2\,\mathcal{L}\,\sigma_{b\bar{b}}\,f^2(b\to\Lambda_b) \BR^2(\Lambda_b\to X_c\mu^-\bar\nu_\mu) \BR(\Lambda_c^+ \to \Lambda X) \BR(\Lambda \to p\pi^-)\, \epsilon_{b,2}\, \epsilon_{z,2}\, \epsilon_\Lambda \;, \\
    N_{b\bar{b}}^{ie} &= 2\,\mathcal{L}\,\sigma_{b\bar{b}}\,f^2(b\to\Lambda_b) \BR^2(\Lambda_b\to X_c\mu^-\bar\nu_\mu) \BR(\Lambda_c^+ \to {\rm reco.})\, \epsilon_{b,2}\, \epsilon_{z,2}\, \epsilon_{\rm reco} \;, \\
    N_{b\bar{b}}^{se} &= 2\,\mathcal{L}\,\sigma_{b\bar{b}}\,f^2(b\to\Lambda_b) \BR^2(\Lambda_b\to X_c\mu^-\bar\nu_\mu) \BR(\Lambda_c^+ \to \Lambda X) \BR(\Lambda \to p\pi^-) \nonumber\\
    &\quad \times \BR(\Lambda_c^+ \to {\rm reco.})\, \epsilon_{b,2}\, \epsilon_{z,2}\, \epsilon_\Lambda\, \epsilon_{\rm reco}
\end{align}
for the inclusive/semi-inclusive, inclusive/exclusive and semi-inclusive/exclusive selections, respectively. We included a factor of two since different selections can be applied to each of the jets interchangeably and contribute the same. The resulting numbers are included in tables~\ref{fig:b num with pt frag} and \ref{fig:b num with pt frag hi lumi}.

\subsubsection{CMS Parked Data}
\label{sub:parked data}

It is also interesting to consider the CMS \emph{parked} $b$ dataset, which was collected during part of Run~2 using a special data acquisition strategy~\cite{CMS-DP-2019-043,Bainbridge:2020pgi,CMS:2024syx}.\footnote{Similar data parking was done during part of Run~3~\cite{CMS:2023gfb}. We are unaware of the specifics of data parking that might be planned for the HL-LHC, so we will not address that case. We hope that the analyses we propose in this paper, for the $b$, $c$, and $s$ quarks, will add to the motivation for such data streams.} The dataset contains about $N_0 = 10^{10}\,$ $pp \to b\bar b$ events, in which one side of the event includes a displaced muon that triggered the event, with a $p_T$ threshold that varied between $7$ and $12$~GeV. We will consider using these semileptonic decays with muons for the $b$ and $\bar b$ polarization measurements. We will also consider spin correlation measurements, in which we will demand a semileptonic decay with a muon also on the other side of the event.

The number of events available for the polarization measurements with the inclusive selection is
\begin{equation}
    N_b^i \simeq \frac12\,f(b\to\Lambda_b)\,N_0 \approx 3.5 \times 10^8 \,,
\end{equation}
where we accounted for the fact that only half of the triggering decays come from a $b$ (while the other half are from a $\bar b$) and took the approximation that the inclusive semileptonic branching ratios and muon kinematics are approximately the same for the different $b$ hadrons. We neglect the contribution of muons from the $b \to c \to \mu$ chain relative to the contribution of the direct $b \to \mu$ process since the more energetic muons from the direct decay pass the trigger threshold much more easily, similar to what we have seen in figure~\ref{fig:Z dist} (left). For the semi-inclusive and exclusive selections, we compute the number of events as follows:
\begin{align}
    N_b^s &= N_b^i\, \BR(\Lambda_c^+ \to \Lambda X) \BR(\Lambda \to p\pi^-)\, \epsilon_\Lambda \;, \\
    N_b^e &= N_b^i\, \BR(\Lambda_c^+ \to {\rm reco.})\, \epsilon_{\rm reco} \;.
\end{align}
For the semi-inclusive selection, our rough estimate for the $\Lambda$ reconstruction efficiency, $\epsilon_\Lambda$, will be as follows. Since the muon that triggers the event, with a $p_T$ threshold that varied between $7$ and $12$~GeV~\cite{CMS-DP-2019-043,Bainbridge:2020pgi,CMS:2024syx}, is produced in the $\Lambda_b$ decay together with the $\Lambda_c^+$, which subsequently decays to the $\Lambda$ and additional particles, we estimate $p_T^\Lambda$ to be around a few GeV. As a result, the $\Lambda$ baryons will usually decay sufficiently early to be reconstructed by an analogue of the LRT algorithm~\cite{ATL-PHYS-PUB-2017-014,ATL-PHYS-PUB-2019-013,ATLAS:2023nze}, but not at the shortest distances where the efficiency is maximal. We will therefore take $\epsilon_\Lambda \simeq \epsilon_{\rm track}^2$ with $\epsilon_{\rm track} \approx 65\%$ as a ballpark figure.

Our rough estimate for the number of events for the spin correlation measurement with the inclusive selection on both sides will be
\begin{equation}
    N_{b\bar b}^{ii} \simeq f^2(b\to\Lambda_b)\BR(\Lambda_b \to X_c\mu^-\bar\nu_\mu)\,{\cal A}\epsilon\,N_0 \approx 1.1 \times 10^6 \,,
\end{equation}
where we took a ballpark figure of ${\cal A}\epsilon \sim 0.2$ for the acceptance times efficiency on the non-triggering side of the event (including the acceptance in $\eta$). The contribution of $b \to c \to \mu$ decays from that side of the event can be eliminated based on the muon charge (compared with the triggering muon). For the other selections, we have
\begin{align}
    N_{b\bar b}^{ss} &= N_{b\bar b}^{ii}\, \BR^2(\Lambda_c^+ \to \Lambda X) \BR^2(\Lambda \to p\pi^-)\, \epsilon_\Lambda^2 \;, \\
    N_{b\bar b}^{ee} &= N_{b\bar b}^{ii}\, \BR^2(\Lambda_c^+ \to {\rm reco.})\, \epsilon_{\rm reco}^2 \;, \\
    N_{b\bar b}^{is} &= 2N_{b\bar b}^{ii}\, \BR(\Lambda_c^+ \to \Lambda X) \BR(\Lambda \to p\pi^-)\, \epsilon_\Lambda \;, \\
    N_{b\bar b}^{ie} &= 2N_{b\bar b}^{ii}\, \BR(\Lambda_c^+ \to {\rm reco.})\, \epsilon_{\rm reco} \;, \\
    N_{b\bar b}^{se} &= 2N_{b\bar b}^{ii}\, \BR(\Lambda_c^+ \to \Lambda X) \BR(\Lambda \to p\pi^-) \BR(\Lambda_c^+ \to {\rm reco.})\, \epsilon_\Lambda\, \epsilon_{\rm reco} \;.
\end{align}

\subsubsection{Expected Precision}

The expected precision of the $b\bar b$ polarization and spin correlation measurements for either a muon or neutrino spin analyzer can be computed using eqs.~\eqref{eq:b err}--\eqref{eq:c err}. As indicated in table~\ref{tab:decay scheme}, the neutrino in the $\Lambda_b$ decay has a larger spin analyzing power than the muon and therefore has the potential to provide higher precision. Since the neutrino is not observed directly, a reconstruction is required, and it requires certain approximations, as discussed at the end of section~\ref{sec:decay chains}. However, the decay reconstruction is needed anyway, even if the muon is used as the spin analyzer. We will therefore present the statistical uncertainties that can be obtained with the neutrino. The analogous results for the muon can be obtained by dividing the polarization uncertainties by $|\alpha_{\mu^-}|$ and the spin correlation uncertainties by $\alpha_{\mu^-}^2$, where $\alpha_{\mu^-} \approx -0.26$.

\begin{table}
\centering
\begin{tabular}{c|ccccc}\hline\hline
\qquad channel $\to$ & inclusive & \multicolumn{2}{c}{inclusive/inclusive} & \multicolumn{2}{c}{inclusive/exclusive} \\\hline
$m_{jj}$ cut~[GeV]   & $r_i\Delta b^\pm_i$ & $r_i^2\Delta c_{ii}$ & $r_ir_j\Delta c_{ij(\ell)}$ & $r_i^2\Delta c_{ii}$ & $r_ir_j\Delta c_{ij(\ell)}$ \\\hline
    \mbox{no cut} & 0.003 &0.14 & 0.10 & 0.11 & 0.079\\  
    100 & 0.004 & 0.18 & 0.13 & 0.15 & 0.10\\
    300 & 0.013 & &  & & \\ 
    500 & 0.036 &&&&\\ 
    750 & 0.093 &&&&\\ 
    1000 & 0.19 &&&&\\ 
    \hline
    parked data & 0.0003 & 0.039 & 0.027 & 0.031 & 0.022\\
    \hline\hline
\end{tabular}
\vskip 4mm
\centering
\begin{tabular}{c|ccccc}\hline\hline
\qquad channel $\to$ & semi-inclusive & \multicolumn{2}{c}{semi-inclusive/semi-inclusive} & \multicolumn{2}{c}{semi-inclusive/inclusive} \\\hline
$m_{jj}$ cut~[GeV]& $r_i\Delta b^\pm_i$ & $r_i^2\Delta c_{ii}$ & $r_ir_j\Delta c_{ij(\ell)}$ & $r_i^2\Delta c_{ii}$ & $r_ir_j\Delta c_{ij(\ell)}$ \\\hline
    \mbox{no cut}  & 0.005 & 0.36 & 0.25 & 0.16 & 0.11 \\ 
    100 & 0.006 & 0.47 &0.33 & 0.21 & 0.15\\ 
    300 & 0.022 & & & & \\
    500 & 0.072 &&&&\\ 
    750 & 0.21 &&&&\\ 
    \hline
    parked data &0.0004 &0.050 &0.035 &0.031 &0.022\\
    \hline\hline
\end{tabular}
\vskip 4mm
\centering
\begin{tabular}{c|ccccc}\hline\hline
\qquad channel $\to$ & exclusive & \multicolumn{2}{c}{exclusive/exclusive} & \multicolumn{2}{c}{exclusive/semi-inclusive} \\\hline
$m_{jj}$ cut~[GeV]& $r_i\Delta b^\pm_i$ & $r_i^2\Delta c_{ii}$ & $r_ir_j\Delta c_{ij(\ell)}$ & $r_i^2\Delta c_{ii}$ & $r_ir_j\Delta c_{ij(\ell)}$ \\\hline
    \mbox{no cut} & 0.003 & 0.18 & 0.11 & 0.18 & 0.13\\ 
    100 & 0.004 & 0.23 & 0.16 & 0.23 & 0.16 \\ 
    300 & 0.015 &  &      &  &  \\ 
    500 & 0.040\\ 
    750 & 0.10\\ 
    1000 & 0.21\\ 
    \hline
    parked data &0.0004 & 0.049 & 0.034 & 0.034 & 0.024\\
    \hline\hline
\end{tabular}
\caption{Expected statistical uncertainties for the polarization and spin correlation measurements in $b\bar{b}$ in Run~2 with a neutrino spin analyzer for different selection channels. The shorthand $\Delta c_{ij(\ell)}$ denotes the uncertainties in $c_{ij}$ and $c_\ell$, which are of the same size.}
\label{tab:bbbar neutrino}
\end{table}

Table~\ref{tab:bbbar neutrino} shows the expected uncertainties for Run~2, for both the standard datasets and the CMS parked data. The results are promising for the polarization, and with the parked data also for the spin correlations. Analogous results for the HL-LHC are presented in table~\ref{tab:bbbar neutrino hi lumi}.
	
\begin{table}
\centering
\begin{tabular}{c|ccccc}\hline\hline
\qquad channel $\to$ & inclusive & \multicolumn{2}{c}{inclusive/inclusive} & \multicolumn{2}{c}{inclusive/exclusive} \\\hline
$m_{jj}$ cut~[GeV]   & $r_i\Delta b^\pm_i$ & $r_i^2\Delta c_{ii}$ & $r_ir_j\Delta c_{ij(\ell)}$ & $r_i^2\Delta c_{ii}$ & $r_ir_j\Delta c_{ij(\ell)}$ \\\hline
	\mbox{no cut} & 0.0004 & 0.015 & 0.011 & 0.012 & 0.0086\\  
        100 & 0.0005 & 0.025 & 0.017 & 0.020 & 0.014\\
	300 & 0.0022 & 0.13  & 0.091  & 0.10  & 0.071\\ 
	500 & 0.0063 & 0.36  & 0.26   & 0.29  & 0.20\\ 
	750 & 0.016 &&&&\\ 
	1000 & 0.032 &&&&\\ 
	1500 & 0.11 &&&&\\ 
        2000 & 0.27 &&&&\\ 
	\hline\hline
\end{tabular}
\vskip 4mm
\centering
\begin{tabular}{c|ccccc}\hline\hline
\qquad channel $\to$ & semi-inclusive & \multicolumn{2}{c}{semi-inclusive/semi-inclusive} & \multicolumn{2}{c}{semi-inclusive/inclusive} \\\hline
$m_{jj}$ cut~[GeV]& $r_i\Delta b^\pm_i$ & $r_i^2\Delta c_{ii}$ & $r_ir_j\Delta c_{ij(\ell)}$ & $r_i^2\Delta c_{ii}$ & $r_ir_j\Delta c_{ij(\ell)}$ \\\hline
	\mbox{no cut}  & 0.0004 & 0.018 & 0.013 & 0.012 & 0.0084 \\ 
	100 & 0.0005 & 0.029 & 0.021 & 0.019 & 0.013 \\ 
	300 & 0.0027 & 0.21 & 0.15 & 0.12 & 0.082 \\
	500 & 0.0091 &&&&\\ 
	750 & 0.026 &&&&\\ 
	1000 & 0.058 &&&&\\
        1500 & 0.24 &&&&\\
	\hline\hline
\end{tabular}
\vskip 4mm
\centering
\begin{tabular}{c|ccccc}\hline\hline
\qquad channel $\to$ & exclusive & \multicolumn{2}{c}{exclusive/exclusive} & \multicolumn{2}{c}{exclusive/semi-inclusive} \\\hline
$m_{jj}$ cut~[GeV]& $r_i\Delta b^\pm_i$ & $r_i^2\Delta c_{ii}$ & $r_ir_j\Delta c_{ij(\ell)}$ & $r_i^2\Delta c_{ii}$ & $r_ir_j\Delta c_{ij(\ell)}$ \\\hline
	\mbox{no cut} & 0.0004 & 0.019 & 0.013& 0.013 & 0.0093\\ 
        100 & 0.0005 & 0.031 & 0.022 & 0.021 & 0.015 \\ 
        300 & 0.0025 & 0.16  & 0.11  & 0.13  & 0.091 \\ 
        500 & 0.0070\\ 
        750 & 0.018\\ 
        1000 & 0.037\\ 
        1500 & 0.13\\ 
        2000 & 0.32\\ 
	\hline\hline
\end{tabular}
\caption{Expected statistical uncertainties for the polarization and spin correlation measurements in $b\bar{b}$ at the HL-LHC with a neutrino spin analyzer for different selection channels.}
\label{tab:bbbar neutrino hi lumi}
\end{table}

\FloatBarrier

\section{Summary and Discussion}
\label{sec:summ and concl}

In this work, we analyzed the feasibility of measuring quark polarization and spin correlations in high-energy $pp\to q\bar{q}$ events in ATLAS or CMS, where the quark $q$ is $b$, $c$, or $s$. Such measurements can be done using the decay angular distributions of baryons produced from these quarks. Our main results are as follows:
\begin{itemize}
\item[\ding{70}]
In the $b$-quark case, Run~2 data allows measuring the polarization in a number of channels of semileptonic $\Lambda_b$ decays, with dijet invariant mass cuts up to $\sim 1$~TeV. Particularly precise measurements are possible with the CMS parked data. Spin correlation measurements are also feasible in multiple channels, although the statistical uncertainties will be sizable with the standard Run~2 datasets. An opportunity to do more precise measurements is again offered by the Run~2 CMS parked dataset.
\item[\ding{70}]
For the $c$ quark, polarization measurements with $\Lambda_c^+$ decays using either the semileptonic or mixed channel are possible with the Run~2 dataset, and they can be extended to dijet invariant masses above 1~TeV at the HL-LHC using the same channels or the hadronic channel. Spin correlation measurements will be feasible at the HL-LHC in the semileptonic and mixed channels.
\item[\ding{70}]
The $s$-quark case is challenging in terms of the statistics that can be collected with the standard triggers. Polarization measurements become only borderline feasible with the statistics of the HL-LHC, and systematic uncertainties may pose challenges as well due to the low purity of the sample. There will not be enough statistics for $s\bar s$ spin correlation measurements even at the HL-LHC.
\end{itemize}
In table~\ref{tab:conclusion}, we provide a more detailed summary of the measurement channels we have considered, showing in which of them a measurement is expected to be possible.

\begin{table}
\centering
\begin{tabular}{c|c|cc|c} \hline\hline
    &\multicolumn{4}{c}{Polarization} \\\hline
    \multirow{2}{*}{Quark} & \multirow{2}{*}{Channel} & \multicolumn{2}{c|}{Run 2} & \multirow{2}{*}{HL-LHC} \\
    & & standard & parked & \\\hline
    $s$ & & & & (\halfcheckmark)\\\hline
    \multirow{3}{*}{$c$}
        & hadronic        & (\halfcheckmark) & & \checkmark\\
	& semileptonic    & \checkmark       & & \checkmark\\
	& mixed           & (\checkmark)     & & \checkmark\\\hline
    \multirow{3}{*}{$b$}
         & inclusive      & (\checkmark) & (\checkmark) & (\checkmark) \\
         & semi-inclusive & \checkmark   & \checkmark   & \checkmark \\
         & exclusive      & \checkmark   & \checkmark   & \checkmark
    \\\hline\hline
\end{tabular}
\vskip 4mm
\centering
\begin{tabular}{c|c|cc|c} \hline\hline
    &\multicolumn{4}{c}{Spin Correlations}\\\hline
    \multirow{2}{*}{Quark} & \multirow{2}{*}{Channel} & \multicolumn{2}{c|}{Run 2} & \multirow{2}{*}{HL-LHC} \\
    & & standard & parked & \\\hline
    $s$ & & & \\\hline
    \multirow{3}{*}{$c$}
        & hadronic     & & & \\
	& semileptonic & & & \checkmark \\
	& mixed        & & & \checkmark \\\hline
    \multirow{6}{*}{$b$}
	 & inclusive/inclusive           & (\halfcheckmark) & (\checkmark) & (\checkmark) \\
	 & semi-inclusive/semi-inclusive & \halfcheckmark   & \checkmark   & \checkmark \\
	 & exclusive/exclusive           & \halfcheckmark   & \checkmark   & \checkmark \\
	 & inclusive/exclusive           & (\halfcheckmark) & (\checkmark) & (\checkmark) \\
	 & inclusive/semi-inclusive      & (\halfcheckmark) & (\checkmark) & (\checkmark) \\
	 & exclusive/semi-inclusive      & \halfcheckmark   & \checkmark   & \checkmark \\
    \hline\hline
\end{tabular}
\caption{The prospects in Run~2 and HL-LHC datasets for the different analysis channels examined. The top table is for polarization and the bottom is for spin correlations. Check marks indicate that a measurement is expected to be possible, and crossed check marks indicate borderline cases. Parentheses around a check mark indicate low sample purity, under 10\%. The ``parked'' column refers to the CMS parked $b$-hadron dataset~\cite{CMS-DP-2019-043,Bainbridge:2020pgi,CMS:2024syx}.}
\label{tab:conclusion}
\end{table}

We would like to note that doing this broad survey of the different possible analysis channels required us to use various assumptions and approximations, as we described along the way. Therefore, our predicted sensitivities should be viewed as rough estimates. The actual performance of the corresponding analyses may be worse because of neglected backgrounds or effects of unfolding. On the other hand, improvements in the reconstruction of electrons within jets to make them comparable to muons would enhance the performance. Advanced machine learning techniques that have been entering the field recently can also boost the performance by improving object identification and reconstruction and/or background rejection. Finally, new trigger paths that could be deployed in future runs of the LHC have the potential to improve the statistics of all of the proposed analyses. The most promising channels would benefit from relaxing the $p_T$ thresholds of the single- and double-muon triggers. For $b\bar b$ this can be aided by requiring the muon track to have a significant impact parameter (as was done in the CMS parked data). Requiring the presence of tracks consistent with an energetic $\Lambda \to p\pi^-$ decay or one of the reconstructible $\Lambda_c^+$ decays in the relevant $b\bar b$ or $c\bar c$ channels might also be useful for extending the triggering to softer muons. In addition, an energetic $\Lambda \to p\pi^-$ or $\Lambda_c^+ \to pK^-\pi^+$ decay requirement could be used to lower the jet trigger threshold for the $s\bar s$ and the hadronic $c\bar c$ channels, respectively.

Since the angular distribution coefficients $C_{ii}$ and $C_{ij}^\pm$ that will be measured in the spin correlation analyses depend on the polarization retention factors $r_L$ and $r_T$ via eq.~\eqref{eq:B,Cii,Cij+-}, the analyses proposed in this paper can be used to determine these factors for $c$ and $b$ quarks. By relying on spin correlation components allowed to be large by the symmetries of QCD (specifically $c_{kk}$, $c_{nn}$, $c_{rr}$, and $c_{rk}$),\footnote{In practice, $c_{rk}$ happens to be small when $p_T$ cuts are applied, as can be seen in tables~\ref{tab:bb cc spin correlations} and \ref{tab:bb cc spin correlations hi lumi}.} one can extract $r_L$ and $r_T$ using the (partly redundant) relations
\begin{equation}
    r_L^2 = \frac{C_{kk}}{c_{kk}\alpha_+\alpha_- f}\;,\quad
    r_T^2 = \frac{C_{nn}}{c_{nn}\alpha_+\alpha_- f}\;,\quad
    r_T^2 = \frac{C_{rr}}{c_{rr}\alpha_+\alpha_- f}\;,\quad
    r_L r_T = \frac{C_{rk}^+}{2c_{rk}\alpha_+\alpha_- f}\;.		
\end{equation}
The measurements of $r_L$ and $r_T$ can lead to further understanding of the polarization transfer in QCD fragmentation and act as a test for different models of low-energy QCD phenomenology (e.g., ref.~\cite{Adamov:2000is}). It will also be interesting to compare the values of $r_L$ that will be obtained in these spin correlation measurements with those that will hopefully be obtained in polarization measurements of $b$ and $c$ quarks in $t\bar t$~\cite{Galanti:2015pqa} and $W$+$c$~\cite{Kats:2015zth} samples. The estimates in the current work (using the $c_{kk}$ component, whose value with the relevant cuts is close to $-1$) and those in ref.~\cite{Galanti:2015pqa} predict comparable statistical uncertainties, of order $10\%$, for the $b$-quark $r_L$ measurement using the Run~2 dataset, while part of the systematic effects will be different. With higher statistics, doing both types of analyses, which cover different kinematic ranges, may also be helpful for measuring the renormalization group evolution of the polarized fragmentation functions.

The polarization and spin correlation measurements can in principle also be utilized to probe for BSM contributions to $pp \to q\bar q$ processes, especially using the numerous components in which the SM contribution is expected to be small. Whether such searches can be competitive with other probes of the same BSM scenarios is a subject for another study.

\FloatBarrier

\acknowledgments

This research was supported in part by the Israel Science Foundation (grants no.~780/17 and 1666/22) and the United States~- Israel Binational Science Foundation (grant no.~2018257).

\appendix

\section{Statistical Uncertainty Formulas}
\label{app:uncertainty}

The angular distributions in eqs.~\eqref{eq:B dist}, \eqref{eq:Cii dist} and~\eqref{eq:Cij dist} are all of the form
\begin{equation}
    \frac{1}{\sigma}\frac{d\sigma}{dX} = \frac12\left(1 + c X\right)f(|X|) \,,
\end{equation}
where
\begin{equation}
   X \;=\; \cos\theta^\pm_i\,,\quad
   \cos\theta^+_i\cos\theta^-_j\,,\quad
   \cos\theta^+_i\cos\theta^-_j \pm \cos\theta^+_j\cos\theta^-_i
\end{equation}
and
\begin{equation}
    f(|X|) \;=\; 1\,,\quad\;\;
    \ln\left(\frac{1}{|X|}\right),\quad\;\;
    \cos^{-1}(|X|)\,, \qquad\qquad\qquad
\end{equation}
respectively. We imagine extracting the value of $c$ from the data with a $\chi^2$ minimization procedure for a binned distribution of $X$. The $\chi^2$ function is
\begin{equation}
    \chi^2(c) = \sum_i\frac{\left[n_i - \frac12\left(1 + c X_i\right)f(|X_i|) N\Delta X\right]^2}{\sigma_i^2} \;,
\end{equation}
where the summation is over bins of size $\Delta X$, $n_i$ is the number of events observed in bin $i$, $X_i$ is the value of $X$ in the middle of that bin, $N$ is the total number of events, and
\begin{equation}
    \sigma_i^2 \simeq n_i \simeq \frac12\left(1 + c X_i\right)f(|X_i|)N\Delta X
\end{equation}
is the squared uncertainty of the expected number of events in bin $i$. Solving the minimization condition $d\chi^2/dc = 0$ for $c$ and then propagating the statistical uncertainties of $n_i$, we find for the uncertainty of $c$:
\begin{equation}
    \Delta c = \frac{1}{S(c)\sqrt N} \;,
\end{equation}
with
\begin{equation}
    S^2(c) \simeq \frac{\Delta X}{2} \sum_i\frac{X_i^2}{1 + c X_i}\,f(|X_i|)\,,
\end{equation}
where we assumed $n_i \gg 1$. In the limit of a large number of bins we can approximate the sum with an integral and obtain
\begin{equation}
    S^2(c) \simeq \frac{1}{2} \int_{-1}^1 \frac{X^2}{1 + c X}\,f(|X|)\,dX
    = \int_0^1 \frac{X^2}{1 - c^2 X^2}\,f(X)\,dX \,.
\end{equation}
The uncertainties on the parameters $B^\pm_i$ ($\equiv B$), $C_{ii}$, and $C_{ij}^\pm/2$ are then given by
\begin{align}
    \Delta B &= \frac{1}{S_B(B)\sqrt{N}} \;,
    \label{eq: B error}\\
    \Delta C_{ii} &= \frac{1}{S_{C_{ii}}(C_{ii})\sqrt{N}} \;,
    \label{eq: Cii error}\\
    \Delta (C_{ij}^\pm/2) &= \frac{1}{S_{C_{ij}^\pm/2}(C_{ij}^\pm/2)\sqrt{N}} \;,
    \label{eq: Cpm error}
\end{align}
where
\begin{align}
    S_B^2(B) &\simeq \int_0^1\frac{X^2}{1 - B^2 X^2}\,dX
    = \frac{\arctanh B - B}{B^3} \;,\label{S2B} \\
    S_{C_{ii}}^2(C_{ii}) &\simeq \int^0_1\frac{X^2}{1 - C_{ii}^2 X^2} \ln\left(\frac{1}{X}\right) dX
    = \frac{1}{9} + \frac{1}{4}\,C_{ii}^2\,\Phi(C_{ii}^2,2,\tfrac52) \;,\\
    S_{C_{ij}^\pm/2}^2(c) &\simeq \int^0_1\frac{X^2}{1 - c^2 X^2} \cos^{-1}(X)\,dX \nonumber\\
    &= \frac{1}{4ic^3}\left[2\,\mbox{Li}_2\left(\frac{1}{\left(ic+\sqrt{1-c^2}\right)^2}\right) - 8\,\mbox{Li}_2\left(\sqrt{1-c^2}-ic\right) \right.
    \label{S2C+-} \\
    &\qquad\qquad \left. +\, 2 \left(\pi + 2i\ln\left(\frac{\sqrt{1-c^2}-1}{c}\right)\right)\arcsin c - 4ic + \pi^2 \right] \;,\nonumber
\end{align}
where the function $\Phi$ is the Lerch transcendent, $\mbox{Li}_2$ is the dilogarithm, and $c$ in eq.~\eqref{S2C+-} stands for $C_{ij}^\pm/2$.

\begin{figure}
\centering
\includegraphics[width=0.8\linewidth]{"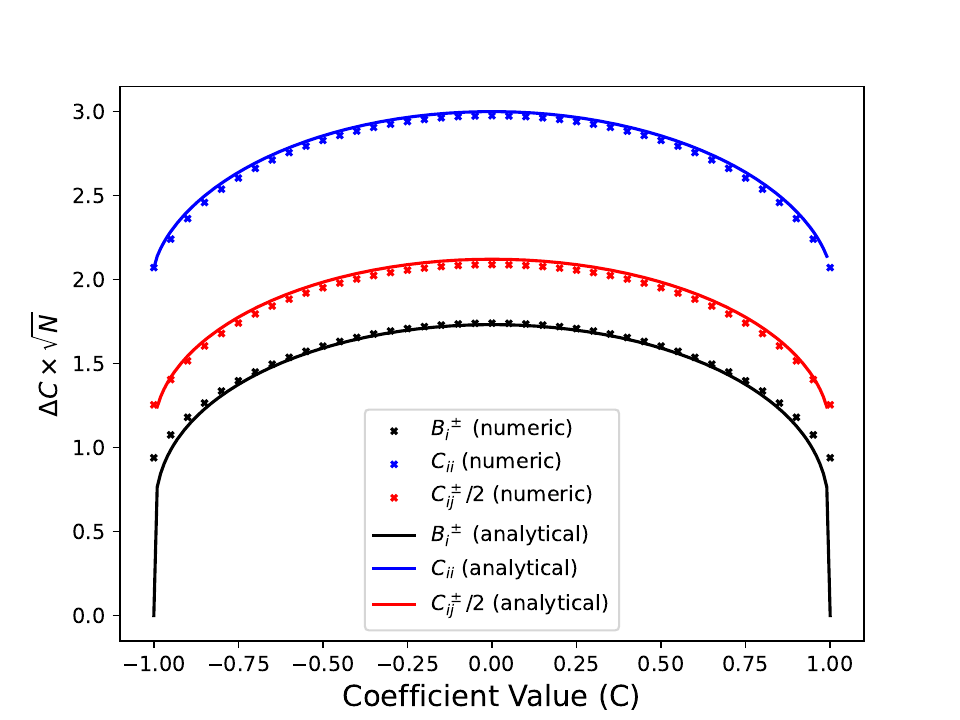"}
\caption{Statistical uncertainty (multiplied by $\sqrt N$, where $N$ is the number of events) of the angular distribution coefficients $B^\pm_i$ (black), $C_{ii}$ (blue), and $C_{ij}^\pm/2$ (red) as a function of the coefficient values, calculated both analytically (solid curves) and numerically (dots) The analytical results are in the limit of an infinite number of bins, while the numerical (Monte Carlo) results are for 10 bins.}
\label{fig:shape coff}
\end{figure}

In figure~\ref{fig:shape coff} we plot the uncertainties of the coefficients as a function of the coefficient values based on the analytic formulas in eqs.~\eqref{S2B}--\eqref{S2C+-}, which correspond to the limit of a large number of bins. We also show the results of a Monte Carlo simulation of a $\chi^2$ minimization procedure with 10 bins. There is a good agreement between the two sets of results, except very close to $B = -1$ or $1$, where the analytical formula, which assumes infinitesimally small (but at the same time highly populated) bins, gives vanishing uncertainty. This happens because bins near $X = 1$ or $-1$ in these cases contain a vanishingly small fraction of events, making the uncertainty in those bins vanishingly small too, which in turn leads to a vanishingly small uncertainty in $B$. This situation is unphysical because measurements of $X$ always have some uncertainty, $\Delta X$, which prevents bins much smaller than $\Delta X$ from being useful.

While the polarization and spin correlation components ($b^\pm_i$, $c_{ij}$, $c_\ell$) can in principle be anywhere in the range $[-1,1]$, the polarization retention factors $r_i$ and in many cases also the purity $f$ or the spin analyzing powers $\alpha_\pm$ in eq.~\eqref{eq:B,Cii,Cij+-} will limit the angular distribution coefficients $B^\pm_i$, $C_{ii}$, and $C_{ij}^\pm/2$ to a smaller range of values. This, together with the flatness of the curves in figure~\ref{fig:shape coff} in the vicinity of zero, implies that the uncertainties can usually be approximated by their values at zero. We have
\begin{equation}
    S_B(0) =\frac{1}{\sqrt3}\;,\qquad
    S_{C_{ii}}(0) = \frac13\;,\qquad
    S_{C_{ij}^\pm/2}(0) = \frac{\sqrt2}{3}\;,
\end{equation}
which leads to eq.~\eqref{uncert-at-0}.

\bibliographystyle{utphys}
\begingroup
\setstretch{1.095}
\bibliography{references}

\providecommand{\href}[2]{#2}\begingroup\raggedright\begin{thebibliography}{100}

\bibitem{Bernreuther:2001rq}
W.~Bernreuther, A.~Brandenburg, Z.~G. Si, and P.~Uwer, ``{Top-Quark Spin
  Correlations at Hadron Colliders: Predictions at Next-to-Leading Order
  QCD},'' \href{http://dx.doi.org/10.1103/PhysRevLett.87.242002}{{\em Phys.
  Rev. Lett.} {\bfseries 87} (2001) 242002},
  \href{http://arxiv.org/abs/hep-ph/0107086}{{\ttfamily arXiv:hep-ph/0107086}}.

\bibitem{Bernreuther:2004jv}
W.~Bernreuther, A.~Brandenburg, Z.~G. Si, and P.~Uwer, ``{Top quark pair
  production and decay at hadron colliders},''
  \href{http://dx.doi.org/10.1016/j.nuclphysb.2004.04.019}{{\em Nucl. Phys. B}
  {\bfseries 690} (2004) 81--137},
  \href{http://arxiv.org/abs/hep-ph/0403035}{{\ttfamily arXiv:hep-ph/0403035}}.

\bibitem{Mahlon:2010gw}
G.~Mahlon and S.~J. Parke, ``{Spin correlation effects in top quark pair
  production at the LHC},''
  \href{http://dx.doi.org/10.1103/PhysRevD.81.074024}{{\em Phys. Rev. D}
  {\bfseries 81} (2010) 074024},
  \href{http://arxiv.org/abs/1001.3422}{{\ttfamily arXiv:1001.3422 [hep-ph]}}.

\bibitem{Baumgart:2011wk}
M.~Baumgart and B.~Tweedie, ``{Discriminating top-antitop resonances using
  azimuthal decay correlations},''
  \href{http://dx.doi.org/10.1007/JHEP09(2011)049}{{\em JHEP} {\bfseries 09}
  (2011) 049}, \href{http://arxiv.org/abs/1104.2043}{{\ttfamily arXiv:1104.2043
  [hep-ph]}}.

\bibitem{Baumgart:2012ay}
M.~Baumgart and B.~Tweedie, ``{A new twist on top quark spin correlations},''
  \href{http://dx.doi.org/10.1007/JHEP03(2013)117}{{\em JHEP} {\bfseries 03}
  (2013) 117}, \href{http://arxiv.org/abs/1212.4888}{{\ttfamily arXiv:1212.4888
  [hep-ph]}}.

\bibitem{Baumgart:2013yra}
M.~Baumgart and B.~Tweedie, ``{Transverse top quark polarization and the $t\bar
  t$ forward-backward asymmetry},''
  \href{http://dx.doi.org/10.1007/JHEP08(2013)072}{{\em JHEP} {\bfseries 08}
  (2013) 072}, \href{http://arxiv.org/abs/1303.1200}{{\ttfamily arXiv:1303.1200
  [hep-ph]}}.

\bibitem{Bernreuther:2013aga}
W.~Bernreuther and Z.-G. Si, ``{Top quark spin correlations and polarization at
  the LHC: Standard model predictions and effects of anomalous top chromo
  moments},'' \href{http://dx.doi.org/10.1016/j.physletb.2013.06.051}{{\em
  Phys. Lett. B} {\bfseries 725} (2013) 115--122},
  \href{http://arxiv.org/abs/1305.2066}{{\ttfamily arXiv:1305.2066 [hep-ph]}}.
  [Erratum: \emph{Phys.\ Lett.\ B} \textbf{744} (2015) 413--413].

\bibitem{Bernreuther:2015yna}
W.~Bernreuther, D.~Heisler, and Z.-G. Si, ``{A set of top quark spin
  correlation and polarization observables for the LHC: Standard Model
  predictions and new physics contributions},''
  \href{http://dx.doi.org/10.1007/JHEP12(2015)026}{{\em JHEP} {\bfseries 12}
  (2015) 026}, \href{http://arxiv.org/abs/1508.05271}{{\ttfamily
  arXiv:1508.05271 [hep-ph]}}.

\bibitem{Behring:2019iiv}
A.~Behring, M.~Czakon, A.~Mitov, A.~S. Papanastasiou, and R.~Poncelet,
  ``{Higher Order Corrections to Spin Correlations in Top Quark Pair Production
  at the LHC},'' \href{http://dx.doi.org/10.1103/PhysRevLett.123.082001}{{\em
  Phys. Rev. Lett.} {\bfseries 123} no.~8, (2019) 082001},
  \href{http://arxiv.org/abs/1901.05407}{{\ttfamily arXiv:1901.05407
  [hep-ph]}}.

\bibitem{Afik:2020onf}
Y.~Afik and J.~R.~M. de~Nova, ``{Entanglement and quantum tomography with top
  quarks at the LHC},''
  \href{http://dx.doi.org/10.1140/epjp/s13360-021-01902-1}{{\em Eur. Phys. J.
  Plus} {\bfseries 136} no.~9, (2021) 907},
  \href{http://arxiv.org/abs/2003.02280}{{\ttfamily arXiv:2003.02280
  [quant-ph]}}.

\bibitem{Severi:2022qjy}
C.~Severi and E.~Vryonidou, ``{Quantum entanglement and top spin correlations
  in SMEFT at higher orders},''
  \href{http://dx.doi.org/10.1007/JHEP01(2023)148}{{\em JHEP} {\bfseries 01}
  (2023) 148}, \href{http://arxiv.org/abs/2210.09330}{{\ttfamily
  arXiv:2210.09330 [hep-ph]}}.

\bibitem{ATLAS:2014abv}
{ ATLAS} Collaboration, G.~Aad {\em et~al.}, ``{Measurement of Spin Correlation
  in Top-Antitop Quark Events and Search for Top Squark Pair Production in $pp$
  Collisions at $\sqrt{s}=8$~TeV Using the ATLAS Detector},''
  \href{http://dx.doi.org/10.1103/PhysRevLett.114.142001}{{\em Phys. Rev.
  Lett.} {\bfseries 114} no.~14, (2015) 142001},
  \href{http://arxiv.org/abs/1412.4742}{{\ttfamily arXiv:1412.4742 [hep-ex]}}.

\bibitem{ATLAS:2016bac}
{ ATLAS} Collaboration, M.~Aaboud {\em et~al.}, ``{Measurements of top quark
  spin observables in $t\bar t$ events using dilepton final states in
  $\sqrt{s}=8$~TeV $pp$ collisions with the ATLAS detector},''
  \href{http://dx.doi.org/10.1007/JHEP03(2017)113}{{\em JHEP} {\bfseries 03}
  (2017) 113}, \href{http://arxiv.org/abs/1612.07004}{{\ttfamily
  arXiv:1612.07004 [hep-ex]}}.

\bibitem{CMS:2016piu}
{ CMS} Collaboration, V.~Khachatryan {\em et~al.}, ``{Measurements of $t\bar t$
  spin correlations and top quark polarization using dilepton final states in
  $pp$ collisions at $\sqrt s = 8$~TeV},''
  \href{http://dx.doi.org/10.1103/PhysRevD.93.052007}{{\em Phys. Rev. D}
  {\bfseries 93} no.~5, (2016) 052007},
  \href{http://arxiv.org/abs/1601.01107}{{\ttfamily arXiv:1601.01107
  [hep-ex]}}.

\bibitem{CMS:2019nrx}
{ CMS} Collaboration, A.~M. Sirunyan {\em et~al.}, ``{Measurement of the top
  quark polarization and $t\bar t$ spin correlations using dilepton final
  states in proton-proton collisions at $\sqrt{s}=13$~TeV},''
  \href{http://dx.doi.org/10.1103/PhysRevD.100.072002}{{\em Phys. Rev. D}
  {\bfseries 100} no.~7, (2019) 072002},
  \href{http://arxiv.org/abs/1907.03729}{{\ttfamily arXiv:1907.03729
  [hep-ex]}}.

\bibitem{ATLAS:2019zrq}
{ ATLAS} Collaboration, M.~Aaboud {\em et~al.}, ``{Measurements of top-quark
  pair spin correlations in the $e\mu$ channel at $\sqrt{s} = 13$ TeV using
  $pp$ collisions in the ATLAS detector},''
  \href{http://dx.doi.org/10.1140/epjc/s10052-020-8181-6}{{\em Eur. Phys. J. C}
  {\bfseries 80} no.~8, (2020) 754},
  \href{http://arxiv.org/abs/1903.07570}{{\ttfamily arXiv:1903.07570
  [hep-ex]}}.

\bibitem{ATLAS:ttbar-entanglement}
{ ATLAS} Collaboration, G.~Aad {\em et~al.}, ``{Observation of quantum
  entanglement in top-quark pairs using the ATLAS detector},''
  \href{http://arxiv.org/abs/2311.07288}{{\ttfamily arXiv:2311.07288
  [hep-ex]}}.

\bibitem{Galanti:2015pqa}
M.~Galanti, A.~Giammanco, Y.~Grossman, Y.~Kats, E.~Stamou, and J.~Zupan,
  ``{Heavy baryons as polarimeters at colliders},''
  \href{http://dx.doi.org/10.1007/JHEP11(2015)067}{{\em JHEP} {\bfseries 11}
  (2015) 067}, \href{http://arxiv.org/abs/1505.02771}{{\ttfamily
  arXiv:1505.02771 [hep-ph]}}.

\bibitem{Kats:2015cna}
Y.~Kats, ``{Measuring polarization of light quarks at ATLAS and CMS},''
  \href{http://dx.doi.org/10.1103/PhysRevD.92.071503}{{\em Phys. Rev. D}
  {\bfseries 92} (2015) 071503},
  \href{http://arxiv.org/abs/1505.06731}{{\ttfamily arXiv:1505.06731
  [hep-ph]}}.

\bibitem{Kats:2023gul}
Y.~Kats, ``{Kinked tracks from \ensuremath{\Sigma}$^{+}$ baryons as a probe of
  light quark polarizations},''
  \href{http://dx.doi.org/10.1007/JHEP07(2023)018}{{\em JHEP} {\bfseries 07}
  (2023) 018}, \href{http://arxiv.org/abs/2301.06188}{{\ttfamily
  arXiv:2301.06188 [hep-ph]}}.

\bibitem{Kats:2015zth}
Y.~Kats, ``{Measuring $c$-quark polarization in $W$+$c$ samples at ATLAS and
  CMS},'' \href{http://dx.doi.org/10.1007/JHEP11(2016)011}{{\em JHEP}
  {\bfseries 11} (2016) 011}, \href{http://arxiv.org/abs/1512.00438}{{\ttfamily
  arXiv:1512.00438 [hep-ph]}}.

\bibitem{Mannel:1991bs}
T.~Mannel and G.~A. Schuler, ``{Semileptonic decays of bottom baryons at
  LEP},'' \href{http://dx.doi.org/10.1016/0370-2693(92)91864-6}{{\em Phys.
  Lett. B} {\bfseries 279} (1992) 194--200}.

\bibitem{Ball:1992fw}
F.~E. Close, J.~G. {K\"orner}, R.~J.~N. Phillips, and D.~J. Summers, ``{Report
  of the b-fragmentation working group. Section 5: b polarization at LEP},''
\href{http://dx.doi.org/10.1088/0954-3899/18/10/013}{{\em J. Phys.} {\bfseries
  G18} (1992) 1703}.

\bibitem{Falk:1993rf}
A.~F. Falk and M.~E. Peskin, ``{Production, decay, and polarization of excited
  heavy hadrons},'' \href{http://dx.doi.org/10.1103/PhysRevD.49.3320}{{\em
  Phys. Rev. D} {\bfseries 49} (1994) 3320--3332},
  \href{http://arxiv.org/abs/hep-ph/9308241}{{\ttfamily arXiv:hep-ph/9308241}}.

\bibitem{ALEPH:1995aqx}
{ ALEPH} Collaboration, D.~Buskulic {\em et~al.}, ``{Measurement of the
  $\Lambda_b$ polarization in Z decays},''
  \href{http://dx.doi.org/10.1016/0370-2693(95)01433-0}{{\em Phys. Lett. B}
  {\bfseries 365} (1996) 437--447}.

\bibitem{OPAL:1998wmk}
{ OPAL} Collaboration, G.~Abbiendi {\em et~al.}, ``{Measurement of the average
  polarization of $b$ baryons in hadronic $Z^0$ decays},''
  \href{http://dx.doi.org/10.1016/S0370-2693(98)01387-2}{{\em Phys. Lett. B}
  {\bfseries 444} (1998) 539--554},
  \href{http://arxiv.org/abs/hep-ex/9808006}{{\ttfamily arXiv:hep-ex/9808006}}.

\bibitem{DELPHI:1999hkl}
{ DELPHI} Collaboration, P.~Abreu {\em et~al.}, ``{$\Lambda_b$ polarization in
  $Z^0$ decays at LEP},''
  \href{http://dx.doi.org/10.1016/S0370-2693(99)01431-8}{{\em Phys. Lett. B}
  {\bfseries 474} (2000) 205--222}.

\bibitem{ALEPH:1996oew}
{ ALEPH} Collaboration, D.~Buskulic {\em et~al.}, ``{Measurement of $\Lambda$
  polarization from Z decays},''
  \href{http://dx.doi.org/10.1016/0370-2693(96)00300-0}{{\em Phys. Lett. B}
  {\bfseries 374} (1996) 319--330}.

\bibitem{ALEPH:1997an}
{ ALEPH} Collaboration, ``{Update of $\Lambda$ polarization from Z decays},''
  Tech. Rep. CERN-OPEN-99-328, CERN, Geneva, 1997.
\newblock \url{http://cds.cern.ch/record/407920}.

\bibitem{OPAL:1997oem}
{ OPAL} Collaboration, K.~Ackerstaff {\em et~al.}, ``{Polarization and
  forward-backward asymmetry of $\Lambda$ baryons in hadronic $Z^0$ decays},''
  \href{http://dx.doi.org/10.1007/s100520050123}{{\em Eur. Phys. J. C}
  {\bfseries 2} (1998) 49--59},
  \href{http://arxiv.org/abs/hep-ex/9708027}{{\ttfamily arXiv:hep-ex/9708027}}.

\bibitem{LHCb:2013hzx}
{ LHCb} Collaboration, R.~Aaij {\em et~al.}, ``{Measurements of the
  $\Lambda_b^0 \to J/\psi \Lambda$ decay amplitudes and the $\Lambda_b^0$
  polarisation in $pp$ collisions at $\sqrt{s} = 7$~TeV},''
  \href{http://dx.doi.org/10.1016/j.physletb.2013.05.041}{{\em Phys. Lett. B}
  {\bfseries 724} (2013) 27--35},
  \href{http://arxiv.org/abs/1302.5578}{{\ttfamily arXiv:1302.5578 [hep-ex]}}.

\bibitem{LHCb:2020iux}
{ LHCb} Collaboration, R.~Aaij {\em et~al.}, ``{Measurement of the
  $\Lambda^0_b\rightarrow J/\psi\Lambda$ angular distribution and the
  $\Lambda^0_b$ polarisation in $pp$ collisions},''
  \href{http://dx.doi.org/10.1007/JHEP06(2020)110}{{\em JHEP} {\bfseries 06}
  (2020) 110}, \href{http://arxiv.org/abs/2004.10563}{{\ttfamily
  arXiv:2004.10563 [hep-ex]}}.

\bibitem{CMS:2018wjk}
{ CMS} Collaboration, A.~M. Sirunyan {\em et~al.}, ``{Measurement of the
  $\Lambda_b$ polarization and angular parameters in $\Lambda_b\to J/\psi\,
  \Lambda$ decays from pp collisions at $\sqrt{s}=$ 7 and 8 TeV},''
  \href{http://dx.doi.org/10.1103/PhysRevD.97.072010}{{\em Phys. Rev. D}
  {\bfseries 97} no.~7, (2018) 072010},
  \href{http://arxiv.org/abs/1802.04867}{{\ttfamily arXiv:1802.04867
  [hep-ex]}}.

\bibitem{Chen:1994ar}
K.~Chen, G.~R. Goldstein, R.~L. Jaffe, and X.-D. Ji, ``{Probing quark
  fragmentation functions for spin-1/2 baryon production in unpolarized
  $e^+e^-$ annihilation},''
  \href{http://dx.doi.org/10.1016/0550-3213(95)00193-V}{{\em Nucl. Phys. B}
  {\bfseries 445} (1995) 380--398},
  \href{http://arxiv.org/abs/hep-ph/9410337}{{\ttfamily arXiv:hep-ph/9410337}}.

\bibitem{Adamov:2000is}
A.~Adamov and G.~R. Goldstein, ``{Excited state contributions to the heavy
  baryon fragmentation functions in a quark-diquark model},''
  \href{http://dx.doi.org/10.1103/PhysRevD.64.014021}{{\em Phys. Rev. D}
  {\bfseries 64} (2001) 014021},
  \href{http://arxiv.org/abs/hep-ph/0009300}{{\ttfamily arXiv:hep-ph/0009300}}.

\bibitem{CMS-DP-2019-043}
{ CMS} Collaboration, ``{Recording and reconstructing 10 billion unbiased b
  hadron decays in CMS},'' Tech. Rep. CMS-DP-2019-043, CERN, Geneva, 2019.
\newblock \url{https://cds.cern.ch/record/2704495}.

\bibitem{Bainbridge:2020pgi}
{ CMS} Collaboration, R.~Bainbridge, ``{Recording and reconstructing 10 billion
  unbiased b hadron decays in CMS},''
  \href{http://dx.doi.org/10.1051/epjconf/202024501025}{{\em EPJ Web Conf.}
  {\bfseries 245} (2020) 01025}.

\bibitem{CMS:2024syx}
{ CMS} Collaboration, A.~Hayrapetyan {\em et~al.}, ``{Test of lepton flavor
  universality in B$^{\pm}$$\to$ K$^{\pm}\mu^+\mu^-$ and B$^{\pm}$$\to$
  K$^{\pm}$e$^+$e$^-$ decays in proton-proton collisions at $\sqrt{s}$ = 13
  TeV},'' \href{http://arxiv.org/abs/2401.07090}{{\ttfamily arXiv:2401.07090
  [hep-ex]}}.

\bibitem{Alonso:2021boj}
R.~Alonso, C.~Fraser-Taliente, C.~Hays, and M.~Spannowsky, ``{Prospects for
  direct CP tests of $hqq$ interactions},''
  \href{http://dx.doi.org/10.1007/JHEP08(2021)167}{{\em JHEP} {\bfseries 08}
  (2021) 167}, \href{http://arxiv.org/abs/2105.06879}{{\ttfamily
  arXiv:2105.06879 [hep-ph]}}.

\bibitem{Metz:2016swz}
A.~Metz and A.~Vossen, ``{Parton Fragmentation Functions},''
  \href{http://dx.doi.org/10.1016/j.ppnp.2016.08.003}{{\em Prog. Part. Nucl.
  Phys.} {\bfseries 91} (2016) 136--202},
  \href{http://arxiv.org/abs/1607.02521}{{\ttfamily arXiv:1607.02521
  [hep-ex]}}.

\bibitem{Stratmann:1996hn}
M.~Stratmann and W.~Vogelsang, ``{Next-to-leading order evolution of polarized
  and unpolarized fragmentation functions},''
  \href{http://dx.doi.org/10.1016/S0550-3213(97)00182-X}{{\em Nucl. Phys. B}
  {\bfseries 496} (1997) 41--65},
  \href{http://arxiv.org/abs/hep-ph/9612250}{{\ttfamily arXiv:hep-ph/9612250}}.

\bibitem{deFlorian:1997zj}
D.~de~Florian, M.~Stratmann, and W.~Vogelsang, ``{QCD analysis of unpolarized
  and polarized $\Lambda$-baryon production in leading and next-to-leading
  order},'' \href{http://dx.doi.org/10.1103/PhysRevD.57.5811}{{\em Phys. Rev.
  D} {\bfseries 57} (1998) 5811--5824},
  \href{http://arxiv.org/abs/hep-ph/9711387}{{\ttfamily arXiv:hep-ph/9711387}}.

\bibitem{Gustafson:1992iq}
G.~Gustafson and J.~Hakkinen, ``{$\Lambda$-polarization in
  $e^+e^-$-annihilation at the $Z^0$ pole},''
  \href{http://dx.doi.org/10.1016/0370-2693(93)91444-R}{{\em Phys. Lett. B}
  {\bfseries 303} (1993) 350--354}.

\bibitem{Sjostrand:2014zea}
T.~Sj\"ostrand, S.~Ask, J.~R. Christiansen, R.~Corke, N.~Desai, P.~Ilten,
  S.~Mrenna, S.~Prestel, C.~O. Rasmussen, and P.~Z. Skands, ``{An introduction
  to PYTHIA 8.2},'' \href{http://dx.doi.org/10.1016/j.cpc.2015.01.024}{{\em
  Comput. Phys. Commun.} {\bfseries 191} (2015) 159--177},
  \href{http://arxiv.org/abs/1410.3012}{{\ttfamily arXiv:1410.3012 [hep-ph]}}.

\bibitem{Albino}
S.~Albino, B.~A. Kniehl, and G.~Kramer, ``{AKK Update: Improvements from New
  Theoretical Input and Experimental Data},''
  \href{http://dx.doi.org/10.1016/j.nuclphysb.2008.05.017}{{\em Nucl. Phys. B}
  {\bfseries 803} (2008) 42--104},
  \href{http://arxiv.org/abs/0803.2768}{{\ttfamily arXiv:0803.2768 [hep-ph]}}.

\bibitem{Workman:2022ynf}
{ Particle Data Group} Collaboration, R.~L. Workman and Others, ``{Review of
  Particle Physics},'' \href{http://dx.doi.org/10.1093/ptep/ptac097}{{\em PTEP}
  {\bfseries 2022} (2022) 083C01}.

\bibitem{Manohar:1993qn}
A.~V. Manohar and M.~B. Wise, ``{Inclusive semileptonic $B$ and polarized
  $\Lambda_b$ decays from QCD},''
  \href{http://dx.doi.org/10.1103/PhysRevD.49.1310}{{\em Phys. Rev. D}
  {\bfseries 49} (1994) 1310--1329},
  \href{http://arxiv.org/abs/hep-ph/9308246}{{\ttfamily arXiv:hep-ph/9308246}}.

\bibitem{Lisovyi:2015uqa}
M.~Lisovyi, A.~Verbytskyi, and O.~Zenaiev, ``{Combined analysis of charm-quark
  fragmentation-fraction measurements},''
  \href{http://dx.doi.org/10.1140/epjc/s10052-016-4246-y}{{\em Eur. Phys. J. C}
  {\bfseries 76} no.~7, (2016) 397},
  \href{http://arxiv.org/abs/1509.01061}{{\ttfamily arXiv:1509.01061
  [hep-ex]}}.

\bibitem{LHCb:2022sck}
{ LHCb} Collaboration, R.~Aaij {\em et~al.}, ``{Amplitude analysis of the
  $\Lambda_c^+ \to p K^- \pi^+$ decay and $\Lambda_c^+$ baryon polarization
  measurement in semileptonic beauty hadron decays},''
  \href{http://dx.doi.org/10.1103/PhysRevD.108.012023}{{\em Phys. Rev. D}
  {\bfseries 108} no.~1, (2023) 012023},
  \href{http://arxiv.org/abs/2208.03262}{{\ttfamily arXiv:2208.03262
  [hep-ex]}}.

\bibitem{Czarnecki:1994pu}
A.~Czarnecki and M.~Jezabek, ``{Distributions of leptons in decays of polarized
  heavy quarks},'' \href{http://dx.doi.org/10.1016/0550-3213(94)90266-6}{{\em
  Nucl. Phys. B} {\bfseries 427} (1994) 3--21},
  \href{http://arxiv.org/abs/hep-ph/9402326}{{\ttfamily arXiv:hep-ph/9402326}}.

\bibitem{ATLAS:2016lxn}
{ ATLAS} Collaboration, M.~Aaboud {\em et~al.}, ``{Measurements of charge and
  CP asymmetries in $b$-hadron decays using top-quark events collected by the
  ATLAS detector in $pp$ collisions at $\sqrt{s}=8$~TeV},''
  \href{http://dx.doi.org/10.1007/JHEP02(2017)071}{{\em JHEP} {\bfseries 02}
  (2017) 071}, \href{http://arxiv.org/abs/1610.07869}{{\ttfamily
  arXiv:1610.07869 [hep-ex]}}.

\bibitem{CMS:2021scf}
{ CMS} Collaboration, A.~Tumasyan {\em et~al.}, ``{A new calibration method for
  charm jet identification validated with proton-proton collision events at
  $\sqrt{s} = 13$~TeV},''
  \href{http://dx.doi.org/10.1088/1748-0221/17/03/P03014}{{\em JINST}
  {\bfseries 17} no.~03, (2022) P03014},
  \href{http://arxiv.org/abs/2111.03027}{{\ttfamily arXiv:2111.03027
  [hep-ex]}}.

\bibitem{CMS:2023aim}
{ CMS} Collaboration, A.~Tumasyan {\em et~al.}, ``{Measurement of the
  production cross section for a W boson in association with a charm quark in
  proton\textendash{}proton collisions at $\sqrt{s} = 13\,\hbox {TeV}$},''
  \href{http://dx.doi.org/10.1140/epjc/s10052-023-12258-4}{{\em Eur. Phys. J.
  C} {\bfseries 84} (2024) 27},
  \href{http://arxiv.org/abs/2308.02285}{{\ttfamily arXiv:2308.02285
  [hep-ex]}}.

\bibitem{Cacciari:2008gp}
M.~Cacciari, G.~P. Salam, and G.~Soyez, ``{The anti-$k_t$ jet clustering
  algorithm},'' \href{http://dx.doi.org/10.1088/1126-6708/2008/04/063}{{\em
  JHEP} {\bfseries 04} (2008) 063},
  \href{http://arxiv.org/abs/0802.1189}{{\ttfamily arXiv:0802.1189 [hep-ph]}}.

\bibitem{Cacciari}
M.~Cacciari, G.~P. Salam, and G.~Soyez, ``{FastJet User Manual},''
  \href{http://dx.doi.org/10.1140/epjc/s10052-012-1896-2}{{\em Eur. Phys. J. C}
  {\bfseries 72} (2012) 1896}, \href{http://arxiv.org/abs/1111.6097}{{\ttfamily
  arXiv:1111.6097 [hep-ph]}}.

\bibitem{ALEPH:2001pfo}
{ ALEPH} Collaboration, A.~Heister {\em et~al.}, ``{Study of the fragmentation
  of b quarks into B mesons at the Z peak},''
  \href{http://dx.doi.org/10.1016/S0370-2693(01)00690-6}{{\em Phys. Lett. B}
  {\bfseries 512} (2001) 30--48},
  \href{http://arxiv.org/abs/hep-ex/0106051}{{\ttfamily arXiv:hep-ex/0106051}}.

\bibitem{DELPHI:2011aa}
{ DELPHI} Collaboration, J.~Abdallah {\em et~al.}, ``{A study of the b-quark
  fragmentation function with the DELPHI detector at LEP I and an averaged
  distribution obtained at the Z Pole},''
  \href{http://dx.doi.org/10.1140/epjc/s10052-011-1557-x}{{\em Eur. Phys. J. C}
  {\bfseries 71} (2011) 1557}, \href{http://arxiv.org/abs/1102.4748}{{\ttfamily
  arXiv:1102.4748 [hep-ex]}}.

\bibitem{OPAL:2002plk}
{ OPAL} Collaboration, G.~Abbiendi {\em et~al.}, ``{Inclusive analysis of the b
  quark fragmentation function in Z decays at LEP},''
  \href{http://dx.doi.org/10.1140/epjc/s2003-01229-x}{{\em Eur. Phys. J. C}
  {\bfseries 29} (2003) 463--478},
  \href{http://arxiv.org/abs/hep-ex/0210031}{{\ttfamily arXiv:hep-ex/0210031}}.

\bibitem{SLD:2002poq}
{ SLD} Collaboration, K.~Abe {\em et~al.}, ``{Measurement of the $b$-quark
  fragmentation function in $Z^0$ decays},''
  \href{http://dx.doi.org/10.1103/PhysRevD.66.079905}{{\em Phys. Rev. D}
  {\bfseries 65} (2002) 092006},
  \href{http://arxiv.org/abs/hep-ex/0202031}{{\ttfamily arXiv:hep-ex/0202031}}.
  [Erratum: \emph{Phys.\ Rev.\ D} \textbf{66} (2002) 079905].

\bibitem{ATLAS:2022miz}
{ ATLAS} Collaboration, G.~Aad {\em et~al.}, ``{Measurements of jet observables
  sensitive to $b$-quark fragmentation in $t\bar t$ events at the LHC with the
  ATLAS detector},'' \href{http://dx.doi.org/10.1103/PhysRevD.106.032008}{{\em
  Phys. Rev. D} {\bfseries 106} no.~3, (2022) 032008},
  \href{http://arxiv.org/abs/2202.13901}{{\ttfamily arXiv:2202.13901
  [hep-ex]}}.

\bibitem{ATLAS:2021agf}
{ ATLAS} Collaboration, G.~Aad {\em et~al.}, ``{Measurement of $b$-quark
  fragmentation properties in jets using the decay $B^{\pm} \to J/\psi K^{\pm}$
  in $pp$ collisions at $ \sqrt{s} $ = 13 TeV with the ATLAS detector},''
  \href{http://dx.doi.org/10.1007/JHEP12(2021)131}{{\em JHEP} {\bfseries 12}
  (2021) 131}, \href{http://arxiv.org/abs/2108.11650}{{\ttfamily
  arXiv:2108.11650 [hep-ex]}}.

\bibitem{CMS-PAS-TOP-18-012}
{ CMS} Collaboration, ``{Measurement of the shape of the b quark fragmentation
  function using charmed mesons produced inside b jets from
  $\mathrm{t}\bar{\mathrm{t}}$ pair decays},'' Tech. Rep. CMS-PAS-TOP-18-012,
  CERN, Geneva, 2021.
\newblock \url{https://cds.cern.ch/record/2771694}.

\bibitem{OPAL:1997edj}
{ OPAL} Collaboration, K.~Ackerstaff {\em et~al.}, ``{Measurement of $f(c \to
  D^{\ast+}X)$, $f(b \to D^{\ast+}X)$ and $\Gamma_{c\bar c}/\Gamma_{\rm had}$
  using $D^{\ast\pm}$ mesons},''
  \href{http://dx.doi.org/10.1007/s100520050095}{{\em Eur. Phys. J. C}
  {\bfseries 1} (1998) 439--459},
  \href{http://arxiv.org/abs/hep-ex/9708021}{{\ttfamily arXiv:hep-ex/9708021}}.

\bibitem{ALEPH:1999syy}
{ ALEPH} Collaboration, R.~Barate {\em et~al.}, ``{Study of charm production in
  $Z$ decays},'' \href{http://dx.doi.org/10.1007/s100520000421}{{\em Eur. Phys.
  J. C} {\bfseries 16} (2000) 597--611},
  \href{http://arxiv.org/abs/hep-ex/9909032}{{\ttfamily arXiv:hep-ex/9909032}}.

\bibitem{Cacciari:2005uk}
M.~Cacciari, P.~Nason, and C.~Oleari, ``{A study of heavy flavoured meson
  fragmentation functions in $e^+e^-$ annihilation},''
  \href{http://dx.doi.org/10.1088/1126-6708/2006/04/006}{{\em JHEP} {\bfseries
  04} (2006) 006}, \href{http://arxiv.org/abs/hep-ph/0510032}{{\ttfamily
  arXiv:hep-ph/0510032}}.

\bibitem{Dambach:2006ha}
S.~Dambach, U.~Langenegger, and A.~Starodumov, ``{Neutrino reconstruction with
  topological information},''
  \href{http://dx.doi.org/10.1016/j.nima.2006.08.144}{{\em Nucl. Instrum. Meth.
  A} {\bfseries 569} (2006) 824--828},
  \href{http://arxiv.org/abs/hep-ph/0607294}{{\ttfamily arXiv:hep-ph/0607294}}.

\bibitem{LHCb:2015eia}
{ LHCb} Collaboration, R.~Aaij {\em et~al.}, ``{Determination of the quark
  coupling strength $|V_{ub}|$ using baryonic decays},''
  \href{http://dx.doi.org/10.1038/nphys3415}{{\em Nature Phys.} {\bfseries 11}
  (2015) 743--747}, \href{http://arxiv.org/abs/1504.01568}{{\ttfamily
  arXiv:1504.01568 [hep-ex]}}.

\bibitem{LHCb:2020ist}
{ LHCb} Collaboration, R.~Aaij {\em et~al.}, ``{First Observation of the Decay
  $B_s^0 \to K^-\mu^+\nu_\mu$ and a Measurement of $|V_{ub}|/|V_{cb}|$},''
  \href{http://dx.doi.org/10.1103/PhysRevLett.126.081804}{{\em Phys. Rev.
  Lett.} {\bfseries 126} no.~8, (2021) 081804},
  \href{http://arxiv.org/abs/2012.05143}{{\ttfamily arXiv:2012.05143
  [hep-ex]}}.

\bibitem{Ciezarek:2016lqu}
G.~Ciezarek, A.~Lupato, M.~Rotondo, and M.~Vesterinen, ``{Reconstruction of
  semileptonically decaying beauty hadrons produced in high energy $pp$
  collisions},'' \href{http://dx.doi.org/10.1007/JHEP02(2017)021}{{\em JHEP}
  {\bfseries 02} (2017) 021}, \href{http://arxiv.org/abs/1611.08522}{{\ttfamily
  arXiv:1611.08522 [hep-ex]}}.

\bibitem{Dambach:2009wda}
S.~Dambach, \href{http://dx.doi.org/10.3929/ethz-a-005811400}{{\em {CMS pixel
  module readout optimization and study of the $B^0$ lifetime in the
  semileptonic decay mode}}}.
\newblock PhD thesis, Heidelberg U., 2009.

\bibitem{Fano:1983zz}
U.~Fano, ``{Pairs of two-level systems},''
  \href{http://dx.doi.org/10.1103/RevModPhys.55.855}{{\em Rev. Mod. Phys.}
  {\bfseries 55} (1983) 855--874}.

\bibitem{Dharmaratna:1996xd}
W.~G.~D. Dharmaratna and G.~R. Goldstein, ``{Single quark polarization in
  quantum chromodynamics subprocesses},''
  \href{http://dx.doi.org/10.1103/PhysRevD.53.1073}{{\em Phys. Rev. D}
  {\bfseries 53} (1996) 1073--1086}.

\bibitem{Bernreuther:1995cx}
W.~Bernreuther, A.~Brandenburg, and P.~Uwer, ``{Transverse polarization of top
  quark pairs at the Tevatron and the Large Hadron Collider},''
  \href{http://dx.doi.org/10.1016/0370-2693(95)01475-6}{{\em Phys. Lett. B}
  {\bfseries 368} (1996) 153--162},
  \href{http://arxiv.org/abs/hep-ph/9510300}{{\ttfamily arXiv:hep-ph/9510300}}.

\bibitem{Alwall:2014hca}
J.~Alwall, R.~Frederix, S.~Frixione, V.~Hirschi, F.~Maltoni, O.~Mattelaer,
  H.~S. Shao, T.~Stelzer, P.~Torrielli, and M.~Zaro, ``{The automated
  computation of tree-level and next-to-leading order differential cross
  sections, and their matching to parton shower simulations},''
  \href{http://dx.doi.org/10.1007/JHEP07(2014)079}{{\em JHEP} {\bfseries 07}
  (2014) 079}, \href{http://arxiv.org/abs/1405.0301}{{\ttfamily arXiv:1405.0301
  [hep-ph]}}.

\bibitem{Artoisenet:2012st}
P.~Artoisenet, R.~Frederix, O.~Mattelaer, and R.~Rietkerk, ``{Automatic
  spin-entangled decays of heavy resonances in Monte Carlo simulations},''
  \href{http://dx.doi.org/10.1007/JHEP03(2013)015}{{\em JHEP} {\bfseries 03}
  (2013) 015}, \href{http://arxiv.org/abs/1212.3460}{{\ttfamily arXiv:1212.3460
  [hep-ph]}}.

\bibitem{Ball:2013hta}
{ NNPDF} Collaboration, R.~D. Ball, V.~Bertone, S.~Carrazza, L.~Del~Debbio,
  S.~Forte, A.~Guffanti, N.~P. Hartland, and J.~Rojo, ``{Parton distributions
  with QED corrections},''
  \href{http://dx.doi.org/10.1016/j.nuclphysb.2013.10.010}{{\em Nucl. Phys. B}
  {\bfseries 877} (2013) 290--320},
  \href{http://arxiv.org/abs/1308.0598}{{\ttfamily arXiv:1308.0598 [hep-ph]}}.

\bibitem{BuarqueFranzosi:2019boy}
D.~Buarque~Franzosi, O.~Mattelaer, R.~Ruiz, and S.~Shil, ``{Automated
  predictions from polarized matrix elements},''
  \href{http://dx.doi.org/10.1007/JHEP04(2020)082}{{\em JHEP} {\bfseries 04}
  (2020) 082}, \href{http://arxiv.org/abs/1912.01725}{{\ttfamily
  arXiv:1912.01725 [hep-ph]}}.

\bibitem{ATL-PHYS-PUB-2019-005}
{ ATLAS} Collaboration, ``{Expected performance of the ATLAS detector at the
  High-Luminosity LHC},'' Tech. Rep. ATL-PHYS-PUB-2019-005, CERN, Geneva, 2019.
\newblock \url{https://cds.cern.ch/record/2655304}.

\bibitem{Collaboration:2759072}
{ CMS} Collaboration, ``{The Phase-2 Upgrade of the CMS Data Acquisition and
  High Level Trigger},'' Tech. Rep. CERN-LHCC-2021-007, CMS-TDR-022, CERN,
  Geneva, 2021.
\newblock \url{https://cds.cern.ch/record/2759072}.

\bibitem{Owen:2302730}
{ ATLAS} Collaboration, R.~E. Owen, ``{The ATLAS Trigger System},''.
  \url{https://cds.cern.ch/record/2302730}.

\bibitem{ATLAS:2020gty}
{ ATLAS} Collaboration, G.~Aad {\em et~al.}, ``{Performance of the ATLAS muon
  triggers in Run 2},''
  \href{http://dx.doi.org/10.1088/1748-0221/15/09/p09015}{{\em JINST}
  {\bfseries 15} no.~09, (2020) P09015},
  \href{http://arxiv.org/abs/2004.13447}{{\ttfamily arXiv:2004.13447
  [physics.ins-det]}}.

\bibitem{ATLAS:2021piz}
{ ATLAS} Collaboration, G.~Aad {\em et~al.}, ``{Configuration and performance
  of the ATLAS $b$-jet triggers in Run 2},''
  \href{http://dx.doi.org/10.1140/epjc/s10052-021-09775-5}{{\em Eur. Phys. J.
  C} {\bfseries 81} no.~12, (2021) 1087},
  \href{http://arxiv.org/abs/2106.03584}{{\ttfamily arXiv:2106.03584
  [hep-ex]}}.

\bibitem{ATLAS:2011xhu}
{ ATLAS} Collaboration, G.~Aad {\em et~al.}, ``{$K_S^0$ and $\Lambda$
  production in $pp$ interactions at $\sqrt{s}=0.9$ and $7$~TeV measured with
  the ATLAS detector at the LHC},''
  \href{http://dx.doi.org/10.1103/PhysRevD.85.012001}{{\em Phys. Rev. D}
  {\bfseries 85} (2012) 012001},
  \href{http://arxiv.org/abs/1111.1297}{{\ttfamily arXiv:1111.1297 [hep-ex]}}.

\bibitem{ATLAS:2012cvl}
{ ATLAS} Collaboration, G.~Aad {\em et~al.}, ``{Measurement of the
  $\Lambda_b^0$ lifetime and mass in the ATLAS experiment},''
  \href{http://dx.doi.org/10.1103/PhysRevD.87.032002}{{\em Phys. Rev. D}
  {\bfseries 87} no.~3, (2013) 032002},
  \href{http://arxiv.org/abs/1207.2284}{{\ttfamily arXiv:1207.2284 [hep-ex]}}.

\bibitem{ATLAS:2014swk}
{ ATLAS} Collaboration, G.~Aad {\em et~al.}, ``{Measurement of the
  parity-violating asymmetry parameter $\alpha_b$ and the helicity amplitudes
  for the decay $\Lambda_b^0\to J/\psi\Lambda^0$ with the ATLAS detector},''
  \href{http://dx.doi.org/10.1103/PhysRevD.89.092009}{{\em Phys. Rev. D}
  {\bfseries 89} no.~9, (2014) 092009},
  \href{http://arxiv.org/abs/1404.1071}{{\ttfamily arXiv:1404.1071 [hep-ex]}}.

\bibitem{ATLAS:2019pqg}
{ ATLAS} Collaboration, G.~Aad {\em et~al.}, ``{Measurement of $K_S^0$ and
  $\Lambda ^0$ production in $t \bar{t}$ dileptonic events in $pp$ collisions
  at $\sqrt{s} = 7$~TeV with the ATLAS detector},''
  \href{http://dx.doi.org/10.1140/epjc/s10052-019-7512-y}{{\em Eur. Phys. J. C}
  {\bfseries 79} no.~12, (2019) 1017},
  \href{http://arxiv.org/abs/1907.10862}{{\ttfamily arXiv:1907.10862
  [hep-ex]}}.

\bibitem{CMS:2011jlm}
{ CMS} Collaboration, V.~Khachatryan {\em et~al.}, ``{Strange Particle
  Production in pp Collisions at $\sqrt{s}=0.9$ and $7$~TeV},''
  \href{http://dx.doi.org/10.1007/JHEP05(2011)064}{{\em JHEP} {\bfseries 05}
  (2011) 064}, \href{http://arxiv.org/abs/1102.4282}{{\ttfamily arXiv:1102.4282
  [hep-ex]}}.

\bibitem{CMS:2019isl}
{ CMS} Collaboration, A.~M. Sirunyan {\em et~al.}, ``{Strange hadron production
  in pp and pPb collisions at $\sqrt{s_\mathrm{NN}} = 5.02$~TeV},''
  \href{http://dx.doi.org/10.1103/PhysRevC.101.064906}{{\em Phys. Rev. C}
  {\bfseries 101} no.~6, (2020) 064906},
  \href{http://arxiv.org/abs/1910.04812}{{\ttfamily arXiv:1910.04812
  [hep-ex]}}.

\bibitem{CMS:2020zzv}
{ CMS} Collaboration, A.~M. Sirunyan {\em et~al.}, ``{Study of excited
  $\Lambda_\mathrm{b}^0$ states decaying to $\Lambda_\mathrm{b}^0\pi^+\pi^-$ in
  proton-proton collisions at $\sqrt{s}=13$~TeV},''
  \href{http://dx.doi.org/10.1016/j.physletb.2020.135345}{{\em Phys. Lett. B}
  {\bfseries 803} (2020) 135345},
  \href{http://arxiv.org/abs/2001.06533}{{\ttfamily arXiv:2001.06533
  [hep-ex]}}.

\bibitem{CMS:2021rvl}
{ CMS} Collaboration, A.~M. Sirunyan {\em et~al.}, ``{Observation of a New
  Excited Beauty Strange Baryon Decaying to $\Xi^-_\mathrm{b} \pi^+ \pi^-$},''
  \href{http://dx.doi.org/10.1103/PhysRevLett.126.252003}{{\em Phys. Rev.
  Lett.} {\bfseries 126} no.~25, (2021) 252003},
  \href{http://arxiv.org/abs/2102.04524}{{\ttfamily arXiv:2102.04524
  [hep-ex]}}.

\bibitem{ATL-PHYS-PUB-2017-014}
{ ATLAS} Collaboration, ``{Performance of the reconstruction of large impact
  parameter tracks in the ATLAS inner detector},'' Tech. Rep.
  ATL-PHYS-PUB-2017-014, CERN, Geneva, 2017.
\newblock \url{https://cds.cern.ch/record/2275635}.

\bibitem{ATL-PHYS-PUB-2019-013}
{ ATLAS} Collaboration, ``{Performance of vertex reconstruction algorithms for
  detection of new long-lived particle decays within the ATLAS inner
  detector},'' Tech. Rep. ATL-PHYS-PUB-2019-013, CERN, Geneva, 2019.
\newblock \url{https://cds.cern.ch/record/2669425}.

\bibitem{ATLAS:2023nze}
{ ATLAS} Collaboration, G.~Aad {\em et~al.}, ``{Performance of the
  reconstruction of large impact parameter tracks in the inner detector of
  ATLAS},'' \href{http://dx.doi.org/10.1140/epjc/s10052-023-12024-6}{{\em Eur.
  Phys. J. C} {\bfseries 83} no.~11, (2023) 1081},
  \href{http://arxiv.org/abs/2304.12867}{{\ttfamily arXiv:2304.12867
  [hep-ex]}}.

\bibitem{ATLAS:2017tny}
{ ATLAS} Collaboration, M.~Aaboud {\em et~al.}, ``{Search for long-lived,
  massive particles in events with displaced vertices and missing transverse
  momentum in $\sqrt{s} = 13$~TeV $pp$ collisions with the ATLAS detector},''
  \href{http://dx.doi.org/10.1103/PhysRevD.97.052012}{{\em Phys. Rev. D}
  {\bfseries 97} no.~5, (2018) 052012},
  \href{http://arxiv.org/abs/1710.04901}{{\ttfamily arXiv:1710.04901
  [hep-ex]}}.

\bibitem{ATLAS:2019kpx}
{ ATLAS} Collaboration, G.~Aad {\em et~al.}, ``{Search for heavy neutral
  leptons in decays of $W$ bosons produced in 13~TeV $pp$ collisions using
  prompt and displaced signatures with the ATLAS detector},''
  \href{http://dx.doi.org/10.1007/JHEP10(2019)265}{{\em JHEP} {\bfseries 10}
  (2019) 265}, \href{http://arxiv.org/abs/1905.09787}{{\ttfamily
  arXiv:1905.09787 [hep-ex]}}.

\bibitem{ATLAS:2019fwx}
{ ATLAS} Collaboration, G.~Aad {\em et~al.}, ``{Search for displaced vertices
  of oppositely charged leptons from decays of long-lived particles in $pp$
  collisions at $\sqrt {s} = 13$~TeV with the ATLAS detector},''
  \href{http://dx.doi.org/10.1016/j.physletb.2019.135114}{{\em Phys. Lett. B}
  {\bfseries 801} (2020) 135114},
  \href{http://arxiv.org/abs/1907.10037}{{\ttfamily arXiv:1907.10037
  [hep-ex]}}.

\bibitem{ATLAS:2019jcm}
{ ATLAS} Collaboration, G.~Aad {\em et~al.}, ``{Search for long-lived neutral
  particles produced in $pp$ collisions at $\sqrt{s} = 13$~TeV decaying into
  displaced hadronic jets in the ATLAS inner detector and muon spectrometer},''
  \href{http://dx.doi.org/10.1103/PhysRevD.101.052013}{{\em Phys. Rev. D}
  {\bfseries 101} no.~5, (2020) 052013},
  \href{http://arxiv.org/abs/1911.12575}{{\ttfamily arXiv:1911.12575
  [hep-ex]}}.

\bibitem{ATLAS:2020xyo}
{ ATLAS} Collaboration, G.~Aad {\em et~al.}, ``{Search for long-lived, massive
  particles in events with a displaced vertex and a muon with large impact
  parameter in $pp$ collisions at $\sqrt{s} = 13$~TeV with the ATLAS
  detector},'' \href{http://dx.doi.org/10.1103/PhysRevD.102.032006}{{\em Phys.
  Rev. D} {\bfseries 102} no.~3, (2020) 032006},
  \href{http://arxiv.org/abs/2003.11956}{{\ttfamily arXiv:2003.11956
  [hep-ex]}}.

\bibitem{ATLAS:2020wjh}
{ ATLAS} Collaboration, G.~Aad {\em et~al.}, ``{Search for Displaced Leptons in
  $\sqrt{s} = 13$~TeV $pp$ Collisions with the ATLAS Detector},''
  \href{http://dx.doi.org/10.1103/PhysRevLett.127.051802}{{\em Phys. Rev.
  Lett.} {\bfseries 127} no.~5, (2021) 051802},
  \href{http://arxiv.org/abs/2011.07812}{{\ttfamily arXiv:2011.07812
  [hep-ex]}}.

\bibitem{ATLAS:2021jig}
{ ATLAS} Collaboration, G.~Aad {\em et~al.}, ``{Search for exotic decays of the
  Higgs boson into long-lived particles in $pp$ collisions at $\sqrt{s} =
  13$~TeV using displaced vertices in the ATLAS inner detector},''
  \href{http://dx.doi.org/10.1007/JHEP11(2021)229}{{\em JHEP} {\bfseries 11}
  (2021) 229}, \href{http://arxiv.org/abs/2107.06092}{{\ttfamily
  arXiv:2107.06092 [hep-ex]}}.

\bibitem{ATLAS:2023oti}
{ ATLAS} Collaboration, G.~Aad {\em et~al.}, ``{Search for long-lived, massive
  particles in events with displaced vertices and multiple jets in $pp$
  collisions at $\sqrt{s} = 13$~TeV with the ATLAS detector},''
  \href{http://dx.doi.org/10.1007/JHEP06(2023)200}{{\em JHEP} {\bfseries 2306}
  (2023) 200}, \href{http://arxiv.org/abs/2301.13866}{{\ttfamily
  arXiv:2301.13866 [hep-ex]}}.

\bibitem{Strebler:2022nkh}
{ ATLAS} Collaboration, T.~Strebler, ``{Expected tracking performance of the
  ATLAS Phase-II Inner Tracker Upgrade},''
  \href{http://dx.doi.org/10.22323/1.414.0665}{{\em PoS} {\bfseries ICHEP2022}
  (2022) 665}.

\bibitem{CMS:2019uws}
{ CMS} Collaboration, A.~M. Sirunyan {\em et~al.}, ``{Production of
  $\Lambda_\mathrm{c}^+$ baryons in proton-proton and lead-lead collisions at
  $\sqrt{s_\mathrm{NN}}=$ 5.02 TeV},''
  \href{http://dx.doi.org/10.1016/j.physletb.2020.135328}{{\em Phys. Lett. B}
  {\bfseries 803} (2020) 135328},
  \href{http://arxiv.org/abs/1906.03322}{{\ttfamily arXiv:1906.03322
  [hep-ex]}}.

\bibitem{CMS:2023frs}
{ CMS} Collaboration, A.~Tumasyan {\em et~al.}, ``{Study of charm hadronization
  with prompt $ {\Lambda}_{\textrm{c}}^{+} $ baryons in proton-proton and
  lead-lead collisions at $ \sqrt{s_{\textrm{NN}}} $ = 5.02 TeV},''
  \href{http://dx.doi.org/10.1007/JHEP01(2024)128}{{\em JHEP} {\bfseries 01}
  (2024) 128}, \href{http://arxiv.org/abs/2307.11186}{{\ttfamily
  arXiv:2307.11186 [nucl-ex]}}.

\bibitem{ATL-PHYS-PUB-2018-041}
{ ATLAS} Collaboration, ``{CP-violation measurement prospects in the $B^0_s \to
  J/\psi\phi$ channel with the upgraded ATLAS detector at the HL-LHC},'' Tech.
  Rep. ATL-PHYS-PUB-2018-041, CERN, Geneva, 2018.
\newblock \url{https://cds.cern.ch/record/2649881}.

\bibitem{ATL-PHYS-PUB-2018-032}
{ ATLAS} Collaboration, ``{Prospects for lepton flavour violation measurements
  in $\tau\rightarrow 3\mu$ decays with the ATLAS detector at the HL-LHC},''
  Tech. Rep. ATL-PHYS-PUB-2018-032, CERN, Geneva, 2018.
\newblock \url{https://cds.cern.ch/record/2647956}.

\bibitem{ATL-PHYS-PUB-2018-035}
{ ATLAS} Collaboration, ``{Evaluation of theoretical uncertainties for
  simplified template cross section measurements of $V$-associated production
  of the Higgs boson},'' Tech. Rep. ATL-PHYS-PUB-2018-035, CERN, Geneva, 2018.
\newblock \url{https://cds.cern.ch/record/2649241}.

\bibitem{CMS-PAS-FTR-18-013}
{ CMS} Collaboration, ``{Measurement of rare $B \to \mu^+\mu^-$ decays with the
  Phase-2 upgraded CMS detector at the HL-LHC},'' Tech. Rep.
  CMS-PAS-FTR-18-013, CERN, Geneva, 2018.
\newblock \url{https://cds.cern.ch/record/2650545}.

\bibitem{CMS-PAS-FTR-18-033}
{ CMS} Collaboration, ``{Study of the expected sensitivity to the
  P$\mathrm{_5'}$ parameter in the $\mathrm{B^0 \to K^{*0}}\mu^+\mu^-$ decay at
  the HL-LHC},'' Tech. Rep. CMS-PAS-FTR-18-033, CERN, Geneva, 2018.
\newblock \url{https://cds.cern.ch/record/2651298}.

\bibitem{ATLAS:2015thz}
{ ATLAS} Collaboration, G.~Aad {\em et~al.}, ``{Performance of $b$-Jet
  Identification in the ATLAS Experiment},''
  \href{http://dx.doi.org/10.1088/1748-0221/11/04/P04008}{{\em JINST}
  {\bfseries 11} no.~04, (2016) P04008},
  \href{http://arxiv.org/abs/1512.01094}{{\ttfamily arXiv:1512.01094
  [hep-ex]}}.

\bibitem{CMS:2017wtu}
{ CMS} Collaboration, A.~M. Sirunyan {\em et~al.}, ``{Identification of
  heavy-flavour jets with the CMS detector in pp collisions at 13 TeV},''
  \href{http://dx.doi.org/10.1088/1748-0221/13/05/P05011}{{\em JINST}
  {\bfseries 13} no.~05, (2018) P05011},
  \href{http://arxiv.org/abs/1712.07158}{{\ttfamily arXiv:1712.07158
  [physics.ins-det]}}.

\bibitem{ATLAS-CONF-2018-055}
{ ATLAS} Collaboration, ``{Calibration of the $b$-tagging eﬃciency on charm
  jets using a sample of $W$+$c$ events with $\sqrt{s}$ = 13 TeV ATLAS data},''
  Tech. Rep. ATLAS-CONF-2018-055, CERN, Geneva, 2018.
\newblock \url{https://cds.cern.ch/record/2652195}.

\bibitem{ATLAS:2022qxm}
{ ATLAS} Collaboration, G.~Aad {\em et~al.}, ``{ATLAS flavour-tagging
  algorithms for the LHC Run 2 $pp$ collision dataset},''
  \href{http://dx.doi.org/10.1140/epjc/s10052-023-11699-1}{{\em Eur. Phys. J.
  C} {\bfseries 83} no.~7, (2023) 681},
  \href{http://arxiv.org/abs/2211.16345}{{\ttfamily arXiv:2211.16345
  [physics.data-an]}}.

\bibitem{ATLAS:2017oal}
{ ATLAS} Collaboration, M.~Aaboud {\em et~al.}, ``{Search for long-lived
  charginos based on a disappearing-track signature in pp collisions at $
  \sqrt{s}=13 $ TeV with the ATLAS detector},''
  \href{http://dx.doi.org/10.1007/JHEP06(2018)022}{{\em JHEP} {\bfseries 06}
  (2018) 022}, \href{http://arxiv.org/abs/1712.02118}{{\ttfamily
  arXiv:1712.02118 [hep-ex]}}.

\bibitem{ATLAS:2022rme}
{ ATLAS} Collaboration, G.~Aad {\em et~al.}, ``{Search for long-lived charginos
  based on a disappearing-track signature using 136 fb$^{-1}$ of pp collisions
  at $\sqrt{s} = 13$~TeV with the ATLAS detector},''
  \href{http://dx.doi.org/10.1140/epjc/s10052-022-10489-5}{{\em Eur. Phys. J.
  C} {\bfseries 82} no.~7, (2022) 606},
  \href{http://arxiv.org/abs/2201.02472}{{\ttfamily arXiv:2201.02472
  [hep-ex]}}.

\bibitem{CMS:2019ybf}
{ CMS} Collaboration, A.~M. Sirunyan {\em et~al.}, ``{Searches for physics
  beyond the standard model with the $M_\mathrm{T2}$ variable in hadronic final
  states with and without disappearing tracks in proton-proton collisions at
  $\sqrt{s}=$ 13 TeV},''
  \href{http://dx.doi.org/10.1140/epjc/s10052-019-7493-x}{{\em Eur. Phys. J. C}
  {\bfseries 80} no.~1, (2020) 3},
  \href{http://arxiv.org/abs/1909.03460}{{\ttfamily arXiv:1909.03460
  [hep-ex]}}.

\bibitem{CMS:2020atg}
{ CMS} Collaboration, A.~M. Sirunyan {\em et~al.}, ``{Search for disappearing
  tracks in proton-proton collisions at $\sqrt{s} =$ 13 TeV},''
  \href{http://dx.doi.org/10.1016/j.physletb.2020.135502}{{\em Phys. Lett. B}
  {\bfseries 806} (2020) 135502},
  \href{http://arxiv.org/abs/2004.05153}{{\ttfamily arXiv:2004.05153
  [hep-ex]}}.

\bibitem{CMS:2023mny}
{ CMS} Collaboration, A.~Hayrapetyan {\em et~al.}, ``{Search for supersymmetry
  in final states with disappearing tracks in proton-proton collisions at
  $\sqrt{s} = 13$~TeV},'' \href{http://arxiv.org/abs/2309.16823}{{\ttfamily
  arXiv:2309.16823 [hep-ex]}}.

\bibitem{CMS:2012feb}
{ CMS} Collaboration, S.~Chatrchyan {\em et~al.}, ``{Identification of b-quark
  jets with the CMS experiment},''
  \href{http://dx.doi.org/10.1088/1748-0221/8/04/P04013}{{\em JINST} {\bfseries
  8} (2013) P04013}, \href{http://arxiv.org/abs/1211.4462}{{\ttfamily
  arXiv:1211.4462 [hep-ex]}}.

\bibitem{ATLAS:2022jbw}
{ ATLAS} Collaboration, G.~Aad {\em et~al.}, ``{Measurement of the top-quark
  mass using a leptonic invariant mass in $pp$ collisions at $\sqrt{s} =
  13$~TeV with the ATLAS detector},''
  \href{http://dx.doi.org/10.1007/JHEP06(2023)019}{{\em JHEP} {\bfseries 06}
  (2023) 019}, \href{http://arxiv.org/abs/2209.00583}{{\ttfamily
  arXiv:2209.00583 [hep-ex]}}.

\bibitem{HFLAV:2014fzu}
{ HFLAV} Collaboration, Y.~Amhis {\em et~al.}, ``{Averages of $b$-hadron,
  $c$-hadron, and $\tau$-lepton properties as of summer 2014},''
  \href{http://arxiv.org/abs/1412.7515}{{\ttfamily arXiv:1412.7515 [hep-ex]}}.

\bibitem{CMS:2023gfb}
{ CMS} Collaboration, A.~Hayrapetyan {\em et~al.}, ``{Development of the CMS
  detector for the CERN LHC Run 3},''
  \href{http://arxiv.org/abs/2309.05466}{{\ttfamily arXiv:2309.05466
  [physics.ins-det]}}.

\end{thebibliography}\endgroup
\endgroup

\end{document}